\documentclass[usenatbib]{mn2e}
\usepackage{graphicx,times}
\newcommand{\feh}{\mbox{[Fe/H]}}
\newcommand{\fefiv}{\mbox{Fe\,5270}}
\newcommand{\fesix}{\mbox{Fe\,5335}}

\newcommand{\hd}{\mbox{H$\delta$}}
\newcommand{\hda}{\mbox{H$\delta_A$}}
\newcommand{\mgb}{\mbox{Mg$_b$}}

\newcommand{\kms}{km~s$^{-1}$}
\newcommand{\oii}{\mbox{[O\,{\scriptsize II}]}}

\def\ga{\mathrel{\mathchoice {\vcenter{\offinterlineskip\halign{\hfil
$\displaystyle##$\hfil\cr>\cr\sim\cr}}}
{\vcenter{\offinterlineskip\halign{\hfil$\textstyle##$\hfil\cr
>\cr\sim\cr}}}
{\vcenter{\offinterlineskip\halign{\hfil$\scriptstyle##$\hfil\cr
>\cr\sim\cr}}}
{\vcenter{\offinterlineskip\halign{\hfil$\scriptscriptstyle##$\hfil\cr
>\cr\sim\cr}}}}}
\def\la{\mathrel{\mathchoice {\vcenter{\offinterlineskip\halign{\hfil
$\displaystyle##$\hfil\cr<\cr\sim\cr}}}
{\vcenter{\offinterlineskip\halign{\hfil$\textstyle##$\hfil\cr
<\cr\sim\cr}}}
{\vcenter{\offinterlineskip\halign{\hfil$\scriptstyle##$\hfil\cr
<\cr\sim\cr}}}
{\vcenter{\offinterlineskip\halign{\hfil$\scriptscriptstyle##$\hfil\cr
<\cr\sim\cr}}}}}

\title[Field Early--Type Galaxies in the FDF and WHDF]{The
Evolution of Field Early--Type Galaxies in the FDF and WHDF\footnotemark[2]}

\author[A. Fritz, A. B\"ohm and B.~L.~Ziegler]
{Alexander Fritz,$^{1,2,3}$\footnotemark[1] Asmus B\"ohm,$^{4}$
 and Bodo L.~Ziegler$^{5}$\\
$^{1}$Gemini Observatory, 670 N.\ A'ohoku Place,
Hilo, HI 96720, USA\\
$^{2}$Department of Physics and Astronomy, University of Pennsylvania,
209 South 33rd Street, Philadelphia, PA 19104-6396, USA\\
$^{3}$Institut f\"ur Astrophysik G\"ottingen, Friedrich-Hund-Platz 1,
37077 G\"ottingen, Germany\\
$^{4}$Astrophysikalisches Institut Potsdam, An der Sternwarte 16,
14482 Potsdam, Germany\\
$^{5}$European Southern Observatory, Karl-Schwarzschild-Strasse 2, 85748
Garching bei M\"unchen, Germany}

\date{\today}
\pagerange{\pageref{firstpage}--\pageref{lastpage}} \pubyear{0000}

\begin{document}

\maketitle

\label{firstpage}

\begin{abstract}
We explore the properties of 24 field early--type galaxies in the
redshift range $0.20<z<0.75$ down to $M_B\le-19.30$ in a sample extracted from
the FORS Deep Field and the William Herschel Deep Field. Target galaxies were
selected on the basis of a combination of luminosity, spectrophotometric
type, morphology and photometric redshift or broad--band colours. High
signal--to--noise intermediate--resolution spectroscopy has been acquired at the
Very Large Telescope, complemented by deep high-resolution imaging
with the Advanced Camera for Surveys onboard the \emph{HST} and additional
ground--based multi--band photometry. All galaxy spectra were observed under
sub--arcsecond conditions and allow to derive accurate kinematics and stellar
population properties of the galaxies.
To clarify the low level of star formation detected in some galaxies, we
identify the amount of active galactic nuclei (AGN) activity in our sample
using archive data of \textit{Chandra} and \textit{XMM-Newton} X-ray
surveys. Non of the galaxies in our sample were identified as secure AGN source 
based on their X-ray emission.
The rest-frame $B$ and $K$-band scaling relations of the Faber-Jackson relation
and the Fundamental Plane display a moderate evolution for the field
early--type galaxies. Lenticular (S0) galaxies feature on average a stronger
luminosity evolution and bluer rest-frame colours which can be explained that
they comprise more diverse stellar populations compared to elliptical galaxies.
The evolution of the FP can be interpreted as an average change in the
dynamical (effective) mass-to-light ratio of our galaxies as
$\langle \Delta \log{(M/L_B)}/z \rangle=-0.74\pm0.08$. 
The $M/L$ evolution of these field galaxies suggests a continuous mass
assembly of field early-type galaxies during the last 5 Gyr, that gets
support by recent studies of field galaxies up to $z\sim1$.
Independent evidence for recent star formation activity is provided by
spectroscopic (\oii\ emission, \hd) and photometric (rest--frame
broad--band colors) diagnostics.
Based on the \hd\ absorption feature we detect a weak residual
star formation for galaxies that accounts for 5\%--10\% in the
total stellar mass of these galaxies.
The co-evolution in the luminosity \emph{and} mass of our galaxies favours a
downsizing formation process. We find some evidence that our galaxies
experienced a period of star formation quenching, possible triggered by AGN
activity that is in good agreement with recent results on 
both observational and theoretical side.
\end{abstract}


\footnotetext[1]{E-mail: afritz@gemini.edu}
%\footnotetext[1]{Current address: Gemini Observatory, 670 N.\ A`ohoku Pl.,
%Hilo, HI 96720, USA; e-mail: afritz@gemini.edu}

\footnotetext[2]{Based on observations collected at the European Southern
Observatory, Cerro Paranal, Chile using the Very Large Telescope
under PIDs. 65.O-0049, 66.A-0547, 68.A-0013, 69.B-0278B and 70.B-0251A,
and on observations made with the NASA/ESA \emph{Hubble Space Telescope},
obtained at the Space Telescope Science Institute, which is operated by the
Association of Universities for Research in Astronomy, Inc., under NASA contract
NAS 5-26555. These observations are associated with programs 9502 and 9908.}

%%%%%%%%%%%%%%%%%%%%%%%%%%%%%%%%%%%%%%%%%%%%%%%%%%%%%%%%%%%%%%%%%%%%%%%%%%%%%%%%

\section{Introduction}\label{sec:intro}

Early--type galaxies (the family consisting of elliptical and lenticular
galaxies, hereafter E+S0 galaxies) represent one of the most important 
diagnostics within the cosmological picture of galaxy formation and evolution. 
E+S0 galaxies and bulges
comprise most of the baryonic (stellar) mass (50\% or more
of all stellar spectral types) in the present-day universe 
\citep[e.g.,][]{BMKW03} and reside in the most
massive dark matter haloes with halo masses up to
$M_{{\rm h}}\sim 4\times 10^{13}h^{-1}M_\odot$ at $z=0$,
hence these systems are the key ingredients for
understanding the star formation and the mass assembly history of the galaxy
population as a whole and for the structure formation starting from the
early primordial stages of the universe.

Early hierarchical structure formation models based on
Cold Dark Matter particles predict different evolutionary paths and assembly
time--scales for the densest environments of clusters of galaxies and the
lowest densities of isolated galaxies 
\citep[e.g.,][]{K96,BCFL98,SP99,CLBF00}.
From typical age variations of 43\% in the overall stellar content
between cluster and field galaxies, a late mass assembly for massive E+S0
galaxies of smaller sub--units is expected with the involved physical processes
(mainly star formation rates, gas cooling and heating)
poorly understood. As a consequence, the simplified baryonic recipes are
interpreted and implemented differently in the simulations \citep{BEM02}.
In this paradigm, field 
early--type galaxies are assembled through merging and accretion events of
smaller clumps and fragments on longer time--scales up to the recent past 
\citep{DKBWSE01,DLKW04}. The environment is therefore suggested to
play an important role to constrain the formation and evolution of E+S0
galaxies and their internal properties.
These predictions have significant imprints for the abundance and
mass distribution of an existing or even fully established
population of old massive systems at
$1\leq z\leq 2$ \citep{Gla04,BEC06,CDR06,Bor06,Con07} and also for recent
claims of the detection of their possible progenitors at high redshifts of
$z>2$ \citep{CSBI04,Krie06,CDR06,MSC07}.

Over the past few years theoretical semi-analytic models of galaxy formation
have been improved dramatically. By including a secondary star burst mode to
the previous quiescent star formation history, i.e. allowing a fraction of gas
to be directly transformed into stars in merging galaxies \citep{KH00,SPF01},
an increasing amount of supernova feedback and/or an
additional effect of Active Galactic Nuclei (AGN) feedback 
\citep{GDZSBD04,CSW06,BBM06,DLSWC06} 
or virial shocks of infalling gas in a spherical halo \citep{Bir07} to suppress
the remaining star formation activity, a much better agreement with the
observations has been achieved. The suppression of star formation in massive
halos using primarily AGN feedback has a strong impact at the top of the
luminosity function (LF) that results in a far less overproduction of massive
galaxies at $z=0$ than earlier simulations, while still providing a good
description of the rest of the observed LF up to $z\simeq1.5$
\citep{BBM06}. 
However, there is still a non-negligible excess of local
luminous galaxies detectable in the local simulated LF, which models adjust by
including additional internal dust obscuration or a relative inefficiency of
the stellar mass of merging galaxies that results in a population of
free-floating intergalactic stars \citep{MMBF06}.

Previous studies have quantified to detect the deviations from passive
evolutionary tracks as predicted by semi-analytic models in the evolution
of the Fundamental Plane \citep[FP;][]{DD87,Dre:87}. 
This relationship is constructed among the basic parameters of E+S0
galaxies, their size (the half-light effective radius $R_{\rm e}$ in kpc),
stellar kinetic energy (central velocity dispersion $\sigma$ in \kms) and
luminosity (average surface brightness within $R_{\rm e}$,
$\langle \mu_{\rm e}\rangle$, in units of
mag arcsec$^{-2}$) and provides a powerful tool to measure the 
the evolution of the mass-to-light ($M/L$) ratio as a function of redshift
without being subject to uncertainties in the photometric model estimates,
such as initial mass function or age and/or metallicity degeneracy. 
Using this observational technique  cluster early--type galaxies
formed the majority of their stars at redshifts $z\ge2$ and their stellar
populations follow a slow passive evolution since then
\citep[e.g.,][]{vDF96,BSZ98,vDS03,FZBSD05,JCFBBD06}.
However, the cosmic formation epoch and evolutionary history of isolated
field galaxies remains unclear. Indeed, galaxy formation models based on a
hierarchical structure growth envision for a given galaxy mass a dependence on
environment \citep{DKBWSE01}.
At a given redshift this environmental effect results in 
explicit age variations between E+S0 galaxies in clusters and   
in low--density regions, where cluster ellipticals show
an average luminosity--weighted age of 9.6~Gyr, whereas field E+S0s being
by $\sim$4~Gyr younger and displaying a broader age spread than their  
counterparts in clusters \citep{CLBF00}.
In addition, brighter (and hence more massive) field galaxies ($M_B<-20.1$)
should feature on average $\sim$2 Gyr younger ages than fainter ones and
comprise more metal rich $0.05\le\feh\,\le 0.1$~dex populations than their
cluster representatives \citep{TMBO05}, but no differences in their
element abundance ratios ($\alpha$/Fe).

So far, for the field galaxy population inconsistent results have been derived.
Some studies suggest an evolution similar to early--type galaxies
in clusters \citep{vDFKI01,Treu01a,vdV03,vDE03,ZTBBFM05},
but others prefer a faster mass assembly over a shorter
time period since $z\sim1$ \citep{TSCMB02,GFKIS03}.
Recently, \cite{T05} performed a comprehensive investigation of early-type
field galaxies within the GOODS-N field and detected evidence for a mass
dependent evolution. Similar results were derived by  \cite{vdW05} and
\cite{SVCL05} from an analysis of field E+S0 galaxies in the CDFS.
However, although these samples were drawn 
from the same field they rather found incomparable galaxy sizes and velocity
dispersions for individual objects (see \S~\ref{sec:litcmp} for more details).
Moreover, the different target selection procedures among the studies 
(e.g., magnitude, colour, morphology) might induce some inconsistency
in the final sample, for
example selecting galaxies with the reddest colours would favour the more
massive systems where variations are less pronounced or
using low signal-to-noise ($S/N$) spectra limits the detection post-starburst
galaxies or the contamination by bulge-dominated spiral galaxies. 
Furthermore, difficulties may arise from the small number statistics 
of some works and
the impact of cosmic variance remains yet to be unknown and hard to
distinguish from the used sample selection
\citep[e.g.,][but see \S~\ref{sec:litcmp}]{vdW05,SVCL05}.
There is clear evidence that the interpretations of measurements are highly
sensitive to selection effects and high-quality spectroscopic data is stringent
necessary to allow a careful analysis of the properties of galaxies
at high redshift.

In the present work, we aim to analyse in detail the assembly history and
the stellar populations of a carefully selected sample of 24
early-type field spheroidal galaxies down to $M_B\le-19.30$
in two deep sky surveys,
the FORS Deep Field and the William Herschel Deep Field, by using a combination
of high signal-to-noise VLT spectroscopy together with
deep ground-based multi--band imaging and high-resolution space-based photometry.
Although our study is restricted to small number statistics, our data is based
on high-quality spectra providing both precise kinematic and stellar population
properties. In a future work, we are aiming to disentangle both difficulties of
high redshift investigations related to small number statistics and cosmic
variance (A. Fritz et al. 2009a, in preparation).

The structure of this paper is as follows. 
In section \S2, the sample selection, observations and the kinematic
analysis of the spectroscopic data are described. The photometric
observations together with the derivation of structural parameters,
galaxy morphologies and spectrophotometric properties are presented in 
section \S3. Here also a discussion on available X-ray data is given.
The scaling relations for field early-type galaxies are presented in \S4,
and the implication of these results on their formation is discussed 
in \S5. A comparison with previous works is given in \S6, followed by a 
discussion in \S7. The main conclusions and results are summarised in \S8.
Throughout this article the concordance cosmology for a flat,
low-density Universe is assumed with $\Omega_{m}=0.3$, $\Omega_{\Lambda}=0.7$
and $H_0=70$\,km\,s$^{-1}$\,Mpc$^{-1}$.
Unless otherwise noted, all magnitudes are given in the Vega-system.

%%%%%%%%%%%%%%%%%%%%%%%%%%%%%%%%%%%%%%%%%%%%%%%%%%%%%%%%%%%%%%%%%%%%%%%%%%%%%%%%

%
\begin{figure}
%\resizebox{\hsize}{!}{\includegraphics[width=14.8cm,clip=]{whdfbcc.eps}}
\includegraphics[width=1.0\linewidth]{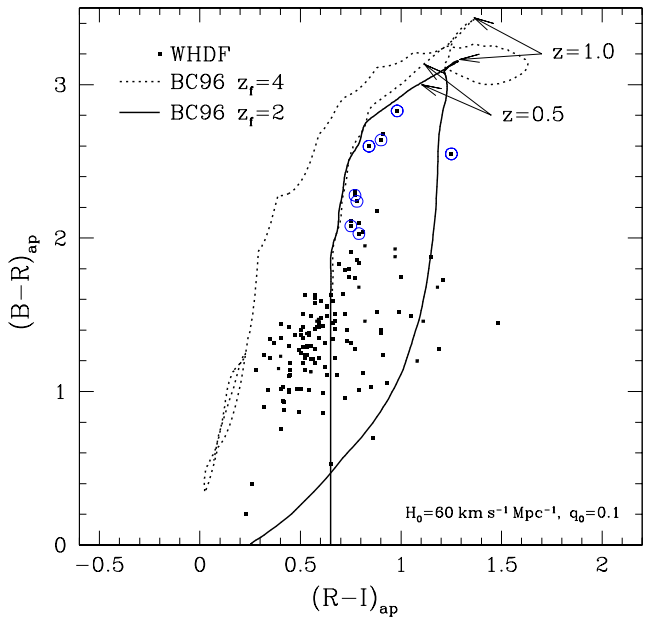}
\caption{\label{fig:cmdbrbi}
$(B-R)$--$(R-I)$ colour-colour diagram for the WHDF field 
galaxies. Spiral candidates (squares) are displayed with the final observed
field early-type galaxies (circles) and compared to evolutionary
tracks for E/S0 with different formation redshifts of $z_{\rm f}=2$
(solid line) and $z_{\rm f}=4$ (dotted line) predicted by the burst BC96 models.
For the BC96 models a cosmology with $H_0=60$\,km\,s$^{-1}$\,Mpc$^{-1}$ and
$q_0=0.1$ was adopted.}
\end{figure}

\section{Spectroscopy}

\subsection{Sample Selection}\label{sec:sel}

Target objects were selected based upon deep $UBgRI$ and $BRI$ multi-band
imaging surveys of the FORS Deep Field \citep[FDF,][]{Hei03} and William
Herschel Deep Field \citep[WHDF,][]{Met01}, respectively. 
The 50\% completeness limits of the FDF and WHDF are 
$B=27.7$ and $B=27.9$, respectively.
Early-type candidates were selected according to their 
(i) total apparent magnitude $R\le22.0^m$, (ii) spectrophotometric
type, (iii) elongated structureless appearance and (iv) photometric redshift
(only available for the FDF).
The constraint in the apparent brightness ensured a sufficient
signal-to-noise of $S/N$$\ge$15 in the absorption
lines, mandatory for a robust determination of velocity dispersions
and line-strengths measurements. For this reason, faint early-type candidates
with apparent magnitudes $R>20.5^m$ were included in more than one MOS
setup and the individual spectra combined after data reduction.

Spectrophotometric types and estimated redshifts
were selected from the FDF photometric redshifts catalogue
of more than 3800 objects \citep{Ben01}.
Only candidates with an early (E/S0) model Spectral Energy
Distribution (SED) were considered and the photometric redshifts were
restricted to $z_{\rm phot}\le0.6$. Based on the photometric redshifts,
the elliptical candidates were spread out on different CCD positions
among the spirals, which were observed for a similar project studying the
Tully--Fisher relation of late--type galaxies, for each MOS setup in the
observed spectral wavelength range such that either the \mgb-feature passband
($\lambda_0\approx5170$\,\AA) or the G-band ($\lambda_0\approx4300$\,\AA)
was included. 
For a reliable star/galaxy separation, targets with insignificant photometric
redshift and large uncertainty (i.e., $z_{\rm phot}-dz_{\rm phot}\le0$)
were discarded.
Field candidates were selected from a deep FORS2 $I$-band reference image
consisting of 10 seeing averaged $I$-band images (between 0.47--0.50\,arcsec
FWHM) with a final integration time of 3000\,sec and a 
Point-Spread-Function (PSF) of 0.49\,arcsec FWHM. In case of the FORS1
observations, the target selection was based on a FORS1 $I$-band reference
image with a PSF of 0.52\,arcsec FWHM. In the WHDF, field objects were drawn
from a deep over five individual exposures averaged $I$-band reference image 
with a total integration time of 1500\,sec and a PSF of 0.6\,arcsec FWHM. 
To avoid misclassifications, detections of the Source Extractor
package \citep[SExtractor;][]{BA96} that exhibited a star
classification parameter \texttt{star}\,$\ge$\,0.9 and 
$a$\,$\approx$\,$b$\,$\approx$\,2.5\,pixel were most likely
bonafide stellar objects and were therefore rejected.

Apart from photometric redshifts, the target selection for the WHDF field
early-type galaxies was performed with the same constraints as for the
FDF field elliptical galaxies. 
Redshifts of field early-type candidates were estimated using
a combination of colour-colour diagrams and apparent magnitudes.
Figure~\ref{fig:cmdbrbi} illustrates the target selection of the WHDF field
elliptical candidates in the $(B-R)$--$(R-I)$ colour-colour diagram. The
total WHDF spectroscopic data and the actually observed WHDF field ellipticals
are compared to evolutionary tracks for E/S0 with different formation redshifts
of $z_{\rm f}=2$ (solid line) and $z_{\rm f}=4$ (dotted line) 
as predicted by the passive evolution models for a simple
(single burst) stellar population of Bruzual \& Charlot
\footnote{For consistency with previous works
\citep[e.g.,][]{vDFKI01}, evolutionary models have been
computed using the Bruzual \& Charlot stellar population synthesis models 
(GISSEL96 version, hereafter BC96). Throughout the analysis a \cite{Sal55} 
initial mass function (IMF), solar metallicity and Kurucz atmosphere models are
adopted.}.
Note, that the BC96 models have been computed for a slightly different
cosmology with $H_0=60$\,km\,s$^{-1}$\,Mpc$^{-1}$ and $q_0=0.1$.
In the $(B-R)$--$(R-I)$ colour-colour diagram elliptical and spiral galaxies
\citep{BZ07}
are well separated up to $z\approx0.5$ (indicated by the arrows) and follow
the predictions of the evolutionary models quite well.
To avoid a selection of high redshift elliptical galaxies with $z>1$,
the apparent $R$ magnitude was used as an additional constraint.
Although the BC96 models represent more a generalisation of the various
stellar populations of early-type galaxies, together with the cut in magnitude 
they provide a good separation between high redshift interlopers and lower
redshift galaxies which we were interested in. More complex model star
formation histories \citep[cf.][]{GBSBS02} are beyond the scope of this paper. 
To allow a better estimate of the star formation histories of the
galaxies in our sample, we will utilise absorption line indices and other
spectral diagnostics in a future work (A. Fritz et al. 2009b, in preparation).
As shown in Figure~\ref{fig:cmdbrbi}, galaxies populating a narrow range
in the colour-colour diagram of $2\ga(B-R)\ga2.9$ and $0.7\ga(R-I)\ga1.1$
have been selected as bonafide field ellipticals and successfully verified.
One object (ID\,14) initially selected as an early-type candidate
with $(R-I)=0.79$, $(B-R)=1.78$ and $R=19.04^{m}$
($z=0.1060$) is an intermediate-type Sc spiral galaxy and was therefore
discarded. It turned out that this galaxy is a spiral galaxy with a clearly
visible disc on the ACS images which were not available at the time of the
target selection.

\subsection{Observations}\label{sec:obs}

Spectroscopic observations of the sample of field early-type galaxies were
performed simultaneously with those of the late-type galaxies which were
subject to the investigation of the evolution of the Tully--Fisher
relation in the FDF \citep{ZBF02,BZS04} and in
the WHDF \citep{BZ07}. As the target selections and instrument
configurations were very similar the two individual data sets are discussed
in combination.

Multi-object spectroscopy (MOS) was conducted at the Very Large Telescope
(VLT) using the FOcal Reducer and low dispersion Spectrograph 2  
\citep[FORS2;][]{App98} instrument in September and October 2000, July and
October 2002 and FORS1 in December 1999 and October 2001. During these
observations, a total of 220 galaxy spectra have been acquired under mostly
excellent seeing conditions. Using the grism \texttt{600R} with
the order separation filter \texttt{GG435$+$81} and slit width of 1.0\,arcsec
resulted in spectral resolutions of $R\approx 1200$ for the FDF observations.
For the WHDF run, the volume phased holographic grism \texttt{600RI} was
chosen, which is even more efficient at redder wavelengths
($>$8000\AA) than the grism \texttt{600R} and has a resolution of
$R\approx 1000$. Both versions of FORS instruments
offer 19 slitlets in the MOS configuration, 9 with slit lengths of 22\,arcsec,
8 with slit lengths of 20\,arcsec, whereas the uppermost and lowermost
slits have lengths of $\sim$\,11\,arcsec. The total integration time for 
each setup was 2.5\,hours. Seeing conditions ranged between 0.43\,arcsec and
0.92\,arcsec FWHM, with a median of 0.76\,arcsec. In order to limit the
corrections of atmospheric absorption, the airmass was constraint to
$A\leq2.0$. FORS was operated at standard resolution setup, i.e. CCD read out
in one-port mode. In low gain mode (using port A) FORS1 offers a gain of
3.51\,$e^-$/ADU and a read-out-noise (RON) of RON=7.21\,$e^-$, resulting
in a spatial scale of 0.2\,arcsec/pixel for the FDF spectra. 
After the FORS2 detector system upgrade which was available for science use
in March 2002, the standard setup offers a gain of 0.70\,$e^-$/ADU and
RON of 2.7\,$e^-$. The spatial sampling in the final WHDF spectra was
0.25\,arcsec/pixel.
More details on the observations can be found in \cite{ZTBBFM05}
and \cite{F06}.
Table~\ref{fdfeobs} summarises the spectroscopic observations of the
field early-type candidates in the FDF. 
Columns give the date of the observation, the mean airmass and 
mean Differential Image Motion Monitor (DIMM) seeing value during the
observations, the number of E/S0 galaxies and anonymous galaxies
(additional fill--up objects with no $z_{\rm phot}$)
for each MOS setup. Galaxies which were observed twice are
counted only once in the total number of objects.
A total of nine MOS setups
yielded a final sample of 39 E+S0 galaxies, regardless of their environment
(see further discussion in \S~\ref{sec:zdis}). 
Table~\ref{whdfeobs} gives an outline of the observations for the
field early-type galaxies in the WHDF.
In total, six different MOS setups with various MOS mask orientations
depending on the interested position angles of the spiral galaxies
were observed, in order to fulfill both requirements for spirals,
disc inclination angles of $i>40^\circ$ and restricted
misalignment angles of $\delta<15^\circ$ between apparent major axis
and slit orientation \citep[cf.][]{BZ07}. 
Each setup was splitted into three single exposures of
3000\,sec with a total integration time of 2.5 hours.
and typically comprised 2--3 early-type candidates.
Two additional objects (one in setup 4 and one in setup 1)
fell by coincidence into a slit of a target galaxy.
One object (ID\,810b) turned out to be a field S0 galaxy
at $z=0.2118$. The other object 
is a background spiral galaxy at $z=0.5569$ and thus was discarded.
Therefore, the total sample of WHDF elliptical galaxies comprises
eleven field early-type galaxies.

% -- FDF ellipses Tab obs ---
%{\scriptsize
\begin{table}
\centering
\caption{\label{fdfeobs}\small{VLT observations of early-type galaxies
in the FDF. Repeat observations are considered only once in the total number
of objects.}}
\vspace{0.3cm}
\begin{tabular}{lcccrc} 
\hline
\noalign{\smallskip}
Tel./Instrument &   Date  & airm. & DIMM  & E+S0 & anon. \cr
                & (m/d/y) &       &       &      &       \cr 
\noalign{\smallskip}
\hline
\noalign{\smallskip}
VLT/FORS2 & 09/27/00 & 1.21 & 0.51\arcsec\ &  5 & 1  \\
          & 09/27/00 & 1.33 & 0.43\arcsec\ &  5 & -  \\
	  & 10/04/00 & 1.40 & 0.81\arcsec\ &  6 & 1  \\
	  & 10/05/00 & 1.36 & 0.80\arcsec\ &  5 & 2  \\
	  & 10/05/00 & 1.28 & 0.74\arcsec\ &  8 & -  \\
	  & 10/06/00 & 1.15 & 0.66\arcsec\ &  7 & 2  \\
\medskip&   $\sum$    &	    &		   & 28 & 6  \\
VLT/FORS1 & 12/00/99 & 1.08 & 0.66\arcsec\ &  1 & 0 \\
          & 10/12/01 & 1.43 & 0.76\arcsec\ &  6 & 1 \\
          & 10/14/01 & 1.07 & 0.89\arcsec\ &  5 & 0 \\
          & 10/12/01 & 1.38 & 0.82\arcsec\ &  5 & 3 \\
          &   $\sum$ &	    &		   & 11 & 4 \\
\noalign{\smallskip}
\noalign{\hrule}
\end{tabular}
\end{table}
%

% -- WHDF ellipses Tab obs ---
%{\scriptsize
\begin{table}
\centering
\caption{\label{whdfeobs}\small{VLT/FORS2 observations of early-type galaxies
in the WHDF. Repeat observations are considered only once in the total number
of objects.}}
\vspace{0.3cm}
\begin{tabular}{clccrc} 
\hline
\noalign{\smallskip}
MOS &	Date  & airm. & DIMM  & E+S0 & anon. \cr
    & (m/d/y) &       &       &      &       \cr 
\noalign{\smallskip}
\hline
\noalign{\smallskip}
 1 & 07/10, 10/03/02  & 1.18 & 0.64\arcsec\ &  1 & 1 \\
 2 & 08/07, 09/11/02  & 1.19 & 0.73\arcsec\ &  2 & - \\
 3 & 08/04, 08/07/02  & 1.51 & 0.92\arcsec\ &  3 & - \\
 4 & 08/04, 10/02/02  & 1.13 & 0.89\arcsec\ &  3 & 1 \\
 5 & 09/10, 09/12/02  & 1.18 & 0.91\arcsec\ &  2 & - \\
 6 & 09/12, 10/4-5/02 & 1.14 & 0.85\arcsec\ &  2 & - \\
   & \hspace{0.6em} $\sum$   &      & 	    & 11 & 1	\\
\noalign{\smallskip}
\noalign{\hrule}
\end{tabular}
\end{table}

\subsection{Spectroscopic Analysis}\label{sec:specan}

\subsubsection{Data Reduction}\label{sec:red}

Image reduction followed the standard procedure of overscan subtraction, 
flat--fielding, sky background subtraction, and wavelength calibration
as described in \cite{FZBSD05} and \cite{F06}. Because spectra particularly 
near to the edges of the MOS frame exhibited spatial distortions, images of
each individual slitlet were extracted from the full frame after bias
subtraction allowing the typical two-dimensional image reduction of long-slit
spectroscopy. 
The night sky background was subtracted by iteratively fitting each CCD column
separately using a polynomial fit of first or zero order. During the wavelength
calibration, for the dispersion relation typical rms values of 0.04\,\AA\ 
at a dispersion of $\sim$1.62\,\AA/pixel were achieved. After the wavelength
calibration, spectra of the respective exposures were summed up and the
spectral rows corresponding to the full-width at half maximum (FWHM)
of an object were averaged along the dispersion direction using
a Horne-based algorithm \citep{Horne86}. We also compared this method to a
different slit extraction technique using equal-weight extractions which
yielded to the same velocity dispersion measurements within their
uncertainties \citep{F06}.
In those cases where a galaxy was observed in two different
MOS setups, the wavelength ranges covered did not match each other exactly
leading to varying final count rates at different wavelengths.

\begin{figure}
\centering
\includegraphics[width=1.0\linewidth]{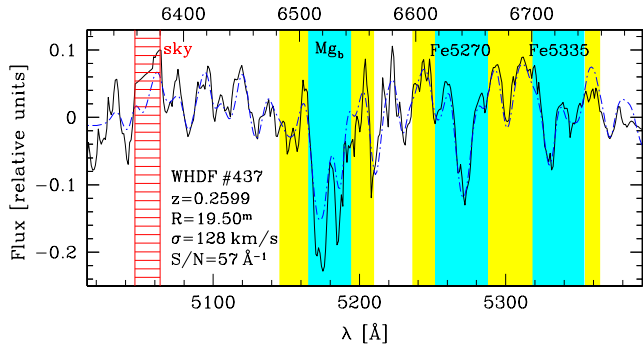}
\caption{\label{fig:fcqgs}\small{Example of a velocity dispersion
measurement for a typical S/N spectrum of a field galaxy in the WHDF.
The galaxy spectrum (solid line) is compared to the stellar template
(blue dot-dashed line) that was smeared to the $\sigma$ of the galaxy.
Absorption lines of \mgb, \fefiv\ and \fesix\ are shown, splitted into
their defined feature passbands (blue and red continuum and central passband),
indicated as shaded regions.
The hatched spectral region is affected by the telluric line at
$\lambda=6367$\,\AA\ and was masked out during the fit. The upper (lower)
abscissa displays the spectrum in observed (rest-frame)} wavelength units,
the inset shows the derived quantities.}
\end{figure}

\subsubsection{Velocity Dispersions}\label{sec:sig}

Internal velocity dispersions ($\sigma$) and radial velocities ($v_{{\rm rad}}$) 
of the galaxies were computed using an updated version of the 
\textit{Fourier Correlation Quotient} (FCQ) method \citep{Ben:90}, as described
in \cite{FZBSD05} and \cite{ZTBBFM05}. A brief summary follows.
The algorithm is based on the deconvolution of the peak of the template-galaxy
correlation function with the peak of the autocorrelation function of a
template star. In comparison to other procedures
\citep{SSBS77,TD79}, the main advantages of the
FCQ method are the possibility to \emph{a posteriori} chose the most
appropriate function to approximate the Line-Of-Sight-Velocity-Distribution
(LOSVD), a very high robustness against  template mismatch and 
a high insensitivity due to metallicity differences in the spectra.
For sufficient
high $S/N$ spectra of 10 per pixel, even velocity dispersions of
$0.5\times \sigma_i$, with $\sigma_i$ being the instrumental resolution, can be
determined with a high degree of accuracy.
For each galaxy, $\sigma$ and $v_{{\rm rad}}$ were measured 
around the spectral wavelength range of the G4300-band (hereafter G-band), 
the \mgb\   and the H$\beta$ feature 
in Fourier space by convolving a template spectrum with a Gaussian to match
the width of absorption lines in the observed galaxy spectrum of interest.
Spectral regions that suffered from strong night sky emission lines, sky subtraction
residuals defects or the A- or B-band absorption caused from the terrestrial
atmosphere were carefully masked out during the fitting process. In 10 out
of 24 cases, at least two different wavelength regimes could be analysed.
For all early-type galaxy candidates the signal-to-noise in the spectra 
was sufficient to yield very stable stellar velocity dispersion measurements
(see Appendix \S~\ref{sigcheck} for a discussion).
Deviations in $\sigma$ from template to template are small and 
ranged from 2 to 20\,\kms.
Since only one good template star (SAO\,162947) was available that was 
observed under the same FORS instrumental configurations as the galaxies, 
three more K giant stars (SAO\,32042, SAO\,80333 \& SAO\,98087) were used 
which had been observed with MOSCA at the 3.5m-telescope at the Calar Alto 
observatory \citep[][A. Fritz et al. 2009c, in preparation]{F06}.
These templates have sufficiently resolved spectra with instrumental
resolution of $\sigma_*\approx55$\,\kms\ around
\mgb\ and $\sigma_*\approx69$\,\kms\ around the G-band.

\begin{figure}
\includegraphics[width=1.0\linewidth]{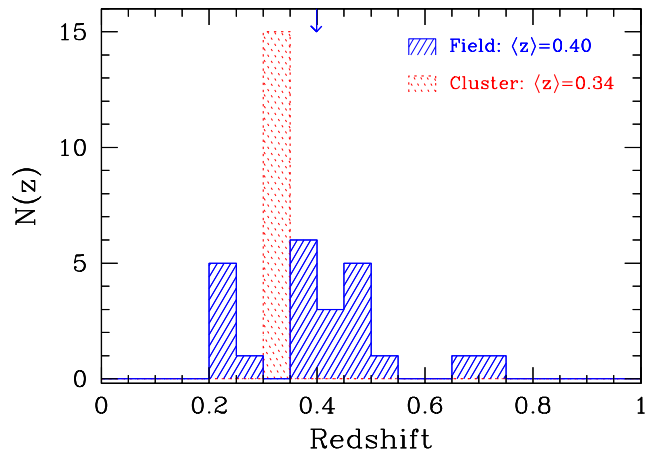}
\caption{\label{fig:fwzdb}\small{
Redshift distribution of the early--type galaxies within the
FDF and WHDF. The median redshift of the total field early--type galaxy
sample is $\langle z\rangle=0.40$ and is indicated with an arrow.
Early--type galaxies within the FDF that were spectroscopically confirmed to be
members of a cluster at $z\approx0.34$ are also shown. Note that these cluster
galaxies are not entering our field sample.}}
\end{figure}

To determine the instrumental resolution of the galaxy spectra,
the widths of 12--14 unblended and not-saturated HgAr/Ne emission lines
of the respective wavelength calibration frames were measured at
wavelengths corresponding to \mgb\ and the G-band separately. 
In case of the FDF observations, the HgCd lamp has
been used additionally to the He, Ar and Ne lamps to warrant better efficiency
in the blue wavelength range which is particularly neccessary for the higher
redshift spiral galaxies. Typical instrumental resolutions for the
observed spectra fall in the range of $\sigma_i\approx85-100$\,\kms.
To ensure a reliable error treatment for the velocity dispersions
and radial velocity measurements and to derived the signal-to-noise of
each galaxy spectra Monte-Carlo simulations for different stellar templates 
over a range of input $\sigma$ and $S/N$ ratios 
were performed \citep{F06}. For $S/N<8$ per \AA\ the velocity dispersions are
systematically underestimated on average by 4\%, but for S/N$\ge$8 per \AA\
the effect becomes negligible.
To check the reliablility and accuracy of our velocity dispersion
measurements, serveral independent tests have been performed. These
verifications are presented in \S~\ref{sigcheck} of the Appendix.

An example of the velocity dispersion measurement 
for a field elliptical galaxy in the WHDF with typical S/N spectrum
at $z=0.2599$ is given in Figure~\ref{fig:fcqgs}. 
Several template star spectra were smoothed to the same resolution as
the galaxy spectrum to derive $\sigma$. 
The velocity dispersion for this galaxy is $\sigma_{f}=128\pm8$~\kms\ 
derived within the extracted spectral region around the \mgb\ feature 
with a signal-to-noise level of $S/N=57$ \AA$^{-1}$ 
(see Figure~\ref{fig:fcqgs}). In this case, the final $\sigma_{f}$
was averaged over three individual measurements $\sigma_{1}=127\pm8$~\kms,
$\sigma_{2}=134\pm8$~\kms\ and $\sigma_{3}=124\pm10$~\kms\ 
which are in excellent agreement as the scatter between the values 
varies only within their uncertainties.

The measured luminosity--weighted average velocity dispersions 
$\sigma_{{\rm ap}}$ were
aperture corrected as $\sigma=(1.15\pm0.04)\sigma_{{\rm ap}}$
\citep[similar to][but adopting a slope of -0.06]{JFK95b} to yield
central velocity dispersions within a circular aperture of radius $R_e/8$.
For each galaxy, its aperture during the observations was individually
computed using the slit width, the number of extracted rows over which the
spectrum was averaged and the respective pixel scale
\citep[see][for a description]{FZBSD05}.

\subsubsection{Redshift and Spatial Distribution}\label{sec:zdis}

\begin{figure}
\includegraphics[width=1.0\linewidth]{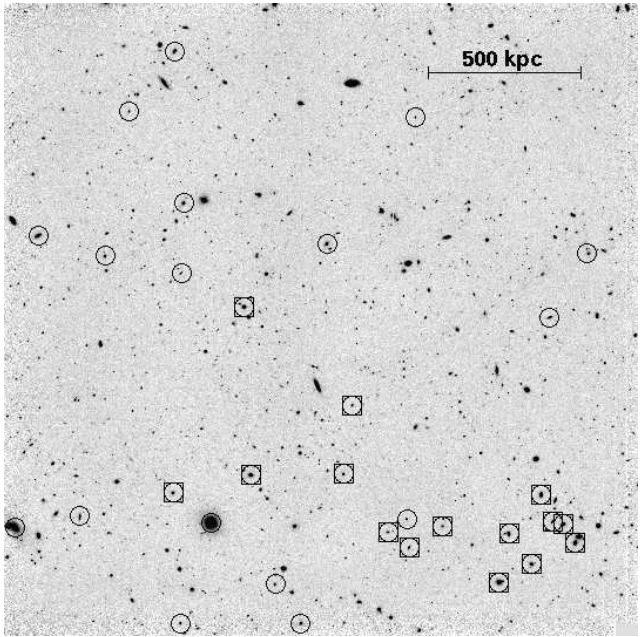}
\caption{\label{fig:fdfed}\small{
Distribution of the field early--type galaxies within the FDF and
possible members of a cluster in the southwestern (lower right) corner.
Field early--type galaxies are shown as circles, cluster 
early--type candidates are denoted additionally by squares.
The bar corresponds to 500\,kpc in projection at the cluster redshift of
$z=0.33$.
}}
\end{figure}

\begin{figure}
\includegraphics[width=1.0\linewidth]{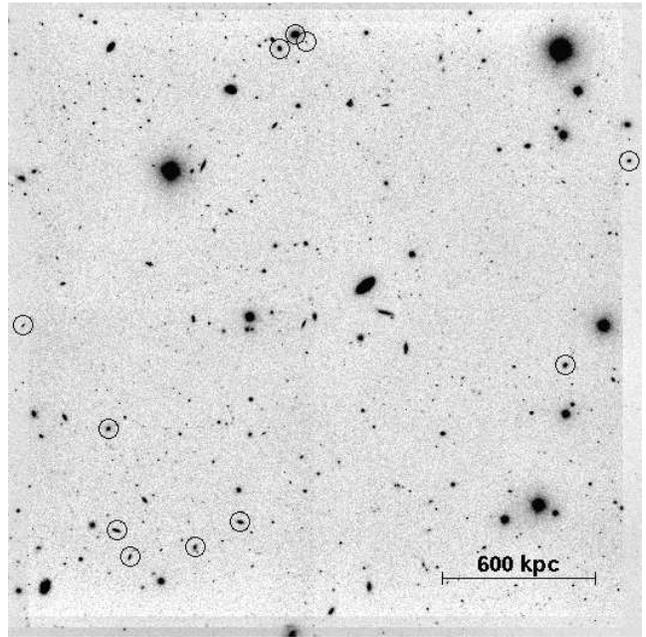}
\caption{\label{fig:whdfed}\small{
Distribution of the field early--type galaxies within the WHDF.
The bar corresponds to 600\,kpc in projection at a redshift of
$z=0.33$. The three galaxies located in
the northern (upper middle) part of the WHDF are not members of a group. 
North is up and east is to the left.
}}
\end{figure}

Figure~\ref{fig:fwzdb} shows the redshift distribution of all spectroscopically
confirmed field early--type galaxies detected within the FDF and WHDF.
Early--type galaxies in the FDF that were spectroscopically identified to be 
members of a cluster at $z\approx0.33$ (see next paragraph) are indicated as
well. The VLT spectra of these cluster E+S0 candidates are of similar high
quality to the spectra obtained for the field galaxies.
The 13 field E+S0 galaxies in the FDF encompass a range in redshifts of 
$0.22\nobreak\le z_{\rm FDF}\nobreak\le\nobreak0.65$ with a mean of
$\overline{z}_{\rm FDF}=0.42\pm0.11$ ($\langle z_{\rm FDF}\nobreak\rangle\nobreak=\nobreak0.41$)
and the 11 field galaxies in the WHDF cover a redshift space of
$0.21\nobreak\le z_{\rm WHDF}\nobreak\le\nobreak0.74$ with a mean of 
$\overline{z}_{\rm WHDF}\nobreak=\nobreak0.36\nobreak\pm\nobreak0.16$
($\langle z_{\rm WHDF}\rangle\nobreak=\nobreak0.40$).
The total field sample of 24 E+S0 galaxies exhibits a mean redshift of
$\overline{z}\nobreak=\nobreak0.39\nobreak\pm\nobreak0.14$ and a
median of $\langle z\rangle\nobreak=\nobreak0.40$.

It turned out that the southwestern corner of the FDF most probably
covers the outskirts of a galaxy cluster at $z=0.335\pm0.002$
\citep{ZTBBFM05}.
Based on the radial velocity measurements for the elliptical galaxy 
candidates in the FDF, the lower limit for the velocity dispersion 
of the cluster is $\sigma_{\rm c}\ga430$\,\kms, which is most likely
only a lower limit as the cluster centre is not located on
the FDF but only its outskirts. Allowing for a spread in redshift space
of $\Delta z=0.04$, which corresponds to $\sim$\,6 times the typical velocity
dispersion of a rich  galaxy cluster ($\sigma_{\rm c}\approx 10^3$\,\kms),
it is found that a total of 13 spiral galaxies and 15 early-type galaxies 
based on their spectroscopic redshifts are likely to be 
members of this cluster. This is also visible through a small 
break in redshift space at $z\approx0.3$ for the 341 galaxies at
$0.1<z<5$ (144 with $z<1.1$) of the total spectroscopic catalog \citep{NMA04}.

The spatial distribution of all early-type field galaxies in the FDF is shown
in Figure~\ref{fig:fdfed}. Potential cluster members are additionally denoted 
by squares. Adding the possible spiral cluster candidates, it is suggestive 
that the cluster has an elongated shape. If the center indeed is positioned to
the southwestern corner of the FDF, the distribution of the galaxies will be
a striking example of the morphology--density relation \citep{Dre80}
with the early--type galaxies mainly populating the dense, inner region
and the late--type galaxies located at larger clustercentric radii. 
Unfortunately at a redshift of $z=0.33$, the Mg\,5170 absorption line is
strongly affected due to the terrestrial absorption of the B band, which 
makes an accurate measurement of the internal galaxy velocity dispersions 
impossible. For this reason, early-type galaxies which are
possible members of a cluster at $z\approx0.33$ will be discarded
from the further analysis. Nevertheless, these E+S0 cluster galaxies
will be used as a reference for comparison reasons at a later stage,
see \S~\ref{stpops}. The coordinates, redshifts and rest-frame $(B-I)$ 
colours of the cluster galaxies are presented in Table~\ref{ccoord}.

\begin{table}
\begin{center}
\caption{\label{fwcoord}\small{Coordinates and redshifts of the field 
early-type galaxies in the FDF and WHDF}}
\vspace{0.1cm} 
\begin{tabular}{lccc}
\hline
\noalign{\smallskip}
ID & z$_{{\rm spec}}$ & $\alpha$  & $\delta$\\
   &   & (J2000.0) & (J2000.0)\\
\noalign{\smallskip}
\hline
\noalign{\smallskip}
FDF-1161 & 0.3959 & 01 05 51.689 & -25 45 45.965 \\ 
FDF-4285 & 0.3956 & 01 06 02.961 & -25 44 55.832 \\ 
FDF-5011 & 0.6538 & 01 06 05.549 & -25 48 48.886 \\ 
FDF-5908 & 0.2237 & 01 06 08.796 & -25 48 06.816 \\ 
FDF-6307 & 0.4513 & 01 06 10.208 & -25 44 28.225 \\ 
FDF-6336 & 0.4593 & 01 06 10.334 & -25 45 16.169 \\ 
FDF-6338 & 0.4100 & 01 06 10.329 & -25 49 15.565 \\ 
FDF-6439 & 0.3955 & 01 06 10.669 & -25 42 44.437 \\ 
FDF-7116 & 0.4567 & 01 06 12.961 & -25 43 25.565 \\ 
FDF-7459 & 0.5386 & 01 06 14.172 & -25 45 04.429 \\ 
FDF-7796 & 0.4089 & 01 06 15.428 & -25 48 02.843 \\ 
FDF-8372 & 0.2290 & 01 06 17.554 & -25 44 50.593 \\ 
FDF-8626 & 0.4081 & 01 06 18.705 & -25 48 10.266 \\ 
WHDF-92   & 0.3988 & 00 22 42.293 & 00 18 27.750  \\
WHDF-111  & 0.7418 & 00 22 39.393 & 00 18 33.610  \\
WHDF-158  & 0.4825 & 00 22 42.873 & 00 18 44.700  \\
WHDF-173  & 0.4507 & 00 22 37.386 & 00 18 50.560  \\
WHDF-318  & 0.3973 & 00 22 43.251 & 00 19 52.580  \\
WHDF-437  & 0.2599 & 00 22 22.977 & 00 20 34.840  \\
WHDF-508  & 0.3977 & 00 22 47.040 & 00 21 01.450  \\
WHDF-749  & 0.2130 & 00 22 20.162 & 00 22 50.460  \\
WHDF-810  & 0.2133 & 00 22 34.984 & 00 24 14.120  \\
WHDF-946  & 0.2123 & 00 22 35.664 & 00 24 04.960  \\
WHDF-810b & 0.2118 & 00 22 34.286 & 00 24 09.670  \\
\noalign{\smallskip}
\noalign{\hrule}
\end{tabular}
\end{center}
%\end{minipage}
\end{table}
%

% Table with Coordinates and Properties of cluster galaxies 
\begin{table}
\begin{center}
\caption{\label{ccoord}\small{Properties of identified cluster early-type 
galaxies in the FDF}}
\vspace{0.1cm} 
\begin{tabular}{lccccc}
\hline
\noalign{\smallskip}
ID & z$_{{\rm spec}}$ & $\alpha$  & $\delta$ & $B_{{\rm tot}}$ & $B-I$\\
   &   & (J2000.0) & (J2000.0)& (mag) & (rest) \\
\noalign{\smallskip}
\hline
\noalign{\smallskip}
FDF-822  & 0.3357 & 01 05 50.371 & -25 48 19.922 & 23.17 & 2.11 \\
FDF-980  & 0.3340 & 01 05 50.986 & -25 48 07.142 & 24.01 & 2.07 \\
FDF-1114 & 0.3345 & 01 05 51.520 & -25 48 05.462 & 22.70 & 2.15 \\
FDF-1265 & 0.3340 & 01 05 52.103 & -25 47 47.219 & 23.33 & 2.12 \\
FDF-1371 & 0.3356 & 01 05 52.579 & -25 48 34.480 & 22.52 & 2.21 \\
FDF-1697 & 0.3365 & 01 05 53.736 & -25 48 13.855 & 22.63 & 2.24 \\
FDF-1836 & 0.3325 & 01 05 54.221 & -25 48 47.082 & 25.16 & 3.84 \\
FDF-2580 & 0.3390 & 01 05 57.080 & -25 48 08.880 & 22.67 & 2.18 \\
FDF-3051 & 0.3335 & 01 05 58.775 & -25 48 23.506 & 22.43 & 2.28 \\
FDF-3365 & 0.3312 & 01 05 59.842 & -25 48 13.029 & 23.92 & 2.04 \\
FDF-3909 & 0.3378 & 01 06 01.661 & -25 46 46.379 & 22.19 & 2.11 \\
FDF-4030 & 0.3360 & 01 06 02.098 & -25 47 33.302 & 23.48 & 2.18 \\
FDF-5362 & 0.3318 & 01 06 06.786 & -25 47 34.088 & 25.22 & 3.80 \\
FDF-5446 & 0.3355 & 01 06 07.131 & -25 45 39.069 & 23.67 & 1.76 \\
FDF-6457 & 0.3360 & 01 06 10.720 & -25 47 46.366 & 25.28 & 1.61 \\
\noalign{\smallskip}
\noalign{\hrule}
\end{tabular}
\end{center}
%\end{minipage}
\end{table}

\begin{table*}
\begin{center}
\caption{\label{phy}\small{Physical Properties of the field 
early-type galaxies in the FDF and WHDF. The velocity dispersions
$\sigma$ are not corrected for aperture. The $S/N$ is per \AA\ (1.6 \AA\ pixel$^{-1}$).
The error on $M/L$ (in solar units) is the same as the
error on the mass $M$.}}
\vspace{0.1cm} 
\begin{tabular}{lccccccccccc}
\hline
\noalign{\smallskip}
ID & z$_{{\rm spec}}$ & $\sigma$ & $S/N$ & $B_{{\rm tot}}$ &
$\log R_{{\rm e}}$  & $R_{{\rm e}}$ & $\langle \mu_{{\rm e}}\rangle_{B}$ & $\log M$ & 
$\log({M/L_B})$ & $\Delta\ln({M/L_B})$ & $B-I$ \\
   &  & (\kms) & (\AA$^{-1}$) & (mag) & (kpc) & (arcsec) & (mag arcsec$^{-2}$) & ($M_{\odot}$) & ($M_{\odot}$/$L_{\odot}$) & & (rest) \\
\noalign{\smallskip}
\hline
\noalign{\smallskip}
FDF-1161 & 0.3959 & 125$\pm$16 &  19 & 23.41 &  0.33$\pm$0.07 &   0.40  &  20.30$\pm$0.12   & 10.65$\pm$0.06 & 0.38 & $-$0.68$\pm$0.15 &   1.87  \\
FDF-4285 & 0.3956 & 159$\pm$16 &  29 & 23.19 &  0.33$\pm$0.07 &   0.40  &  20.30$\pm$0.13   & 10.87$\pm$0.05 & 0.51 & $-$0.30$\pm$0.13 &   1.87  \\
FDF-5011 & 0.6538 &  82$\pm$13 &  15 & 24.84 &  0.54$\pm$0.08 &   0.50  &  20.59$\pm$0.15   & 11.46$\pm$0.04 & 0.87 &    0.00$\pm$0.11 &   1.94  \\
FDF-5908 & 0.2237 & 286$\pm$05 & 127 & 19.71 &  0.95$\pm$0.06 &   2.50  &  21.51$\pm$0.11   & 11.98$\pm$0.04 & 1.03 & $-$0.05$\pm$0.09 &   1.87  \\
FDF-6307 & 0.4513 & 185$\pm$14 &  29 & 23.52 &  0.53$\pm$0.07 &   0.59  &  20.79$\pm$0.08   & 11.21$\pm$0.05 & 0.74 & $-$0.17$\pm$0.12 &   1.91  \\
FDF-6336 & 0.4593 & 140$\pm$43 &   8 & 24.69 &  0.46$\pm$0.08 &   0.50  &  21.79$\pm$0.15   & 10.91$\pm$0.12 & 0.96 &    0.52$\pm$0.00 &   1.99  \\
FDF-6338 & 0.4100 & 164$\pm$19 &  22 & 23.89 &  --            &   --    &    --       &        --      &  --  &        --      &   1.88  \\
FDF-6439 & 0.3955 & 200$\pm$15 &  41 & 22.44 &  0.27$\pm$0.06 &   0.35  &  19.21$\pm$0.09   & 11.01$\pm$0.05 & 0.42 & $-$0.76$\pm$0.12 &   2.07  \\
FDF-7116 & 0.4567 & 123$\pm$21 &  28 & 24.21 &  0.16$\pm$0.07 &   0.25  &  19.68$\pm$0.10   & 10.47$\pm$0.07 & 0.28 & $-$0.77$\pm$0.18 &   1.70  \\
FDF-7459 & 0.5386 & 172$\pm$23 &  20 & 24.28 &  --            &   --    &	--   &        --      &  --  &        --      &   1.85  \\
FDF-7796 & 0.4089 & 106$\pm$33 &  15 & 23.65 &  0.61$\pm$0.08 &   0.75  &  21.80$\pm$0.17   & 10.80$\pm$0.13 & 0.57 & $-$0.34$\pm$0.31 &   2.02  \\
FDF-8372 & 0.2290 & 161$\pm$09 &  57 & 21.62 &  0.17$\pm$0.06 &   0.40  &  19.79$\pm$0.12   & 10.70$\pm$0.04 & 0.46 & $-$0.28$\pm$0.11 &   2.01  \\
FDF-8626 & 0.4081 & 379$\pm$37 &  27 & 21.45 &  0.90$\pm$0.06 &   1.45  &  21.00$\pm$0.12   & 12.05$\pm$0.05 & 0.90 & $-$0.20$\pm$0.12 &   1.95  \\
WHDF-92   & 0.3988 & 183$\pm$13 & 31  &	22.53 &$-$0.02$\pm$0.06 & 0.18  &  17.69$\pm$0.13   & 10.66$\pm$0.03 & 0.15 & $-$1.42$\pm$0.08 &   2.06  \\
WHDF-111  & 0.7418 & 172$\pm$12 & 23  &	22.88 &  0.82$\pm$0.05  &  0.91 &  19.93$\pm$0.13   & 11.46$\pm$0.04 & 0.16 & $-$1.93$\pm$0.11 &   2.05  \\
WHDF-158  & 0.4825 & 134$\pm$14 & 37  &	21.28 &  0.86$\pm$0.07  &  1.28 &  20.81$\pm$0.15   & 11.28$\pm$0.07 & 0.26 & $-$1.57$\pm$0.16 &   1.85  \\
WHDF-173  & 0.4507 & 184$\pm$16 & 37  &	22.51 &  0.70$\pm$0.04  &  0.88 &  20.60$\pm$0.14   & 11.39$\pm$0.02 & 0.60 & $-$0.84$\pm$0.06 &   2.08  \\
WHDF-318  & 0.3973 & 211$\pm$14 & 32  &	22.25 &  1.01$\pm$0.10  &  1.91 &  22.53$\pm$0.11   & 11.81$\pm$0.10 & 1.17 &    0.24$\pm$0.25 &   2.02  \\
WHDF-437  & 0.2599 & 128$\pm$08 & 57  &	21.67 &  0.64$\pm$0.05  &  1.10 &  21.38$\pm$0.10   & 10.99$\pm$0.03 & 0.63 & $-$0.52$\pm$0.09 &   1.89  \\
WHDF-508  & 0.3977 & 138$\pm$39 &  9  &  --   &  1.19$\pm$0.10  &  2.92 &  24.33$\pm$0.20   & 11.64$\pm$0.06 & 1.35 &    0.72$\pm$0.15 &   2.01  \\
WHDF-749  & 0.2130 & 130$\pm$11 & 52  & 21.10 &       --        &  --	&   --              &        --      &  --  &	  --	       &   2.66  \\
WHDF-810  & 0.2133 & 186$\pm$10 & 83  &	20.28 &  0.73$\pm$0.07  &  1.54 &  21.14$\pm$0.09   & 11.40$\pm$0.07 & 0.77 & $-$0.40$\pm$0.16 &   2.00  \\
WHDF-946  & 0.2123 & 107$\pm$05 & 49  &	21.63 &  0.70$\pm$0.07  &  1.47 &  22.40$\pm$0.15   & 10.89$\pm$0.07 & 0.82 & $-$0.05$\pm$0.16 &   2.07  \\
WHDF-810b & 0.2118 & 140$\pm$19 & 20  &	21.59 &  0.13$\pm$0.07  &  0.39 &  19.49$\pm$0.15   & 10.55$\pm$0.02 & 0.46 & $-$0.65$\pm$0.06 &$-$0.47  \\
\noalign{\smallskip}
\noalign{\hrule}
\end{tabular}
\end{center}
%\end{minipage}
\end{table*}

Figure~\ref{fig:whdfed} illustrates the spatial distribution of the
early-type field galaxies in the WHDF. Apart from three galaxies
which are located in the northern part of the image, 
all field ellipticals are homogenously distributed over the entire field.
To test for possible group candidates, radial velocity measurements of the
three galaxies close in projection resulted in a dispersion in radial
velocities of $\Delta v_{\rm rad}\simeq450$ \kms and a group velocity
dispersion of $\sigma_{\rm grp}=190_{-71}^{+63}$ \kms at $z=0.2125$. Looking at
the individual ACS  galaxy images, these objects are not visually associated
to each other through tidally distorted outer structures (e.g. tidal arms,
collisional rings, interaction signatures) or very nearby other companions.
For this reason, both independent approaches give strong evidence that the
three objects are not likely members of a group but isolated field galaxies.

The coordinates and redshifts of the galaxies for which velocity dispersions
were measured are given in Table~\ref{fwcoord}. Velocity dispersions,
S/N ratios, total Johnson $B$-band magnitudes, sizes, surface brightnesses,
dynamical masses, $M/L$ ratios, and rest-frame $(B-I)$ colours of the galaxies 
are presented in Table~\ref{phy}.

\subsubsection{Spectroscopic Star Formation Signatures}\label{sec:ssf}

To estimate the amount of recent star formation present in our field galaxies
we measure the spectral diagnostic features of \oii\,$\lambda$3727  and
\hda $\lambda$4101 in the observed wavelength range of the spectra.
For the indices \oii\,$\lambda$3727 and \hda $\lambda$4101
we adopt the passband definitions by
\cite{BMYCE97} and \cite{WO97}, respectively.
Uncertainties in the line index measurements account for poisson noise
as well as possible noise variations in the spectra. In the subsequent
analysis we always refer to rest-frame equivalent widths in units of \AA.
The spectroscopic measurements of our galaxies will further be
discussed in \S~\ref{sec:sfd}, where the implications of the recent activity
signatures on the stellar population properties will be assessed.

% -- Tab of HST/ACS observations ---
%{\scriptsize
\begin{table}
\centering
\caption{\label{acsobs}\small{HST/ACS F814W observations of the FDF and WHDF}}
\vspace{0.3cm}
\begin{tabular}{llcccc} 
\hline
\noalign{\smallskip}
Field & Date & Area & $T_{\rm tot}$ & M$_{\rm lim}$ \cr
      &	     &  arcmin$^2$   &  [sec] &  $3\sigma$  \cr 
\noalign{\smallskip}
\hline
\noalign{\smallskip}
% xxx xxx:
FDF  & 20/07-29/09/02 & $6.8 \times 6.8$ & 2360 & 25.50 \\ 
WHDF & 14/08-23/08/03 & $6.8 \times 6.8$ & 2450 & 26.30 \\ 
\noalign{\smallskip}
\noalign{\hrule}
\end{tabular}
\end{table}
%

% ######## Figure: ACS/HST surface brightness results  #########################
\begin{figure*}
\centering
\resizebox{0.8\textwidth}{!}{\includegraphics{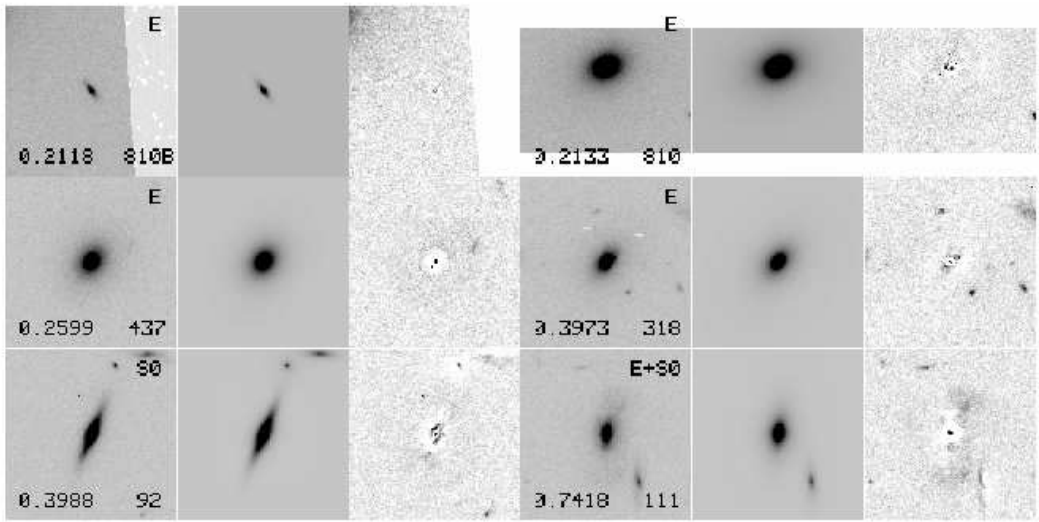}}
\caption{\label{sbfr}\small{Examples of the 
surface brightness modelling using a de Vaucouleur profile
for the field galaxies in our sample in order of increasing redshift.
In case of object ID\,92, a combination of a S\'ersic
plus exponential disc component yielded the best result.
For each galaxy the distortion corrected original image is shown alongside the
best $r^{1/4}$ model and the residuals from the respective model.
Redshift, galaxy ID and morphological type classification are
listed for all objects in the lower left, lower right and upper right corner.
The HST/ACS F814W images are $10''$ on each side, north is up and east
is to the left.}}
\end{figure*}
% ##############################################################################
%
%

\section{Photometry}

\subsection{HST/ACS Observations}

Using the Advanced Camera for Surveys (ACS) Wide Field Channel (WFC), 
\emph{Hubble Space Telescope (HST)}
observations of the FORS and William Herschel Deep Fields were carried out
during cycles 11 and 12.
To cover the $\sim7 \times 7$\,arcmin$^2$ sky areas of the FDF and WHDF,
four WFC pointings were aligned as a $2\times2$ mosaic. A single ACS WFC chip
offers a field--of--view of $\sim$\,$202\times202$\,arcsec$^2$ and a
pixel scale of $\sim$0.049\,arcsec. Each pointing was split into two exposures
and has a total exposure time of 2360\,s (FDF) and 2450\,s (WHDF) through the
F814W ($I$) filter. Table~\ref{acsobs} gives an overview of the 
HST/ACS observations of the FDF and WHDF.
The column $T_{\rm tot}$ represents the total exposure time for each object
and M$_{\rm lim}$ lists the 50\% completeness limit for $3\sigma$ detections of
extended sources within the two deep fields.
Data reduction was performed using the standard CALACS
\footnote{http://www.stsci.edu/hst/acs/analysis}
pipeline and the unprocessed images (including bias and dark subtraction,
flat--fielding, distortion and astrometric correction) were used
for an additional cosmic ray rejection filtering by combining the two
exposures of each visit to produce the final mosaic.

% ######## Figure: ACS/HST thumbnail F814W galaxy images  ######################
\begin{figure*}
\centering
\resizebox{\textwidth}{!}{\includegraphics{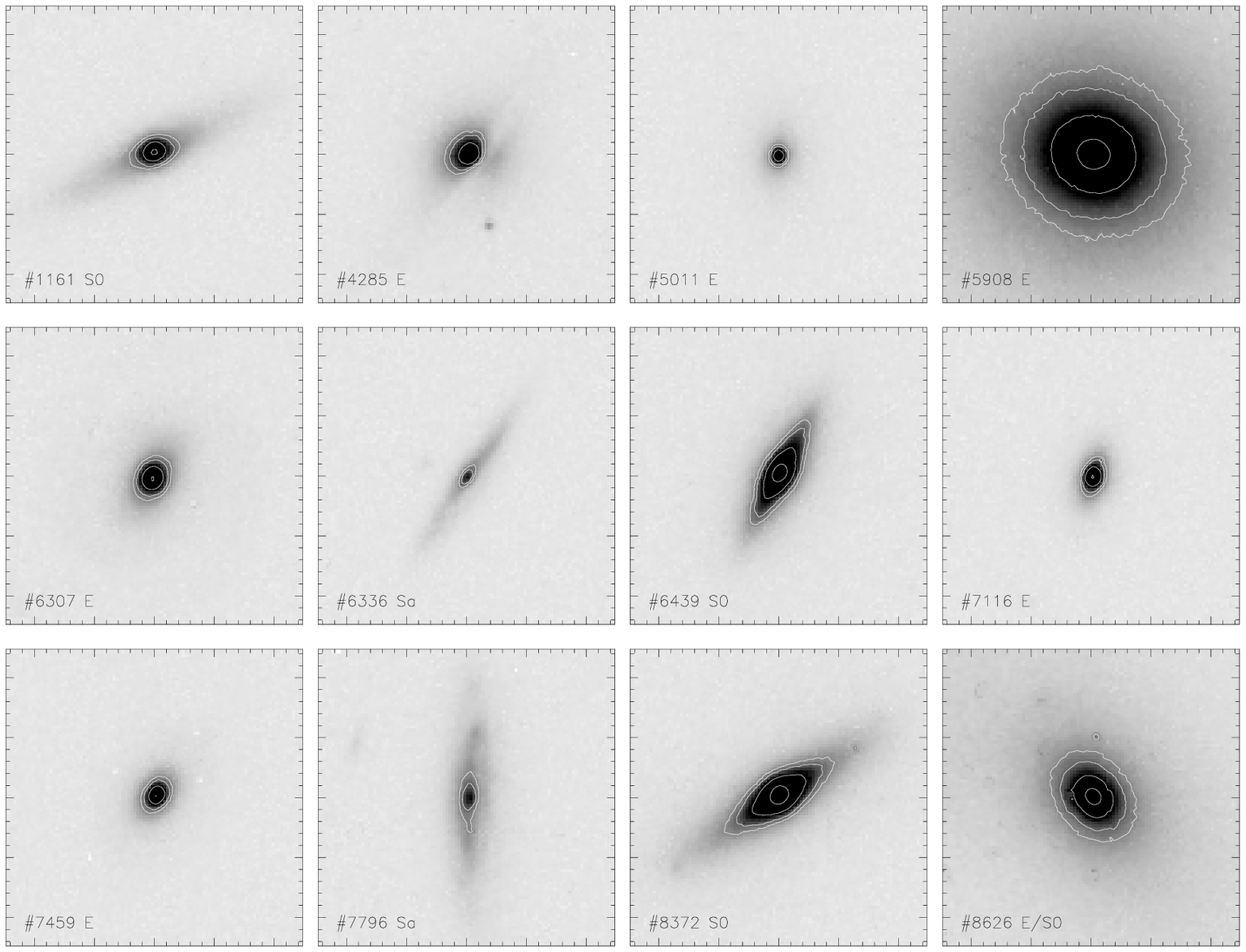}}
\caption{\label{fdfacsthumb}\small{$5''\times5''$ HST/ACS F814W images of
12 FDF galaxies with available spectroscopic information.
North is up and east is to the left. Isophotal contours are logarithmically
displayed across the range $18.80\leq C_{i}\leq 21.20$~mag arcsec$^{-2}$.
The labels give the galaxy ID and the morphological type classification
as listed in Table~\ref{fdfacsrem}.}}
\end{figure*}
% ##############################################################################

% ######## Figure: ACS/HST thumbnail WHDF ellipses F814W galaxy images  ########
\begin{figure*}
\centering
\resizebox{\textwidth}{!}{\includegraphics{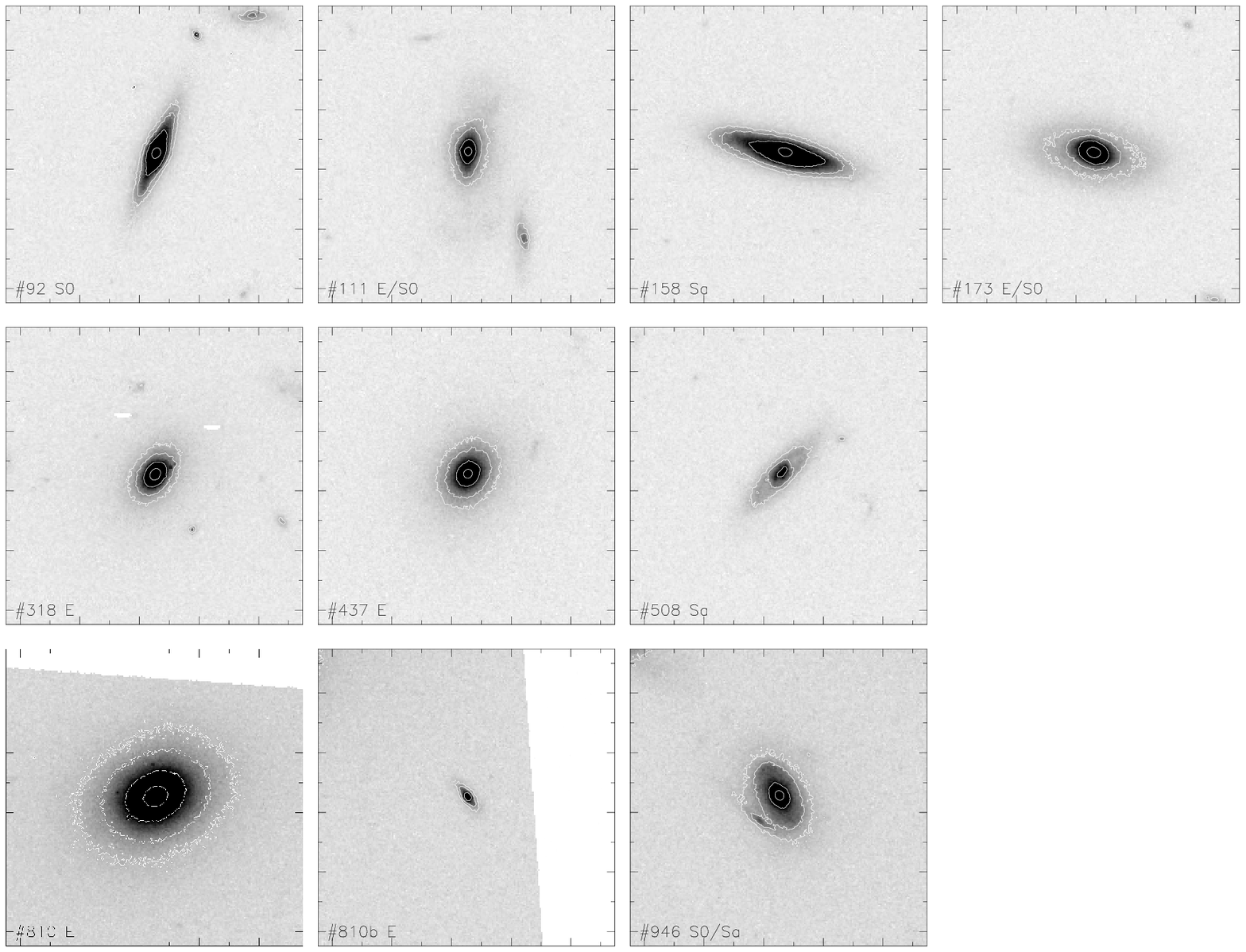}}
\caption{\label{wdhfeacsthumb}\small{$10''\times10''$ HST/ACS F814W images of
10 WHDF galaxies with available spectroscopic information.
North is up and east is to the left. Isophotal contours are logarithmically
displayed across the range $18.80\leq C_{i}\leq 21.20$~mag arcsec$^{-2}$.
The labels give the galaxy ID and the morphological type classification
as listed in Table~\ref{whdfacsrem}.}}
\end{figure*}
% ##############################################################################

\subsection{Surface Photometry}

Structural parameters of
half-light (effective $R_e$) radii, (effective)
surface brightnesses ($\langle \mu_e \rangle$ within $R_e$)
and total magnitudes and shapes of the galaxies were derived with the
GALFIT package \citep{Pen02}. For the convolution of the surface brightness
model profiles, an average PSF was constructed using
$\sim$\,20 unsaturated stars with $I_{814}<23$ which were normalized to the
same central flux and afterwards median--averaged. Two-dimensional surface
brightness profiles of the galaxies were analysed with four different fitting
functions, by a pure classical de Vaucouleurs $r^{1/4}$, a pure 
\cite{Ser:68} with 
variable S\'ersic exponent $n$ ($1\le n\le 4$), and each of those
profiles in
combination with an exponential disc component. The best fit to the observed
light distribution of a galaxy was assessed by comparing the residual images
and reduced $\chi^2$ values. 
Note that by definition a profile consisting of a S\'ersic plus disc component  
results in lower $\chi^2$ numbers than using a single $r^{1/4}$ model alone.
Therefore, preference was given to those model profiles with yielded to the
lowest residual structures in the residual images. 
In 40\% of the cases this was a pure
classical de Vaucouleurs profile, in 60\% of the cases this was a S\'ersic
plus disc component. 
Figure~\ref{sbfr} presents examples of the results of the surface brightness
modelling for our field galaxies. For each galaxy, the original image, the
best 2D model fit and the residuals from an $r^{1/4}$ or $r^{1/n}$+
exponential disc component are shown.
For internal consistency, the surface brightness distribution of
the field galaxies were also modelled with the GIM2D package by \citet{SWVSP02}
and the algorithm by \citet{Sag:97a}. A detailed discussion on differences
between these techniques can be found in \cite{FZBSD05} and \cite{ZTBBFM05}.
Total magnitudes derived with all three methods agree well within the errors
as well as with those values measured on the ground-based $I$-band images of
the FDF and WHDF with the SExtractor package.
Observed magnitudes of the galaxies were calibrated on to the Vega system
using the the synthetic transformation by \cite{SJB05}:
\begin{equation}
\textsc{TMAG} = -2.5\times \mathrm{log} (\mathrm{DN} s^{-1}) + c_0 + 
c_1 \times \textsc{TCOL} + c_2 \times \textsc{TCOL}^2,
\end{equation}
where the first term is the observed magnitude expressed as total count rate per
electrons and seconds in the source system, TMAG and TCOL being the observed
magnitude in the target system and the $c_0$, $c_1$, and $c_2$ the
transformation coefficients. We adopt our $(R-I)$ aperture colors and for the
coefficients $c_0=25.478$, $c_1=0.042$, and $c_2=0.012$.
For the FDF and WHDF a galactic absorption of $E(B-V)\nobreak=\nobreak0.018^m$
and $E(B-V)\nobreak=\nobreak0.025^{m}$ was derived, respectively.
Aperture colours were measured on the ground-based 
$UBgRI$ and $BRI$ band photometry within a 2\arcsec\ diameter aperture.

\subsection{Morphologies}

Galaxy morphologies were classified using a combination between a visual
inspection by one of us (A.F.), the results of the surface brightness
profile analysis (i.e. interpreting the residuals from the
$r^{1/4}$ luminosity profile fitting)
and an independent analysis using GIM2D \citep[see][for a description]{FZBSD05}. The
information of all three independent approaches was combined to yield the best
and most reliable decision on the final morphologies of the galaxies.
Figure~\ref{fdfacsthumb} and Figure~\ref{wdhfeacsthumb} 
display HST/ACS thumbnail images in the F814W filter of the FDF and WHDF
field early-type galaxies, respectively. Table~\ref{fdfacsrem} and
Table~\ref{whdfacsrem} list the morphologies and noteworthy remarks of special
features for the individual galaxies. The first column of each table shows the
galaxy ID, the second indicates the galaxy morphology and the last column gives
additional information about special features to the respective galaxy.
In total, the sample is divided into 13 field ellipticals (E and E/S0) and
nine field lenticulars (S0 to Sa bulges). The two classes comprise the sub-types
of 6 E, 1 E/S0, 3 S0 and 2 Sa bulges in the FDF,
and 4 E, 2 E/S0, 1 S0, 1 S0/Sa and 2 Sa bulges in the WHDF. 
No ACS image is available for the galaxy ID\,6338 in the FDF, based on its
spectrum this object has a morphology of a spiral galaxy (see also
Table~\ref{fdfacsrem}). The findings from the luminosity profile fitting
provided a consistency check for the visual inspection and resulted in the
same classification scheme except for one field galaxy in
the WHDF, ID\,810b: S0 (visual), E (model fit).

%
% Table with noteworthy remarks on HST/ACS FDF objects 
% Morphologies and noteworthy remarks on individual HST/ACS images of the FDF galaxies.
\begin{table}
\begin{center}
\caption{\label{fdfacsrem}\small{HST/ACS morphologies and features of
galaxies in the FDF}}
\vspace{0.1cm}
\begin{tabular}{lcl}
\hline
\noalign{\smallskip}
ID & morp  & Note \\
\noalign{\smallskip}
\hline
\noalign{\smallskip}
1161 &  S0   & \\ 
4285 &  E    & galaxy with dust lane, merger of S0 \& S? \\  
5011 &  E    & \\
5908 &  E    & \\
6307 &  E    & model too boxy, residuum at core \\
6336 &  Sa   & Sb?, peculiar structure, warps \\
6338 &  S    & no ACS image, no analysis, morph based on spec \\
6439 &  S0   & \\
7116 &  E    & hot pixel \\ 
7459 &  E    & no analysis \\ 
7796 &  Sa   & peculiar structure, SF in upper spiral arm?, warps \\
8372 &  S0   & hot pixel \\ 
8626 &  E/S0 & close to edge of chip \\
\noalign{\smallskip}
\noalign{\hrule}
\end{tabular}
\end{center}
%\end{minipage}
\end{table}
%

%
% Table with noteworthy remarks on HST/ACS WHDF objects 
\begin{table}
\begin{center}
\caption{\label{whdfacsrem}\small{HST/ACS morphologies and features of
galaxies in the WHDF}}
\vspace{0.1cm}
\begin{tabular}{lcl}
\hline
\noalign{\smallskip}
ID & morp & Note \\
\noalign{\smallskip}
\hline
\noalign{\smallskip}
\hspace*{1.0ex}92 & S0   &  two background? objects in south\\
 111 & E/S0 &  external halo in upper north part, two objects in north-west \\
 158 & Sa   & \\
 173 & S0   & \\
 318 & E    & close to star, star in center \\
 437 & E    & face on, faint star in north-east, [O\,{\scriptsize II}]\,3727 emission \\ 
 508 & Sa   & peculiar, dust pattern?, rot.?, [O\,{\scriptsize II}]\,3727 emission?, star\\
 749 & E/S0 & no ACS image, morphology  based on spectrum \\
 810 & E    & boxy isophotes, stars in center, [O\,{\scriptsize II}]\,3727 emission \\ 
810b & E    & discy structure, faint, [O\,{\scriptsize II}]\,3727 emission \\ 
946  & S0/Sa & lensed object (maybe in projection) in north-east \\
\noalign{\smallskip}
\noalign{\hrule}
\end{tabular}
\end{center}
\end{table}

\subsection{Spectrophotometric Properties}

\begin{figure}
\includegraphics[width=1.0\linewidth]{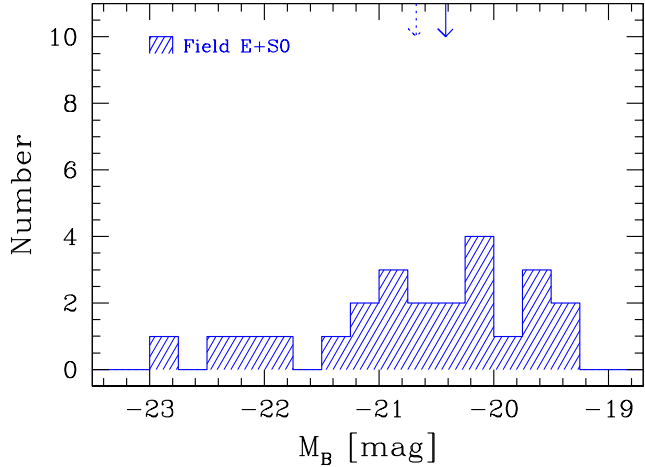}
\caption{\label{mrs3}\small{
Absolute $B$-band magnitude distribution of 24 early-type field galaxies
in the FDF and WHDF. Rest--frame magnitudes were derived based on
the FDF and WHDF photometry. The dashed and the solid arrow give the mean
($\overline{M}_B=-20.68$) and the median
($\langle M_B \rangle=-20.42$) values of the
distribution, respectively.
}}
\end{figure}

$k$-corrections were calculated by convolving respective filter transmission
curves with the redshifted Spectral Energy Distribution (SED) of an
early-type galaxy template from \cite{KCBMSS96} as
described in \cite{FZBSD05} and \cite{F06}. No separation between elliptical
and S0 types was done because both exhibit similar (optical and UV) spectral
shapes, Balmer discontinuities (4000 \AA) and absorption features for
$\lambda>5000$ \AA\ which are indistinguishable from their broadband optical
and near-infrared colours. For internal consistency, synthetic spectral SED
templates of an 12\,Gyr elliptical galaxy which were generated through
evolutionary synthesis modelling by \cite{Moe01} were also utilised.
Typical deviations in the $k$-corrections between the SEDs 
of the two template libraries are small
$\Delta \overline k_{B}\le0.1^m$.
Depending on the redshift of the object the most suitable filter was chosen 
to transform to rest-frame Johnson $B$-band.
The $B_{\rm fors}$ filter was selected for galaxies in the FDF with $z>0.25$,
$g_{\rm fors}$ for $0.25\nobreak\le z\nobreak<\nobreak0.55$ 
and $R_{\rm fors}$ for $0.55\nobreak\le~z\nobreak<\nobreak0.7$.
For field objects in the WHDF with $z>0.25$ a transformation to
$B$ rest-frame was conducted based on the $B$ passband and for
$0.55\le z< 0.75$ the $R_C$ filter was utilised. 
Therefore, the $k$-correction was much less sensitive to spectral type
than using a global filter transformation that is restricted to 
a specific filter, e.g. of $B_{\rm obs} \rightarrow B_{\rm rest}$.
For reasons of consistency, the rest-frame magnitudes were checked  by
transforming the observed F814W-magnitudes into rest-frame Johnson-$B$.
Deviations between magnitudes were small (differences were less
than their errors). Total errors on the absolute $B$-band
magnitudes for the galaxies were computed as the quadratic sum of individual
errors in the respective ground-based photometry filter (either $B$, $R$ or
$I$-band), the error in the $k$-correction and the  
uncertainty in the galactic absorption correction.
Thanks to the very deep imaging of the FDF and WHDF, the photometric errors 
in the ground-based photometry are very small for all 
filter passbands (median uncertainties $<0.015^m$).
For the FDF and WHDF field sample, the total absolute error falls into the range 
$0.08^m\le\delta M_B\le0.20^m$ with a median of 0.15\,mag.
In Figure~\ref{mrs3} the distribution of the early-type field galaxies in the
FDF and WHDF in absolute Johnson $B$-band magnitude are shown.
Only the rest-frame magnitudes derived with the FDF and
WHDF photometry are shown.
The 24 E+S0 field galaxies cover a range in absolute $B$-band magnitude
of  $-22.78\le M_B\le -19.40$ ($\langle M_B \rangle=-20.42$).
The dashed arrow corresponds to the mean value
($\overline{M}_B=-20.68$), whereas the median is 
denoted by the solid arrow.

%
% M_B-distr. FDF+WHDF
%
%

%
% Chandra/XMM WHDF
%
\begin{figure}
\includegraphics[width=1.0\linewidth]{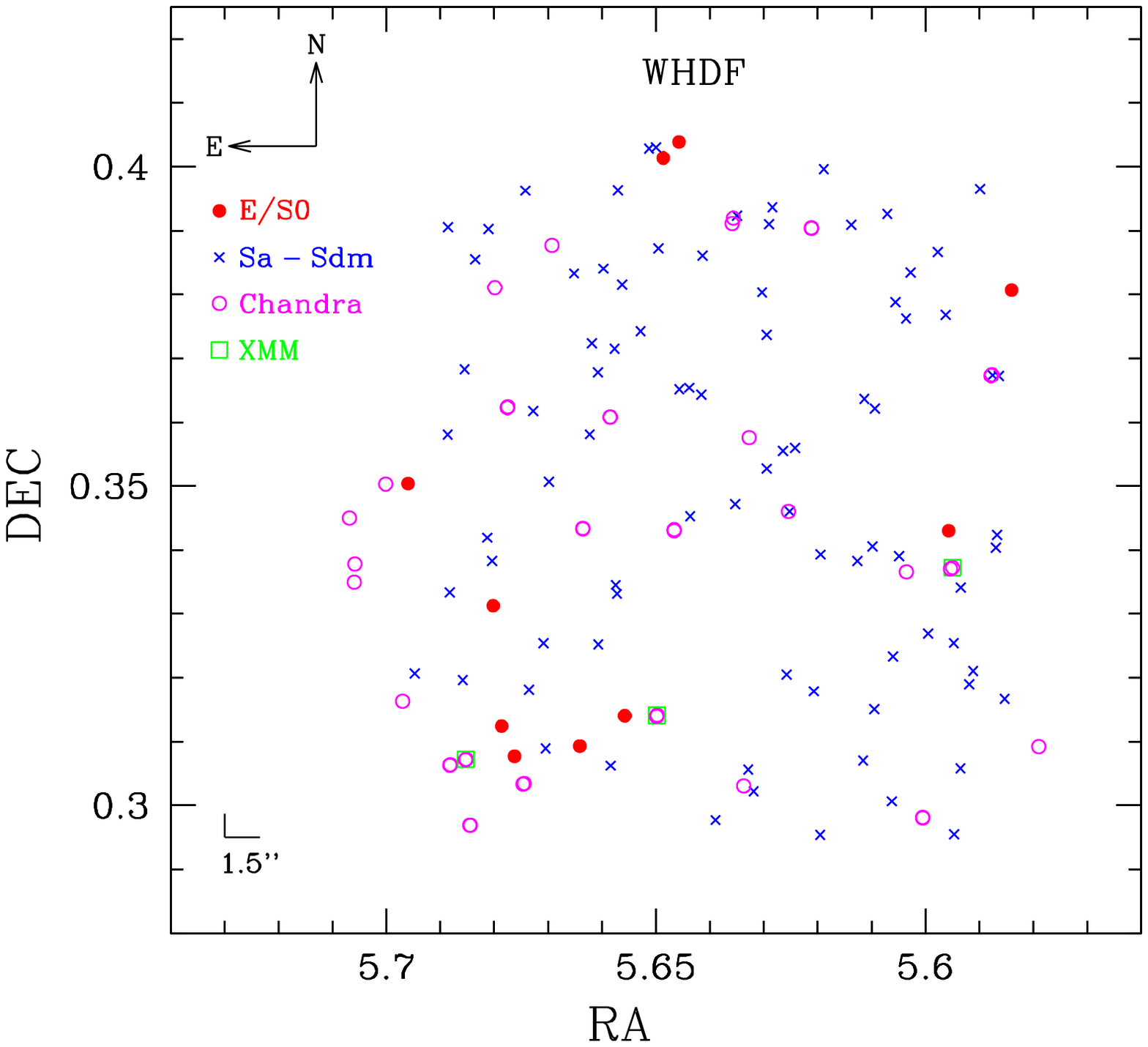}
\caption{\label{fig:chandra}\small{
Distribution of \textit{Chandra} and \textit{XMM-Newton} 
detections within the WHDF. North is up and east is to the left.
Early-type field galaxies in our sample are indicated as red circles.
The error bars in the left correspond to $1\farcs5$ on the respective axes. 
}}
\end{figure}

\subsection{X-Ray Properties}\label{xray}

To acquire independent complementary information 
on nuclear activity in our early-type galaxy sample we checked the archive
for possible X-ray counterparts of our field galaxies.
In the selection process of a suitable sky area for the FDF, one particular
constraint was that the field had to be devoid of strong radio and
X-ray sources to avoid potentially present galaxy clusters at medium redshifts.
The \textit{Chandra} and \textit{XMM-Newton} observatories have not observed
the sky region covered by the FDF; therefore, we cannot look for the presence
of possible active galactic nuclei (AGNs) in these field galaxies.

For the WHDF two \textit{Chandra} surveys are available
(PI: T. Shanks, PID 02900499). A deep,
high resolution ACIS-I imaging survey (72.13 ksec) of the central field
and a shallow survey (10 fields with exposure times of 10 ksec each) with
the ACIS-S around narrow line galaxies that have previously been detected
by ROSAT. The ACIS-I covers the spectral range between 0.2 to 10 keV and
has a spatial resolution of $\sim$0.492\arcsec\ or better, with PSF variations less
than 0.2\arcsec\ across the total 16.9\arcmin$\times$16.9\arcmin\ field-of-view
\citep{Weis02}. Through a cross-correlation with our optical photometry
we identified 28 objects as X-ray point-sources above a 
$2\sigma$ source detection limit (95\% confidence level).
The detected counts for each source were converted into fluxes, assuming a
power law of the
photon spectral flux distribution of $dN/dE\propto E^{-\Gamma}$ with an
intrinsic photon index of $\Gamma=2$ and column density of the galactic
absorption of $nH=1.0\times10^{21}$~cm$^{-2}$. A photon index of $\Gamma=2$
is typical for X-ray emission that has its origin from star formation. 
The on-axis sensitivity limits of the (unabsorbed) flux in the soft ($0.5-2.0$ keV)
and hard ($2.0-8.0$ keV) bands are $7.5\time 10^{-16}$~ergs~cm$^{-2}$~s$^{-1}$
and $1.9\time 10^{-15}$~ergs~cm$^{-2}$~s$^{-1}$, respectively. 
These fluxes correspond to a detection limit in X-ray luminosities emitted in the
soft band of $L_X(0.5-2.0\,{\rm kev})=1.9\times10^{42}$~ergs~s$^{-1}$ and in the
hard band of $L_X(2.0-8.0\,{\rm kev})=4.7\times10^{42}$~ergs~s$^{-1}$.
The \textit{Chandra} exposures are deep enough to both fully resolve the X-ray
background (XRB) at about $1.5\time 10^{-15}$~ergs~cm$^{-2}$~s$^{-1}$ in the 
$2.0-7.0$ keV energy band and detecting all the hard X-ray sources
within the WHDF.

Very recently a shallow \textit{XMM-Newton} survey
(PI: M. Guainazzi, PID 0407030101) targeted a similar sky region 
(27.91 ksec). However, these observations cover only the southern part
of the WHDF and are at a significantly coarser angular resolution of 
$\sim$15\arcsec than the \textit{Chandra} detections.
Using the same photon spectral flux distribution configuration as above 
the sensitivity limits of the (unabsorbed) flux in the
$0.5-2.0$ keV energy band are $9.3\time 10^{-15}$~ergs~cm$^{-2}$~s$^{-1}$ and
in the $2.0-8.0$ keV range $2.6\time 10^{-14}$~ergs~cm$^{-2}$~s$^{-1}$.
The soft and the hard fluxes correspond to detection limits of
$L_X(0.5-2.0\,{\rm kev})=2.3\times10^{43}$~ergs~s$^{-1}$ and 
$L_X(2.0-8.0\,{\rm kev})=6.5\times10^{43}$~ergs~s$^{-1}$.
Figure~\ref{fig:chandra} shows the distribution of \textit{Chandra} and
\textit{XMM-Newton} detections in the WHDF.
Three galaxies that were observed by \textit{Chandra}
were additionally detected by us on the \textit{XMM} images
(see Figure~\ref{fig:chandra}).
Apart from two clear Chandra detections of two Sc spiral galaxies
(ID\,438) at $z=0.0473$ and (ID\,635)
at $z=0.0709$ (no maximum rotation velocity $V_{\rm max}$ could be derived
for both objects), four field early-type galaxies might have associated 
X-ray counterparts. These objects in the WHDF with possible Chandra
and/or XMM counterparts are listed in Table~\ref{xrct}. 
There is only one rather likely candidate for a galaxy with a central
AGN (ID\,437) as this object shows also weak \oii\ emission.
In case of the other candidates either bright (ID\,173) or bright and
extended (ID\,92 or 158) \textit{XMM} sources close to the galaxies are
found that might be associated to the optical counterparts. 
Non of the early-type galaxies in our sample is
surrounded by a diffuse halo-gas, thus no object appears to be an extended
X-ray source, hence the influence of AGN activity for our galaxies is small
or even negligible. 
We will discuss the possible impact of X-ray detections together with our
results in section~\ref{agn}. In Appendix \S~\ref{xrayp}, we assess in more
detail contributions due to low-luminosity AGN (LLAGN) and/or emission from
soft, diffuse X-ray plasma or or low-mass X-ray binaries (LMXBs) in case of
our galaxies.

% -- Tab with Chandra and XMM observations ---
%{\scriptsize
\begin{table}
\centering
\caption{\label{xrct}\small{Galaxies with possible
Chandra/XMM counterparts in the WHDF}}
\vspace{0.3cm}
\begin{tabular}{rccll} 
\hline
\noalign{\smallskip}
ID & morp  & Chandra\,/\,XMM & association & Note \cr
\noalign{\smallskip}
\hline
\noalign{\smallskip}
\hspace*{1.0ex}92 & S0   &  yes?\,/\,no & maybe & XMM bright, extended \\
158 & Sa   &  no\,/\,yes? & not likely & XMM bright, extended \\
173 & S0   &  yes?\,/\,yes? & likely & XMM bright \\
437 & E    &  yes?\,/\,yes? & maybe  & [O\,{\scriptsize II}] emission \\ 
438 & Sc   & yes\,/\,no  & yes   & $z=0.0473$, no $V_{\rm max}$ \\ 
635 & Sc   & yes\,/\,no  & yes   & $z=0.0709$, no $V_{\rm max}$ \\ 
\noalign{\smallskip}
\noalign{\hrule}
\end{tabular}
\end{table}

%%%%%%%%%%%%%%%%%%%%%%%%%%%%%%%%%%%%%%%%%%%%%%%%%%%%%%%%%%%%%%%%%%%%%%%%%%%%%%%%

\section{Scaling Relations of Field Early--Type Galaxies at z=0.4}
\label{scal}

\subsection{\label{local}The Local Reference}

To derive the kinematic and/or the spectrophotometric evolution of
early-type galaxies at intermediate redshift, it is crucial to carefully
select a sample of early-type galaxies at low redshift which can be
utilised as a reference for the purpose of comparison. 
The Coma cluster at $z=0.024$ is one of the best studied local rich 
clusters of galaxies and represents with a look--back time of only
$\sim$0.3 Gyr (corresponding to 2\% of the age of the Universe)
the present-day Universe.
A number of previous works in the literature on both cluster and field
early-type galaxies
\citep[e.g.,][]{vDF96,Treu01a,TSCMB02,T05,vdV03,JCFBBD06}
used this cluster for the purpose of comparison of the Fundamental Plane
at $z>0.1$ and it therefore provides a reliable and widely
accepted local reference when addressing evolutionary questions. 

In order to minimise systematic uncertainties arising from filter
transformations, distance determinations and selection effects, we utilise the 
well defined Coma sample in the Johnson $B$-band by
\cite[][hereafter SBD93]{SBD93} which comprises 39 early-type cluster
galaxies (splitted into 25 E and 14 S0s).
This data compilation is a sub--set of the 
original ``7 Samurai'' sample \citep{FWBDDLT89} 
which contains 59 E+S0 galaxies in the Virgo and
Coma cluster; but for the latter no morphological information is available. 
For deriving the evolution of the $M/L$ with redshift we use  
the FP coefficients as published by \cite{FWBDDLT89}.

\subsection{\label{fjr}The Faber--Jackson Relation}

Early-type galaxies follow a scaling relation between their luminosity and
internal velocity dispersion, also known as the Faber--Jackson relation
\citep{FJ76}. The relationship can be used to constrain the
formation and evolution of E+S0 galaxies but it requires accurate 
velocity dispersion measurements as small uncertainties in the kinematics
transfer into large errors in the magnitudes, hence the measured
luminosity evolution.

\begin{figure}
\includegraphics[width=1.0\linewidth]{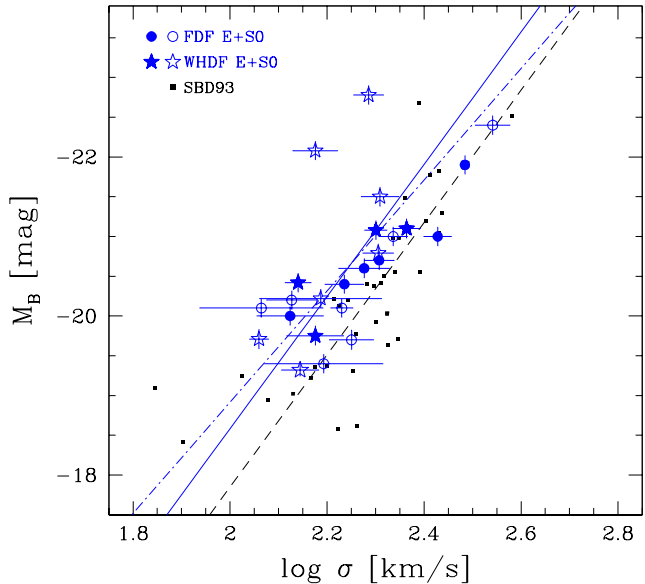}
\caption{\small{\label{fig:fjr}Faber--Jackson relation for the
early-type field galaxies in the FDF and WHDF, compared to the Coma galaxies
of SBD93 (small squares) in rest-frame Johnson $B$-band. The dashed
line shows the $\chi^2$-fit to the local restricted FJR by SBD93
(within selection boundaries defined by the distant galaxies), 
the dot-dashed line indicates an unconstrained $\chi^2$-fit to the distant
sample. The average evolution of the distant field galaxies 
of $\Delta\langle M_B\rangle=-0.44$ mag with respect to
the local galaxies is shown by the solid line.}}
\end{figure}

Figure~\ref{fig:fjr} shows the Faber--Jackson relation (FJR) in the $B$-band
for the field early-type galaxies in the FDF and WHDF, 
compared to the local Coma sample by SBD93. For this scaling relation,
galaxy magnitudes were measured with SExtractor on the deep ground-based
images with $\sim$0.60\,arcsec FWHM in the $I$-band, hence the whole field
sample can be considered. As the median uncertainty in our velocity dispersion
measurements is only 15\,\kms and our magnitude limit reaches
$M_B=-19.30$ (about one magnitude fainter than previous studies),
we can put reliable implications on the evolution of our distant field
galaxies using the FJR.
A linear $\chi^2$-fit to the restricted Coma reference sample yields:
\begin{equation}
M_{B}=-1.22\,{\rm log}\,\sigma-8.31,
\end{equation}
which is indicated as a dashed line in Figure~\ref{fig:fjr}.
Assuming that the local slope holds valid for our distant galaxies
(i.e. there is not evolution in the slope of the FJR), we analyse
the mean residuals from the local FJR, shown as the 
solid line in Figure~\ref{fig:fjr}.
Because of the on average younger light-averaged ages of the stellar
populations in the distant galaxies, for a given velocity dispersion
the field galaxies are on average brighter by $\Delta M_B=-0.65\pm0.16$ mag
(median $\Delta\langle M_B\rangle=-0.44$ mag) than their local counterparts.
The $\pm1\sigma$ scatter of the offsets in the FJR for the
distant galaxies is $\sigma_{r}=0.72^{m}$.
The two objects that are slightly offset from the rest of the distant 
field galaxies show both signs of star formation, one of them being also the
galaxy with the highest redshift in our sample (cf. \S~\ref{sec:sfd}).
The measured evolution is in very good agreement with a slow, passive
evolution of the stellar populations as suggested by    
single-burst simple stellar populations models with a high 
formation redshift of the bulk of the stars \citep{BC03}. For formation
redshift of $z_f=2$ these evolution models predict an average increase in the
Johnson $B$-band of $0.61\pm0.06$~mag at $z=0.4$. 
Similar as for local studies, our distant galaxies show weak evidence for 
a larger scatter at fainter luminosities ($M_B>-20.40$),
that might be indications for an evolution of slope in the FJR.
We interpret this effect as a stronger evolution of the lower-mass
galaxies. However, apart form the mass dependence, an additional small
contribution to the offset of lower-luminosity galaxies might be the result 
of a weak dependence of the velocity dispersion on luminosity of the
early-type galaxies
that was previously found within a large sample of early-type galaxies drawn
from Data Release 4 of the
Sloan Digital Sky Survey \citep[SDSS,][]{DQMW07}. This luminosity dependence
becomes detectable in all FP projections and is most pronounced for
faint galaxies in the SDSS around $M_r=-21.2$ which corresponds to about
$M_B=-20.2$.
At $M_B=-20.3$ our field distant galaxies show a diversity, half of the sample indicates
a strong luminosity evolution whereas the other half a mild evolutionary
trend. An un-constrained linear $\chi^2$-fit to our field galaxy sample yields:
\begin{equation}
M_{B}(z=0.4)=-4.93\,{\rm log}\,\sigma-6.99.
\end{equation}
However, our sample is too small and the dependence on luminosity too weak
to test possible luminosity effects on the slope of the FJR at $z=0.4$ in
greater detail.

% ########## Figure:  FJR for Field Galaxies #################################
% ##############################################################################

% ########## Figure:  FJR for Field Galaxies in K-band #########################
\begin{figure}
\includegraphics[width=1.0\linewidth]{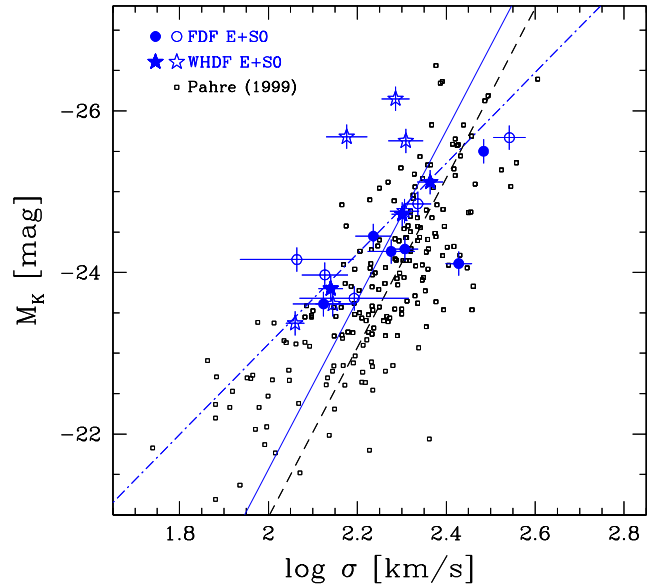}
\caption{\small{\label{fig:fjrk}Faber--Jackson relation for the
early-type field galaxies in the FDF and WHDF in the rest-frame $K$-band.
Large symbols are the field early-type galaxies, open squares are
local cluster early-type galaxies from Pahre (1999).
The Faber-Jackson relation of the distant field early-type galaxies
has a similar scatter as the local relationship.
An average evolution of $\Delta\langle M_K\rangle=-0.59$ mag for the distant
galaxies with respect to the  local reference is found, indicated by the
solid line. An unconstrained $\chi^2$-fit to the distant sample
is shown by the dot-dashed line.}}
\end{figure}
% ##############################################################################

To verify our results and to test the underlying stellar populations in even
more detail, we converted the extinction-corrected $(B-R)$ rest--frame colors
of the field galaxies in our sample into $M/L$ ratios in the $K$-band 
following a similar approach as in \citet{BdJ01}. Absolute $K$-band magnitudes
were derived by using the synthetic stellar population models by \cite{Moe01}.
The FJR in the $K$-band for the field E+S0 galaxies in the FDF and WHDF
is presented in Figure~\ref{fig:fjrk}. The early-type galaxies are compared to
the local relationship of cluster early-type galaxies as found by \cite{P99}.
For a given velocity dispersion the luminosities of the distant field galaxies
(excluding ID\,111 at $z=0.7418$) are on average brighter by
$\Delta M_K=-0.68\pm0.21$ mag (median $\Delta\langle M_K\rangle=-0.59$ mag)
than their local representatives.
The moderate luminosity evolution found for the field E+S0 is in excellent
agreement to the findings derived using the $B$-band magnitudes. Less-massive
galaxies display a larger offset with respect to the local relationship,
confirming the faster evolution of less-massive galaxies in the $B$-band.
Furthermore, lower-luminosity (mass) galaxies show a faster evolution 
in luminosity than higher-luminosity (mass) galaxies. This point will be
addressed in greater detail in \S~\ref{stpops}.

% ########## Figure:  Kormendy relation ########################################
\begin{figure}
\includegraphics[width=1.0\linewidth]{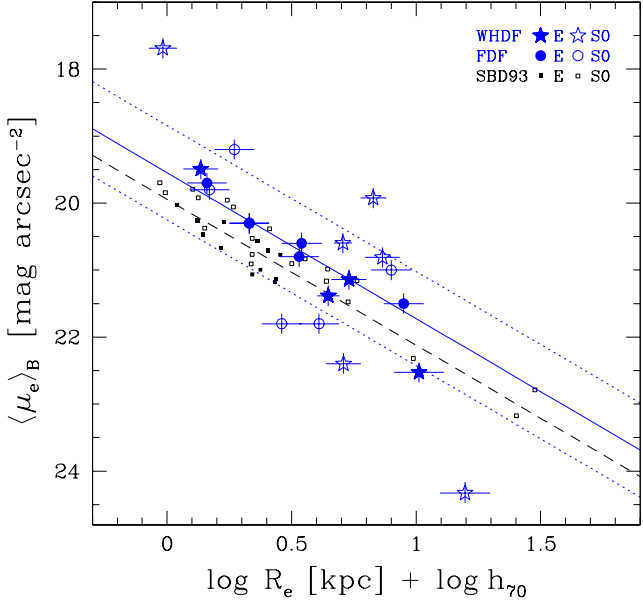}
\caption{\small{\label{fig:krb}Kormendy relation for the
early-type field galaxies in the FDF and WHDF compared to the Coma galaxies
of SBD93 in rest-frame $B$-band. The dashed line shows the $\chi^2$-fit to
the local KR by SBD93. The average evolution of the distant field galaxies 
with its scatter with respect to the local galaxies is shown by the solid
and dotted lines, respectively.}}
\end{figure}
% ##############################################################################

\subsection{\label{KR}The Kormendy Relation}

To study the change of surface brightness evolution for the FP sample at a
fixed size, the magnitude--size relation was constructed. 
The Kormendy relation (KR) represents the projection of the
Fundamental Plane along the velocity dispersion onto the photometric plane.
As only the structural parameters are used, samples have larger scatter in the
KR than in the FP but results should complement and endorse
findings obtained with the FP.

The KR of the field galaxies is presented in Figure~\ref{fig:krb}.
For a given size the surface brightnesses of the distant field galaxies
(excluding the Sa ID\,508) are on average brighter by
$\Delta \mu_{e\ B}=-0.40\pm0.39$ mag arcsec$^{-2}$
(median $\Delta\langle \mu_e\rangle_B=-0.51$ mag arcsec$^{-2}$).
One object (ID\,508) that does not follow the KR is a spiral galaxy,
has a high surface brightness and also shows possible signs of
rotational support in its two-dimensional spectrum (cf. \S~\ref{FP}). 
We confirm the power of the KR to separate passive quiescent
galaxies from galaxies with blue colors as found by \cite{FLCSM05}.
In consistency with passive evolution models, the fading of the stellar
populations implies that the galaxies will evolve to brighter surface
brightnesses with increasing redshift, hence moving away from the local
KR of present day early-type galaxies.

There are no indications of an evolution in the sizes of our field
galaxies, see Figure~\ref{fig:krb}.
We emphasise that it is hard to separate an evolution in the size of the
galaxies from an evolution in the surface brightness as 
the photometric parameters of half-light radius and surface
brightness are correlated and enter the FP and its projections
in combination as $FP_{\rm phot} = \beta\ \mu_{e}+{\rm log}\ R_{e}$,
with $\beta=0.328$ \citep[e.g.,][]{Sag:97a,FZBSD05}.
The measurements and results derived for the KR depend critically
on the selection procedures applied to the sample galaxies. For example,
two studies targeting the same sky region obtained inconsistent results, 
\cite{SVCL05} claims an evolution in the size of galaxies in the CDFS
whereas \cite{vdW05} does not support this finding. Most likely this effect is
due to differences or constraints set in the selection process, where the
former primarily selected fainter objects whereas the latter observed a
more homogenous population of early-type galaxies.
In case of the low-mass galaxies a differential evolution can be
produced by a combined effect of large spectroscopic apertures and 
substantial rotational support in the target objects.
In particular, the combination of the large slit-apertures with 
low S/N galaxy spectra by \cite{SVCL05} can cause 
miss-classifications of spirals with unresolved possible rotational
support as detectable in the emission lines. This effect points to a bias in
the selection procedure. A further discussion on possible selection
effects is presented in \S~\ref{sec:litcmp}.

\subsection{\label{FP}The Fundamental Plane}

Early-type galaxies in the nearby Universe follow a tight fundamental
relationship which is defined by their basic observables as
\begin{equation}\label{fprel}
{\rm log}\,R_{\rm e}\,=\,\alpha\ {\rm log}\,\sigma\,+\,\beta\ 
{\rm log}\,I_{\rm e}\,+\,\gamma,
\end{equation}
where $R_{{\rm e}}$ is the effective (half-light) radius given in kpc, 
$\sigma$ denotes the internal velocity dispersion in \kms and
$I_{\rm e}$ is the surface brightness in mag arcsec$^{-2}$,
related to $\langle \mu_{\rm e}\rangle$ as
${\rm log}\,I_{\rm e}\,=-0.4\langle \mu_{\rm e}\rangle\,$. 
For the FP coefficients we performed a principal component analysis and
derived for the slopes and the intercept $\alpha=1.25$, $\beta=-0.82$ and
$\gamma=-9.0$, which are in good agreement with the FP coefficients by
SBD93; \cite{JFK96}. From the sample of field galaxies of \cite{FWBDDLT89}
we measured that the intercept of the FP is offset by 
$\Delta\,\overline{\gamma}=0.04$ from the relation as given in
equation~\ref{fprel}.

\begin{figure}
\includegraphics[width=1.0\linewidth]{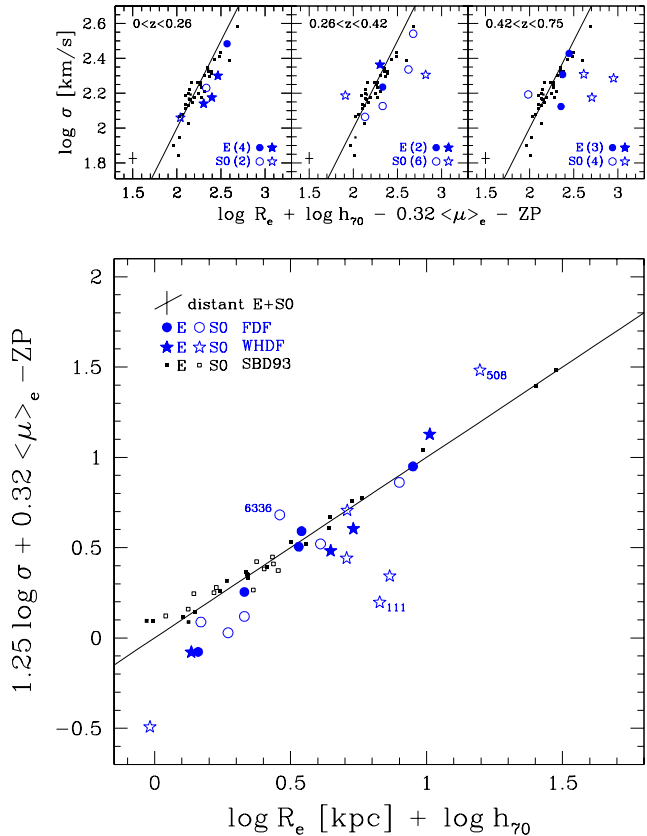}
\caption{\small{\label{FPfw} Fundamental Plane for the field 
early-type galaxies in the FDF and WHDF, compared to the Coma galaxies
of SBD93 (small squares) in rest-frame Johnson $B$-band.
Filled symbols denote ellipticals, open symbols S0 galaxies and Sa bulges.
\emph{Lower panel:} Edge-on view of the FP. The distant field lenticular
galaxies show a stronger evolution and larger scatter than the ellipticals.
\emph{Upper panel:} Evolution of the field FP as a function of redshift.
Field galaxies are binned into different redshift slices,
each compared to the Coma galaxies and shown along the short axis of the
edge-on view. Brackets denote the number of galaxies in a respective
morphological class (E, S0/Sa). The offset of the distant field galaxies from the local FP increases
with redshift, whereas the scatter appears to increase primarily for S0
galaxies with look--back time. Mean error bars are shown in each panel.}}
\end{figure}

Figure~\ref{FPfw} shows the FP for the 21 early-type field galaxies in the FDF
and WHDF in the rest-frame Johnson $B$-band, compared to the 39 early-type
galaxies in the Coma cluster by SBD93. The local reference is indicated with
small squares, whereas the distant field galaxies are shown as the large
circles and stars. Filled symbols denote ellipticals, open symbols S0
galaxies and Sa bulges.
The 21 FDF and WHDF early--type field galaxies are morphologically classified
into 9 E and 12 S0. Five ellipticals and six S0s originate from the FDF and
four Es and six S0s are located in the WHDF. Two galaxies are not visible on
the ACS images and for one elliptical galaxy in the FDF no structural
parameters could be derived.
The distant field galaxies cover a redshift range of
$0.21\le z_{\rm Field \ E}\le0.74$ with a median of
$\langle z_{\rm Field \ E}\rangle=0.40$.
Two outliers in the FP, ID\,6336 (${\rm log}~R_{{\rm e}}=0.46$)
and ID\,508 (${\rm log}~R_{{\rm e}}=1.20$), reveal on the
ACS images a significant disc component and early spiral morphology. 
Galaxy ID\,6336 has an extraordinary
strong H$\delta_{\rm F}$ absorption, which points to  a young
luminosity-weighted average model age under assumption of a single stellar
population. On the ACS images, ID\,508 shows a peculiar structure which
might indicate that this object is an ongoing merger. Both galaxies were
classified as Sa bulges hence these early-type spirals offer at least a 
low level of star formation which increases their total luminosity and
displaces their position beyond the tight plane as established by
early--type galaxies.

% ########## Figure:  FP for Field Galaxies #################################
% ##############################################################################

Rejecting the two outliers, for the total sample of 19 early-type field
galaxies in the FDF and WHDF an average luminosity evolution of 
$\Delta M_{B}=-0.53\pm0.13^{m}$ with a median of 
$\Delta\langle M_{B}\rangle=-0.39$ is found.
Similar results are measured for 17 cluster ellipticals at
$\langle z\rangle\sim0.4$ \citep{ZTBBFM05,FZBBBSD06} with similar look-back
times ($\sim$5 Gyr), which show an average brightening of their stellar
populations of $\Delta M_{B}=-0.44\pm0.18^{m}$, compared to their local
counterparts.
Overall, the distribution along the FP and the measured luminosity evolution 
is similar for cluster and field galaxies at $\langle z\rangle\sim0.4$
and within the cluster sample no differences between ellipticals and S0
galaxies were found. However, the sample of distant cluster galaxies lacks
galaxies with small sizes and has a magnitude limit of $M_B\le-20.50$, whereas
the field galaxy sample comprises no ellipticals with large sizes
(i.e. cD galaxies) and is more than one magnitude fainter ($M_B\le-19.30$).
A comparison of the sizes and luminosities of these two data sets is
therefore not adequate and would cause misleading results.
Table~\ref{fptabf} gives a summary of the results for the FP
of our early-type field galaxies.

%{\scriptsize
\begin{table}
\begin{center}
\caption{\label{fptabf}\small{Evolution of the FP in Johnson $B$-band
as derived for the early-type field galaxies in the FDF and WHDF.
$N$ shows the number of galaxies and $\Delta\,\overline{\gamma}$ indicates the
mean FP zero-point offset. In the fourth and fifth column,
the median FP zero-point evolution $\Delta\,\langle \gamma\rangle$ and the 
median evolution in the FP $\Delta\,\langle \mu_{\rm e}\rangle$ [in mag] are
listed. The last column gives the $\pm$1$\sigma$ scatter
of the mean offsets.}}
\vspace{0.1cm}
\begin{tabular}{lrcccc} 
\noalign{\smallskip}
\hline
\noalign{\smallskip} 
Sample & $N$ & $\Delta\,\overline{\gamma}$  & 
$\Delta\,\langle \gamma\rangle$ & $\Delta\,\langle \mu_{\rm e}\rangle$ &
$\sigma_{\gamma}$ \cr
\noalign{\smallskip}
\hline
\noalign{\smallskip}
FDF              & 11 & 0.066 & 0.075 &$-$0.235 & 0.136 \cr
WHDF             & 10 & 0.199 & 0.214 &$-$0.667 & 0.289 \cr
FDF+WHDF         & 21 & 0.130 & 0.089 &$-$0.279 & 0.227 \cr
FDF+WHDF$^{a}$   & 19 & 0.170 & 0.126 &$-$0.393 & 0.197 \cr
\hline
E                &  9 & 0.075 & 0.075 &$-$0.230 & 0.121 \cr
S0               & 12 & 0.171 & 0.211 &$-$0.643 & 0.281 \cr
S0$^{a}$         & 10 & 0.255 & 0.241 &$-$0.735 & 0.219 \cr
S0$^{b}$         &  9 & 0.214 & 0.211 &$-$0.643 & 0.185 \cr
\hline
low lum.$^{c}$   & 11 & 0.056 & 0.082 &$-$0.255 & 0.180 \cr
high lum.        & 10 & 0.211 & 0.165 &$-$0.515 & 0.254 \cr
high lum.$^{d}$  &  9 & 0.164 & 0.126 &$-$0.393 & 0.219 \cr
\hline
low mass$^{e}$   & 11 & 0.099 & 0.089 &$-$0.279 & 0.221 \cr
high mass        & 10 & 0.163 & 0.126 &$-$0.393 & 0.240 \cr
\noalign{\smallskip}
\noalign{\hrule}
\end{tabular}
\end{center}
\begin{flushleft}
\small{$^{a}$ Omitting ID\,6336 and ID\,508, both Sa bulges.}\\
\small{$^{b}$ Omitting ID\,6336 and ID\,508 and ID\,111.}\\
\small{$^{c}$ lower-luminosity: $M_{B}>-21.404$, higher-luminosity: $M_{B}<-21.404$.}\\
\small{$^{d}$ Omitting ID\,111 at $z=0.74$.}\\
\small{$^{e}$ less-massive: log\,$\sigma<2.275$, more-massive: log\,$\sigma>2.275$.}\\
\end{flushleft}
\end{table}
%}

The thickness of the FP appears to vary between the individual field galaxy
samples. For the FDF an average rms scatter of 0.136 is detected, whereas the
WHDF show a larger 1$\sigma$ dispersion of 0.289. This effect is partly caused
by three high redshift objects $0.4\le z\le0.74$ in the WHDF, which are all
disc-dominated galaxies, without these object the rms scatter is 0.202.
Dividing the field sample with respect to velocity
dispersion at log\,$\sigma=2.275$ or luminosity at $M_{B}=-21.404$, no
strong variations in the amount of evolution between the sub--samples 
are detected. However, the comparisons rely on the assumed cutoffs in
$\sigma$ and $M_{B}$ because outliers, such as early Sa spiral galaxies,
can strongly influence the derived luminosity evolution.
In the next section, possible differences between the stellar 
populations of elliptical and lenticular galaxies will be addressed.

\subsection{Galaxy Types: Ellipticals versus Lenticulars}\label{eso}

Looking at differences between the morphological types of \textit{field}
early-type galaxies, the S0 galaxies display a stronger evolution than
the elliptical galaxies. Assuming that the slope of the local reference holds
valid for the distant galaxies, the 9 field ellipticals show in edge-on
projection of the FP a zero-point offset 
\begin{equation}
\langle\Delta \gamma^{{\rm f}}_{{\rm E}}\,(z=0.4)\rangle=0.08\pm0.06,
\end{equation}
which corresponds to a brightening in their stellar populations of
$\Delta M^{{\rm E}}_{B}=-0.23\pm0.18^{m}$.
By comparison, the nine field lenticulars exhibit an offset in the zero-point 
with respect to the local Coma galaxies
\begin{equation}
\langle\Delta \gamma^{{\rm f}}_{{\rm S0}}(z=0.4)\rangle=0.21\pm0.09,
\end{equation}
which corresponds to an average evolution of
$\Delta M^{{\rm S0}}_{B}=-0.64\pm0.27^{m}$. 
Note, that this FP offset is very similar if we include the two Sa bulges
and results in the same amount of evolution if also the highest redshift
lenticular galaxy is accounted for, see Table~\ref{fptabf} for details.
Errors on the zero-points of
elliptical and S0 galaxies were individually derived as
\begin{equation}\label{fpgerr}
\delta {\rm ZP}^{2}=\delta {\rm FP}_{B}^{2}+\delta {\rm BS}^{2}
\end{equation}
where $\delta {\rm FP}_{B}$ denotes the total error which enters the FP
in the rest-frame $B$-band and $\delta {\rm BS}$ is the uncertainty computed
through an iterative bootstrap re--sampling of the data points 100 times
\citep{FZBSD05}.
Limiting the S0 galaxies to redshifts $z<0.7$ and rejecting the two Sa bulges,
the same results are derived (see Table~\ref{fptabf}). In addition, the field
ellipticals obey a tight FP with a small 1$\sigma$ scatter of 0.121,
whereas the S0 types have a larger dispersion of 0.219 (omitting the
two Sa bulges). This gives further evidence that the lenticular galaxies
have a larger range of different stellar populations. 

A faster evolution of the $M/L$ ratios (see next section \S~\ref{stpops} 
and equation~\ref{dml} for a description of the connection between the FP and 
$M/L$ ratios) of the S0 galaxies with respect to elliptical galaxies
was also found for \textit{cluster} early-type galaxies at $z\sim0.2$
\citep{FZBSD05}, which are in agreement with a low formation
redshift $1~\le~z_{\rm form}~\le~2$. As the slope of the $M/L-M$ relation for
these distant cluster galaxies is steeper than those for the Coma cluster
galaxies, this can be interpreted as a dependence of the evolution on
galaxy mass, and hence a downsizing effect in the evolution of these galaxies.

The evolution of the FP for the early-type FDF and WHDF field galaxies in the
rest-frame Johnson $B$-band is illustrated in Figure~\ref{FPfw}. 
Early-type galaxies were binned in redshift space to investigate their location
within and along the edge-on FP as a function of redshift and to test the
effects of possible outliers. Ellipticals are represented as filled, S0s and
Sa bulges as open symbols. For each morphological type the number of galaxies
within a redshift bin is indicated in the brackets. Small squares denote the
Coma galaxies and the straight line is a principal component fit to
the local sample \citep{ZTBBFM05}. Figure~\ref{FPfw} clearly shows the
evolution of the FP for the distant galaxies with respect to the local Coma FP.
The offsets of the distant field galaxies increase with redshift but the scatter appears to
be mainly amplified for lenticular galaxies. The two 
galaxies with a ``positive'' evolution in the last two panels are the
Sa bulges ID\,6336 and ID\,508 and show a clear disc on the ACS images.
These objects can mimic on average a weaker evolution in luminosity for the
whole galaxy population (cf. Table~\ref{fptabf}). Regardless of the Sa bulges,
the S0s exhibit a larger scatter at higher redshift $0.26<z<0.75$ than the
ellipticals which suggests that the stellar populations of E and S0 types are
different and that lenticulars are a more heterogeneous group. An explanation
could be that these S0 galaxies resemble post--starburst galaxies as
no strong emission lines were detected in their spectra. 
This statement will be further addressed in \S~\ref{sec:sfd}.

\section{Stellar Population Ages}\label{stpops}

\subsection{Constraints on the Formation Epoch of Early-Type Galaxies}\label{sec:zf}

The observed evolution of the $M/L$ ratio as derived from the FP
depends on the age of the stellar population of the galaxies. In general, the
luminosity of a young stellar population becomes rapidly fainter when the
massive and bright stars which have a short lifetime disappear. For an old
population comprising mainly low mass stars the dimming of the luminosity
proceeds on a more gradual evolutionary path. As a consequence of this, a
stellar population formed at lower  redshift will evolve faster than one
generated at high redshift. In the following, the observed evolution of the
$M/L$ ratio will be compared to simple stellar population models of a single
burst formed at redshift $z_{\rm form}$. The models have been generated
following the description of the analytic models by \cite{vD01b}. 

\begin{figure}
\includegraphics[width=1.0\linewidth]{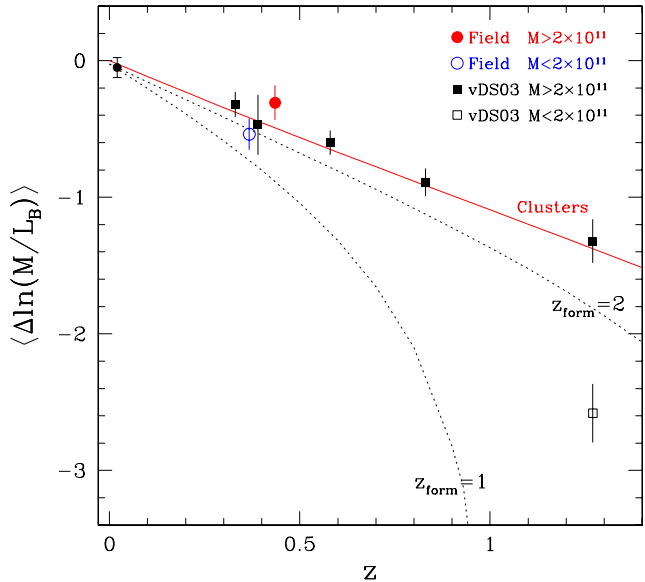}
\caption{\small{\label{dmmlrbz}
The average evolution of the $M/L$ ratio of the \textit{field}
elliptical and S0 galaxies in the FDF and WHDF and the 
cluster early--type galaxies taken from van Dokkum \& Stanford (2003),
in the rest--frame $B$-band.
Dotted lines are model tracks for a single stellar population with
formation redshift 1 (lower line) and 2 (upper line). 
The model tracks are forced to go through the small dot which represents
the average $M/L$ ratio of the field galaxies at $z=0.02$
by Faber et al. (1989). The solid line is the best fitting formation redshift
for the evolution of massive cluster galaxies with masses  
$M>2\times10^{11}M_{\sun}$.
Massive field galaxies with $M>2\times10^{11}M_{\sun}$ show a similar 
slow evolution as cluster galaxies, whereas less--massive field galaxies
with $M<2\times10^{11}M_{\sun}$ evolve faster, similar to less--massive cluster
galaxies.}}
\end{figure}

\begin{figure}
\includegraphics[width=1.0\linewidth]{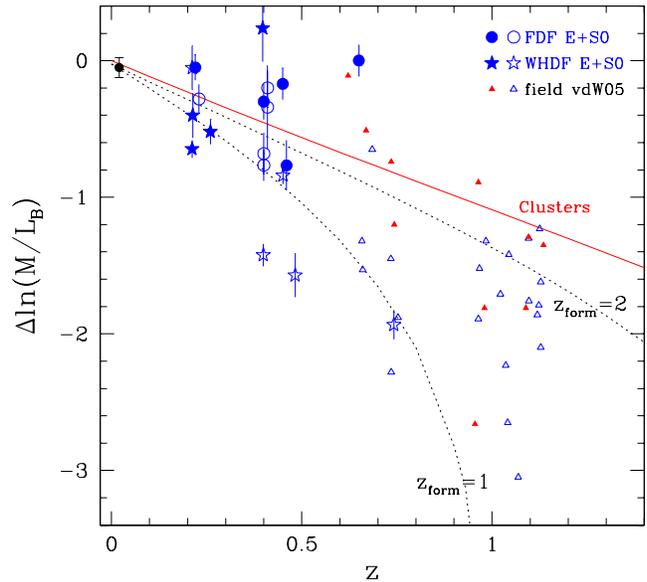}
\caption{\small{\label{dmlrbz}
The evolution of the $M/L$ ratio of the \textit{field}
elliptical and S0 galaxies in the FDF and WHDF and the field 
early-type galaxies at $\langle z\rangle=1$ taken from
van der Wel et al. (2005), in the rest--frame $B$-band.
Dotted lines are model tracks for a single stellar population with
formation redshift 1 (lower line) and 2 (upper line). 
The model tracks are forced to go through the small dot which represents
the average $M/L$ ratio of the field galaxies at $z=0.02$
by Faber et al. (1989). The solid line is the best fitting formation redshift
for the evolution of cluster galaxies with masses  $M>2\times10^{11}M_{\sun}$.
The scatter in the $M/L$ ratios is large but there is a clear
trend for field S0 galaxies to evolve faster. Field ellipticals appear to form 
two separate groups with respect to $M/L$ ratios.}}
\end{figure}

The luminosity evolution of a single-age stellar population can be described by
a power law as
\begin{equation}
L\propto\frac{1}{(t-t_{\rm form})^{\kappa}},
\end{equation}
where $t_{\rm form}$ is the stellar formation time which corresponds to a
formation redshift of $z_{\rm form}$ \citep[e.g.,][]{Tin80}.
The coefficient $\kappa$ depends on the IMF, the
metallicity and the chosen passband in which the luminosity is measured.
In the stellar population models by \cite{BC03}, a normal IMF
with \cite{Sal55} slope, solar metallicity and $\kappa_B=0.96$ for
the rest--frame $B$-band was adopted.
Model predictions are shown in Figures~\ref{dmmlrbz} to
\ref{dmlrbz} for the field elliptical
and S0 galaxies in the rest--frame $B$-band. Note that the
predicted evolution of the models is independent of $H_0$ as the age
dependence of the $M/L$ ratio is a power law. Two different model tracks for
a single stellar population with formation redshift $z_{\rm form}=1$ 
(lower line) and  $z_{\rm form}=2$ (upper line) are indicated as the dotted
lines in Figures~\ref{dmmlrbz} to \ref{dmlrbz}.

To assess the evolution of the stellar populations in the 
early--type galaxies, the evolution of the FP zero-point $\gamma$ for the distant
galaxies can be directly converted into an evolution of the average 
$M/L$ ratio as
\begin{equation}
\langle\Delta{\rm log}(M/L)\rangle=-\frac{\langle\Delta \gamma\rangle}{2.5\beta},
\end{equation}\label{dml}
where the average evolution of $\Delta{\rm log}(M/L)$ was
derived via the average zero-point offset of the distant FP
$\langle\Delta \gamma\rangle$ from the local Coma FP 
\citep{Treu01a}.

In a recent study of 27 field early--type galaxies between
$0.6<z<1.15$ a mass--dependent evolution was found \citep{vdW05}.
This sample of high--redshift galaxies showed that the evolution of low--mass 
field galaxies with a characteristic mass of $M<2\times10^{11}M_{\sun}$ 
is faster than their more massive counterparts.
Similar results have been drawn from field early--type galaxies
at $z\sim1$ from the GOODS-N sky region \citep{T05} and
from the K20 survey \citep{SVCL05}.
The 21 field early--type galaxies in our sample are located at a lower
redshift range of $0.2<z<0.75$. But it is very interesting, if already at a
look--back time of $\sim$5~Gyr a possible evidence for a mass--dependent
evolution or at least a trend with galaxy mass can be detected.
For this reason, the average evolution of the $M/L$ ratio of the field
elliptical and S0 galaxies in the FDF and WHDF was derived.

Figure~\ref{dmmlrbz} displays the average evolution of the $M/L$ ratios for the
field galaxies in the rest--frame $B$-band compared to the early--type cluster
galaxies taken from \cite{vDS03}. For a compilation of massive early--type
galaxies with $M>2\times10^{11}M_{\sun}$ in rich clusters of galaxies taken
from the literature these authors deduced 
an average  formation redshift of $z_{\rm form}\approx3$,
which is indicated as the  solid line.
All calculated model tracks are forced
to go through the average $M/L$ ratio of the field galaxies at $z=0.02$
by \cite{FWBDDLT89}, shown as the small dot in Figures~\ref{dmmlrbz}
and \ref{dmlrbz}. The field FDF and WHDF sample was
splitted according to the characteristic mass at $M=2\times10^{11}M_{\sun}$,
which gives 13 lower--mass and 8 higher--mass galaxies. The average evolution of
the higher--mass field galaxies is 
$\langle\Delta \ln({M/L_B})\rangle=-0.31\pm0.06$ at a redshift of 
$\langle z\rangle=0.44$.
Lower--mass galaxies indicate on average a stronger evolution of  
$\langle\Delta \ln({M/L_B})\rangle=-0.54\pm0.07$ at $\langle z\rangle=0.38$. 
The slight deviation in the redshift is solely due to the not uniform number of
galaxies in each sub--sample. Massive field galaxies are in compliance with a
slow evolution as derived for the massive cluster galaxies and a high
formation redshift for the bulk of the stars of $z_{\rm form}\approx3$. 
In contrast, less--massive field galaxies in the FDF and WHDF with 
$M<2\times10^{11}M_{\sun}$  evolve faster and follow an evolutionary track 
between $1\la z_{\rm form}\la2$ which is similar to the less--massive cluster
galaxies from \cite{vDS03}, which is indicated as the open
square in Figure~\ref{dmmlrbz}. This result gives evidence that already at a 
look--back time of $\sim$5~Gyr differences in the measured $M/L$ evolution
for the FDF and WHDF  field galaxies can be revealed.

% ########################### Figure: dmlrbz B #################################
% ##############################################################################

% ########################### Figure: dmlrbz B #################################
% ##############################################################################

To translate the evolution in the $M/L$ ratios into a specific formation 
redshift $z_{\rm form}$ for the field galaxies,
the evolution of the $M/L$ ratios for the individual field
early--type galaxies are investigated. Figure~\ref{dmlrbz} illustrates 
the offset from the local FP in the rest--frame $B$-band for 
elliptical and S0 galaxies in the FDF and WHDF.
A comparison sample of 27 field early-type galaxies, splitted into
less-massive $M<2\times10^{11}M_{\sun}$ (blue, open symbols)
and more-massive field galaxies $M>2\times10^{11}M_{\sun}$ (red,
filled symbols), at high redshift $0.6<z<1.15$ by \cite{vdW05} is also shown.
Again, model tracks for a single-age
stellar population with formation redshift 1 (lower line) and 2 (upper line)
are displayed as dotted lines.  
All model tracks were normalised to
the average $M/L$ ratio of the field galaxies at $z=0.02$
by \cite{FWBDDLT89}, indicated as the small black dot.
The evolution of cluster galaxies with masses  $M>2\times10^{11}M_{\sun}$
can be best approximated assuming a formation redshift of 
$z_{\rm form}\approx3$, which is denoted as the solid line in 
Figure~\ref{dmlrbz}.
For our field early-type galaxies we measure an evolution of
$\Delta \ln({M/L_B})=-(1.71\pm0.18)\,z$.
Although the scatter in the offsets from the local FP is large, a clear
trend for a faster evolution of field S0 galaxies is found. On the other hand,
field elliptical galaxies appear to form two separate groups with
respect to $M/L$ ratios.
To assess the formation redshift $z_{\rm form}$, we split our field galaxy
sample into a low- and high-mass sample at $M=2\times10^{11}M_{\sun}$. 
An unrestricted linear $\chi^2$-fit to the low-mass early-type field galaxies
with $M<2\times10^{11}M_{\sun}$ gives
\begin{equation}
\Delta \ln({M/L_B})=-(1.25\pm0.56)\,z-(0.08\pm0.41).
\end{equation}
This would correspond to a formation redshift of $z_{\rm form}=1.9\pm0.5$.
The error on $z_{\rm form}$ denotes the observed (1$\sigma$) scatter in the
$\Delta \ln({M/L_B})-z$ relation.
For the higher-mass galaxies with $M>2\times10^{11}M_{\sun}$ a linear
$\chi^2$-fit yields
\begin{equation}
\Delta \ln({M/L_B})=-(0.89\pm0.75)\,z-(0.02\pm0.25).
\end{equation}
For the massive early-type field galaxies, the evolution of the $M/L$ 
predicts a mean star formation epoch for the bulk of their stellar
populations of $z_{\rm form}=3.5\pm1.3$. 
In comparison to the lower-mass counterparts, a higher formation redshift is 
significant on the 2$\sigma$ level based on the uncertainty of the 
high-mass galaxies. 
The lack of field galaxies with masses of
$M>2\times10^{11}M_{\sun}$ introduces a larger uncertainty on the 
$z_{\rm form}$ value. By comparing the derived formation epoch of the
high--mass cluster galaxies with numbers for cluster galaxies in
the literature a good agreement is found. Based on a compilation of clusters
out to redshift of $z=1.27$, \cite{vDS03} deduce 
$z_{\rm form}=2.6^{+0.9}_{-0.4}$. This study gives an upper limit
for the formation redshift of massive cluster galaxies of
$z_{\rm form}=3.5$~(1$\sigma$). The result obtained for the massive field
galaxies here is within this upper limit, which suggests that these field 
galaxies have old passively evolving stellar populations similar to their
massive counterparts in clusters.
Besides the scatter of the individual galaxies, lower-mass field galaxies have a
stronger evolution in their $M/L$ ratio. The mean stellar population ages of
the less-massive field galaxies are younger, with lower 
formation redshift of $1<z_{\rm form}\le2$.
This suggests a \emph{mass--dependent luminosity evolution} with a
stronger evolution for lower--mass galaxies with $M<2\times10^{11}M_{\sun}$. \\

\begin{figure}
\includegraphics[width=1.0\linewidth]{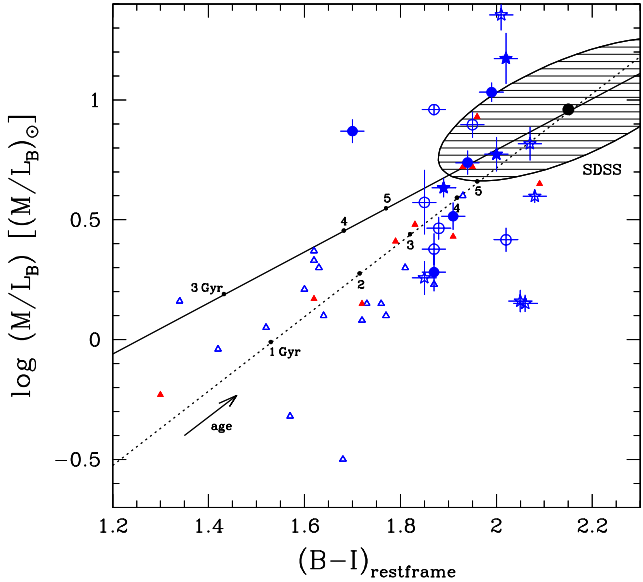}
\caption{\label{fig:biml}\small{
$M/L_B$ as a function of the rest-frame $(B-I)$ colour 
for early-type field galaxies in the FDF and WHDF. Symbol notations as in
Figure~\ref{dmlrbz}. 
The median scatter in the $M/L_B$ and $(B-I)$ colour distribution of massive
field early-type galaxies ($M>2\times 10^{11}$) in the SDSS is indicated
by the shaded ellipse. For comparison, two different Bruzual \& Charlot model
predictions with solar metallicity are shown, for a single stellar population
(dotted line) and for an exponentially declining star formation with
$\tau=1$\,Gyr (solid line). Both model tracks are forced to match the local
values as defined by the SDSS comparison sample.    
}}
\end{figure}

\begin{figure}
\includegraphics[width=1.0\linewidth]{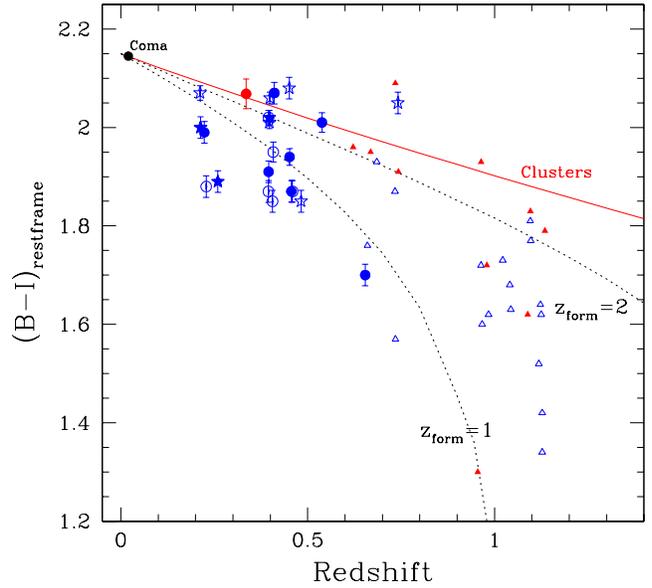}
\caption{\label{fig:bizw}\small{
Evolution of the rest-frame $(B-I)$ colour with redshift for early-type
field galaxies in the FDF and WHDF. Symbol notations as in
Figure~\ref{dmlrbz}. The solid (red) point shows the median $(B-I)$ colour
of early-type cluster galaxies in the FDF.}}
\end{figure}

The rest-frame optical colours offer a powerful tool to investigate the
origin of the scatter in $\Delta \ln({M/L_B})$.
Figure~\ref{fig:biml} displays the $M/L$ ratios as a function of the
rest-frame $(B-I)$ colour for our early-type field galaxies.
As expected from stellar population model predictions, galaxies with lower 
$M/L$ ratios should have bluer colors than galaxies with higher $M/L$.
Indeed, lower-mass galaxies (with low $M/L$ ratios) are on average bluer
than higher-mass galaxies (with high $M/L$ ratios). Galaxies
with low $M/L$ ratios follow the predictions of a single stellar population
with solar metallicity by \cite{BC03}, indicated as  the dotted line in
Figure~\ref{fig:biml} and form an extension to the higher redshift sample by
\cite{vdW05}. Field galaxies with higher $M/L$ ratios populate a transition
zone to the color and $M/L$ ratios regime as defined by the local SDSS sample.

In Figure~\ref{fig:bizw} the rest-frame $(B-I)$ colour as a function of
redshift is displayed for our early-type field galaxies.
As a comparison, we also indicate the median $(B-I)$ colour derived for
early-type cluster galaxies in the FDF (see also \S~\ref{sec:zdis}).
If we compare the Figures~\ref{dmlrbz} and \ref{fig:bizw}, 
strong similarities can be found. The strong correlation between
$M/L$ and galaxy colour and a similar evolution in $M/L$ and in colour
indicates that the observed evolution and scatter in $M/L$ are intrinsic to
the galaxies and due to differences in their stellar populations.
Note, that in this Figure also galaxies for which no $M/L$ ratios could be
derived are shown. If the sample is equally split according to the
characteristic mass separation as introduced in Figure~\ref{dmmlrbz},
lower-mass galaxies with $M<10^{11}M_{\sun}$ have on average bluer colours
(median colour of $\langle(B-I)\rangle=1.88$ than more-massive galaxies
with $M>10^{11}M_{\sun}$ which feature redder colours of
$\langle(B-I)\rangle=2.00$.

% ########## Figure: M/L ratio for FDF and WHDF ###############################
\begin{figure*}
\includegraphics[width=1.0\linewidth]{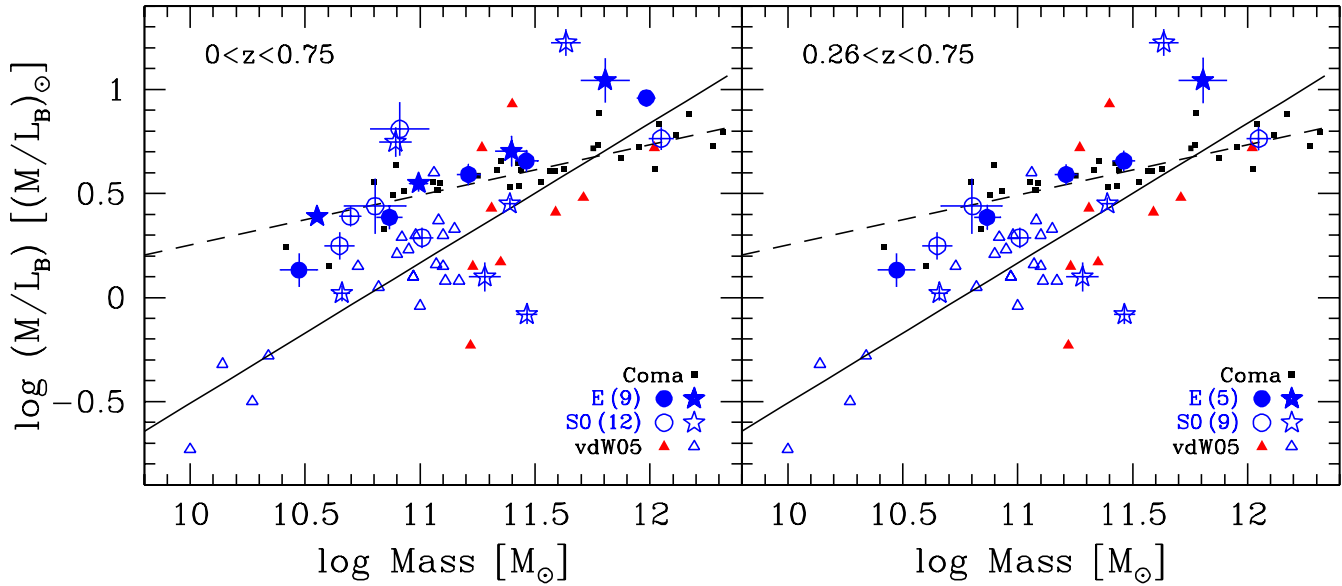}
\caption{\small{\label{MLMfw}$M/L_B$ ratio as a function of mass $M$
in solar units as derived from the FP
for the early-type field galaxies in the FDF and WHDF,
divided into two different redshift bins. Filled symbols denote ellipticals,
open symbols S0 galaxies and Sa bulges. In brackets the respective
number of a morphological class is listed. Errors for the distant sample
are shown in both panels. The nearby Coma cluster sample 
is indicated with small dots (dashed line is a $\chi^2$-fit to Coma) and 
the  high-redshift field E+S0 galaxies at $z\sim1$
by van der Wel et al. (2005) are shown as triangles (solid line is a
$\chi^2$-fit). All data points have been corrected for maximum progenitor
bias $\Delta \ln({M/L_B}) \approx0.2\,z$.
\textit{Left:} $M/L_B-M$ for the whole sample of 21 field
E+S0 galaxies $0<z<0.75$. \textit{Right:} $M/L_B-M$ for 14 higher-redshift 
field E+S0s with $0.26<z<0.75$. Ellipticals follow the local relation whereas
the S0 galaxies have on average lower $M/L$ ratios, a larger scatter and match
the intermediate--mass range of the $M/L_B-M$ relation of the $z\sim1$ field
E+S0 galaxies.}}
\end{figure*}
%
% ##############################################################################

To test the detected mass--dependent evolution even further, Figure~\ref{MLMfw}
displays the $M/L_B$ ratio of the early-type field galaxies 
as a function of mass $M$. The galaxies were divided according to their redshift
into two different bins. The left panel shows the  $M/L_B-M$ for all
field galaxies, whereas in the right panel the field objects are restricted to
a sub-sample of 14 high-redshift galaxies.
As a comparison, the nearby Coma cluster sample (small dots)
and the  high-redshift field E+S0 galaxies at $z\sim1$
of \cite{vdW05} (triangles) are indicated together with linear
$\chi^2$-fits to both samples. All data points have been corrected for 
maximum progenitor bias $\Delta \ln({M/L_B}) \approx0.2\,z$.
Ellipticals, in particular below $z\la0.26$,
follow the local relation whereas the S0 galaxies have on average lower
$M/L$ ratios and match the intermediate--mass range of the $z\sim1$ field E+S0
galaxies. Lenticular galaxies show a larger scatter and the two
outliers in the FP represent the two Sa bulges (ID\,6336 and ID\,508)
that are clearly offset from the rest of  the sample.   
The difference between field elliptical and S0 galaxies found in the FP can be
explained by a difference in their formation epochs. Ellipticals, especially
massive ones, were generated at earlier times and follow rather moderate 
quiescent evolutionary tracks. On the contrary, lenticular and Sa bulges show 
more diverse stellar populations which are younger and have been formed at
recent times with formation redshifts $1<z_{\rm form}\le2$.

\subsection{Independent Diagnostics of Star Formation}\label{sec:sfd}

In Figure~\ref{fig:biml} we have shown that the mass-to-light ratio correlates
well with the rest-frame $B-I$ color. As independent diagnostic tools
for star formation, we also compare the equivalent widths of \oii\ and \hda\
to stellar population models.

Using the \oii\ and \hda\ star formation indicators (see \S~\ref{sec:ssf})
allows us to probe different time scales of star formation episodes in our
sample galaxies. Assuming a passive evolution, for systems with old stellar
populations (average age of $>$1~Gyr) with
no emission lines in their spectra, the equivalent widths of Balmer absorption
lines decrease continuously over time. Within the first $\sim 10^{8}$ yr of the
generation of new stars in these galaxies the equivalent width of H$\delta$
increases rapidly up to its maximum value around 10\AA. Afterwards the Balmer
line decreases steadily until the stellar population reaches an age of 2 Gyr.
This can be interpreted that substantial star formation events
($\sim$30 $M_{\sun}$ yr$^{-1}$) occurred in the first $\sim 10^{8}$ yr of the
systems with subsequent weak star formation thereafter. In case of the
\oii\-emission line, it offers an diagnostic that is sensitive to ongoing or
very recent star formation episodes ($\la$50 Myr).

\begin{figure}
\includegraphics[width=1.0\linewidth]{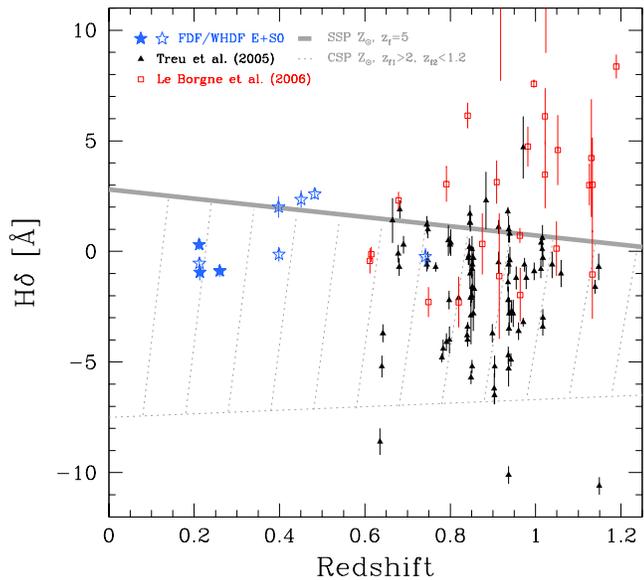}
\caption{\label{fig:hdel}\small{
Evolution of the \hda\ line strength as a function of redshift for
a set of two-burst stellar population models consisting of a main star burst
forming 90\% of the total stellar mass at $z_{\rm f}=5$, plus a secondary burst
forming 10\% of the stellar mass at $z_{{\rm f2}}<z_{{\rm f1}}$ with
$z_{{\rm f2}}<1.2$ (models from Bruzual \& Charlot 2003).
\hda\ measurements for our field E+S0 galaxies 
are denoted by the large symbols, the \hda\ measurements by Treu et al. (2005)
and H$\delta$-strong galaxies in the GDDS by Le Borgne et al. (2006) 
are shown as the triangles and open squares, respectively.
}}
\end{figure}

The evolution of the \hda\ absorption line strength with redshift of our field
galaxies is presented in Figure~\ref{fig:hdel}. We compare
our \hda\ measurements to the \hda\ determinations of field galaxies up to
$z\sim1$ by \cite{T05} and galaxies with strong H$\delta$ absorption 
in the Gemini Deep Deep Survey (GDDS) by
\cite{GDDSLB06}. For consistency with previous studies we define systems
with $\hda\,>4$~\AA\ as being galaxies with strong H$\delta$ absorption.
Two sets of different stellar population models 
by \cite{BC03} are also indicated. Compared to the H$\delta$-strong galaxies,
the field galaxies in the WHDF and FDF exhibit lower \hda\ absorption 
features. Non of our systems exhibit very strong \hda\ equivalent width, in
particular all galaxies gave $\hda\,\leq 2.60\pm0.23$~\AA. Therefore,
they are most likely not systems with strong H$\delta$ absorption. They follow
the evolution track of the composite stellar evolution model consisting of 
an initial burst of SF at $z_{{\rm f1}}>2$ and a secondary SF burst at
$z_{{\rm f2}}<1.2$. A single stellar population model consisting of a
main star burst forming 90\% of the total stellar content at $z_{\rm f}=5$
is consistent with three galaxies. All these galaxies show both stronger \hda\
absorption and additional \oii\,3727 emission in their spectra 
and have experienced a very recent episode of star formation (see next
section~\ref{sec:rsf} for further details).

\subsubsection{Recent Star Formation}\label{sec:rsf}

Six galaxies in our sample feature \oii\,3727 emission lines in their spectra.
Two of these are spiral bulges (ID\,158 and ID\,508), three are of
intermediate type S0/Sa (ID\,946) and E/S0 (ID\,111 and ID\,173)
and one is an elliptical galaxy (ID\,437). For these objects, the main
questions of interest are the star formation rate (SFR), the mass fraction
involved in the star formation and the duration of the star formation episode.

In oder to convert the \oii\,3727 equivalent width to star formation
rates, we assume that the equivalent width measured within the aperture is a
representative diagnostic for the global value of the galaxy. The observed
luminosity $L$(\oii)$_{\rm obs}$ of the \oii\ emission line can be derived
as:
\begin{equation}
L(\oii)_{\rm obs}=(1.4\pm0.3)\times10^{29}\frac{L_B}{L_{B,\odot}}\,
{\rm EW}(\oii) \ [{\rm ergs\ s}^{-1}],
\end{equation}
with $L_B/L_{B,\odot}=10^{0.4\,(M_{B,\odot}-M_B})$ and $M_{B,\odot}=5.48$
\citep{Ken92b,BMYCE97}. 
The observed (measured) $L$(\oii)$_{\rm obs}$ is related to the intrinsic 
$L$(\oii)$_{\rm int}$ luminosity as
\begin{equation}\label{LiLo}
L(\oii)_{\rm int}=3.11\times10^{-20}\,L(\oii)^{1.495}_{\rm obs}.
\end{equation}
Equation \ref{LiLo} takes into account the intrinsic reddening.
To derive the total SFR in units of
solar masses per year, we use the relation established by
\cite{KGJ04}:
\begin{equation}
{\rm SFR}=6.58\times10^{-42}\,L(\oii)_{\rm int}\ \ [M_{\odot}\,{\rm yr}^{-1}].
\end{equation}
Since we do not know the oxygen abundances of our galaxies, we have chosen to
adopt the average calibration by Kewley et al.

Using our observed \oii\ equivalent widths, we derive for the six
emission-line galaxies a median SFR of 6.2 $M_{\odot}\,{\rm yr}^{-1}$.
For the four E+S0 galaxies we find a median SFR of 
3.7 $M_{\odot}\,{\rm yr}^{-1}$, whereas the SFR for two Sa bulges is
15.5 $M_{\odot}\,{\rm yr}^{-1}$. Instead, transforming our observed \oii\
equivalent widths to rest-frame and assuming the relationship reported by
\cite{BMYCE97} with a global extinction correction of E(H$\alpha$)=1~mag,
we find for the whole sample a median SFR of 2.5 $M_{\odot}\,{\rm yr}^{-1}$.
The SFRs for the E+S0 and Sa galaxies are 2.5 and 9.1 $M_{\odot}\,{\rm yr}^{-1}$,
respectively. If we correct the absolute $B$-band magnitude of our galaxies
using the luminosity difference of the distant galaxies and the low-redshift
reference sample as derived from the FP, we get a 
mean mass of these galaxies of $1.48\times10^{11}\,M_{\odot}$ (median
$1.98\times10^{11}\,M_{\odot}$). Assuming an 1~Gyr burst of star formation with
the mean SFR found for the \oii\ line would have involved between
3\% to 10\% (median about 5\%) of the stellar mass of the field galaxies.
We would expect that the involved mass in the SF burst should appear to be
strongly dependent on the galaxy mass. However, our sample statistics of
galaxies with recent SF are too small and the galaxies are spread among
different redshifts to give further constraints.

To further constrain the duration and the mass fraction involved in the star
formation episode we compared our measurements to sets of `toy' models that
consist a mix of two stellar populations \citep{BC03}.
Line emission are either caused by a small mass fraction (3\%) involved in the
star formation burst suggesting $z_{\rm form}=1.1$, or if a larger mass
fraction (10\%) is adopted the episode should have been started earlier
$z_{\rm form}=1.7$.

In general, line emission is evidence for star formation activity or 
an AGN which increases the average luminosity of the galaxy and
thereby causing the system
to be offset from the local FP relation. In contrast to what is expected,
the galaxies with detected ongoing SF in our sample indicate no offset along
the edge-on projection of the FP but a tight relation. Based on our
measurements we estimate that the stellar mass fraction involved in the SF
episode must be in order of 25\% to result in a visible effect within the FP. 
The emission-line galaxies in our sample indicate no signs of
interactions or merging events that could trigger the star formation episodes
in these objects. We therefore find no evidence for a connection between star
formation activity and interaction signatures or mergers.

\begin{table*}
\begin{center}
\caption{\label{pres}\small{Compilation of previous results on field
galaxies}}
\vspace{0.1cm}
\begin{tabular}{lccccccl}
\hline
\noalign{\smallskip}
Reference & $\Delta \ln{(M/L_B)}/z$ & $\Delta \ln{(M/L_B)}/z$ & $\Delta \ln{(M/L_B)}/z$ & $\langle\log (M/M_{\odot})\rangle$ & $\langle z\rangle$ & $N$ & Method; Selection \\
          & reported                & high mass               & low mass                &                        &       &      & \\
  (1)     &  (2)                    & (3)                     &   (4)                   &       (5)              &   (6) &   (7) & (8) \\
\noalign{\smallskip}
\hline
Treu et al. (2001)         & $-1.64\pm0.12$ & $-1.55\pm0.27$ & $-1.47\pm1.89$ & $2.3\times 10^{11}$ & 0.29 & 19  & $\sigma_{\star}$; $T$, magnitude, colour\\
van Dokkum et al. (2001)   & $-1.35\pm0.35$ & $-1.29\pm0.39$ & $-1.94\pm0.33$ & $1.5\times 10^{11}$ & 0.42 & 18  & $\sigma_{\star}$; $T$, magnitude \\
Treu et al. (2001, 2002)   & $-1.66\pm0.37$ & $\cdot$ $\cdot$ $\cdot$ & $\cdot$ $\cdot$ $\cdot$ & $2.5\times 10^{11}$ & 0.38 & 29 & $\sigma_{\star}$; $T$, magnitude, colour\\
van Dokkum \& Ellis (2003) & $-1.25\pm0.25$ & $-1.41\pm0.29$ & $-1.77\pm0.24$ & $1.3\times 10^{11}$ & 0.56 & 27  & $\sigma_{\star}$; $T$, magnitude \\
Gebhardt et al. (2003)     & $-1.84$ to $-2.21$ & $-1.59\pm0.49$ & $-2.05\pm0.17$ & $8.5\times 10^{10}$ & 0.64 & 21  & $\sigma_{\star}$; magnitude, absorption lines \\
van de Ven et al. (2003)   & $-1.43\pm0.30$ & $-1.13\pm0.31$ & $-1.71\pm0.30$ & $2.3\times 10^{11}$ & 0.54 & 21 & lensing isothermal, lenses \\
Rusin \& Kochanek (2005)   & $-1.15\pm0.44$ & $\cdot$ $\cdot$ $\cdot$ & $\cdot$ $\cdot$ $\cdot$ & $\cdot$ $\cdot$ $\cdot$ & 0.54 & 21 & lensing complex; lenses \\ 
van der Wel et al. (2005)  & $-1.75\pm0.16$ & $-1.20\pm0.18$ & $-1.97\pm0.16$ & $1.6\times10^{11}$ & 0.90 & 27 & $\sigma_{\star}$; $T$, magnitude \\
Treu et al. (2005)         & $-1.66\pm0.11$ & $\cdot$ $\cdot$ $\cdot$ & $\cdot$ $\cdot$ $\cdot$ & $1.3\times10^{11}$ & 0.56 & 141 & $\sigma_{\star}$; $T$, magnitude \\
This work                  & $-1.71\pm0.18$ & $-0.89\pm0.19$ & $-1.25\pm0.30$ & $1.0\times 10^{11}$ & 0.40 & 24 & $\sigma_{\star}$; $T$, magnitude, colour \\
\noalign{\smallskip}
\noalign{\hrule}
\end{tabular}
\end{center}
\begin{flushleft}
\small{
Column 2 gives the values of the evolution of the $M/L$ ratios as reported in
the listed studies. Column 3 shows the evolution of galaxies more massive than
$M=2\times 10^{11}M_{\odot}$, and column 4 lists the evolution of less massive
galaxies. In columns 5 to 7, average masses, redshifts, and sample sizes ($N$)
of the data sets are given. The last column shows the method of deriving the
velocity dispersion, '$\sigma_{\star}$' using stellar kinematics , 'lensing
isothermal' $\sigma$ estimate from image separation using an isothermal model
and 'lensing complex' uses a more complex model including statistical variations.
When two references are given, the results are based on data of both studies.
The reported values are all uncorrected for selection effects.
Gebhardt et al. (2003) do not give an error and the average evolution
corresponds to $0.8<z<1.0$.}
\end{flushleft}
\end{table*}

\section{Comparison with Previous Results}\label{sec:litcmp}

Over the past few years, several groups have measured the $M/L$ evolution of
field early--type galaxies at different redshifts, either using direct mass
measurements that involve the internal stellar kinematics of the galaxies
\citep{Treu01a,TSCMB02,vDFKI01,vDE03,GFKIS03,vdW05,T05}
or modelling the stellar kinematic structure and mass profile from the
separation of multiple gravitationally lensed images of background galaxies
\citep{vdV03,RK05}. 
In the following we will
compare these previous results in the literature with our findings, 
and comment on possible inconsistencies that exist between the datasets.
Table~\ref{pres} gives a compilation of the $M/L$ evolution as reported
by various groups based on different datasets. 
For comparison, we also list the methods and selection criteria of each of the
studies, where $T$ corresponds to a selection based on morphologies. In the methods
we present the two most popular techniques in measuring the internal kinematics
of E+S0 galaxies, either using the stellar velocity dispersion $\sigma_{\star}$
or using separation of gravitational lensed images via an isothermal model
('lensing isothermal') or more complex algorithms ('lensing complex').  
Besides the magnitude limited samples, there are two studies that analysed
the same sample of lensing galaxies \citep{vdV03,RK05}. The predictions based
on lensing properties of early-type galaxies agree with one another to within
their quoted uncertainties. Most of these studies report that early-type lens
galaxies appear to have nearly isothermal mass density profiles \citep[e.g.,][]{vdV03}.
\cite{RK05} took a different approach and described the mass
distribution using statistical model assumptions where the galaxy populations
are described by a range of star formation redshifts allowing for a scatter
created by deviations from homology or spread in formation epochs.
Both methods of lensing and direct mass measurements agree quite well within
their uncertainties, although lensing appears to predict a slightly slower
$M/L$ evolution of the galaxies. From comparison with other results, we found
that differences in the fitting technique can change the amount of evolution
in $\ln{(M/L_B)}/z$ for each data sample by up to 26\%.

Using direct kinematic measurements, two groups have 
drawn samples of field galaxies from the CDFS to study their evolution at
$0.6\la z<1$ \citep{vdW05,SVCL05}.
However, these different teams found rather inconsistent galaxy sizes and
velocity dispersions for both individual and same objects. Discrepancies arise
most likely because these studies used different fitting algorithms
for both galaxy parameters (sizes and kinematics).
\cite{SVCL05} argue that their smaller sizes are due to a
selection effect of compact galaxies. These objects are rather faint
and have low velocity dispersions (1/3 of the galaxies have $\sigma<100$~\kms)
and thus could be either biased to early-type spirals or
low-luminosity early-type galaxies containing faint discs that are
unresolved. This argument gets support by their small measured
S\'{e}rsic indices.
In addition, the large spectroscopic apertures used in their work may
cause to underestimate a substantial rotational support 
in the case of faint low-luminosity galaxies. 
The most likely explanation between the differences of \cite{SVCL05} and
\cite{vdW05} might be the selection of the target objects and the different
analysis tools, where the former primarily selected fainter galaxies whereas
the latter included also higher-luminosity (hence more-massive) galaxies.

Additional selection effects may arise from small number statistics, noise
introduced by cosmic variance or the use of shallow imaging data which results
in a contamination by bulge-dominated spiral galaxies. Further, low-$S/N$
spectra cannot disentangle galaxies with post-starburst spectra
(i.e., H$\delta$-strong galaxies) from spectra of normal passive galaxies. 
Unfortunately our current analysis is limited with respect to the sample
size. However, effects due to imperfections of data quality
can completely be ruled out in case of our investigation, as our
sample is based on deep imaging photometry and high $S/N$ galaxy spectra.
Apart form the work of \cite{GFKIS03}, our study comprises the field sample 
with the lowest average masses of galaxies. Therefore, the lower-mass range
where effects of differences in the evolution should be more pronounced can
be investigated in greater detail.
In the nearby future, we will address both important
restrictions of number statistics and cosmic variance that impact high
redshift studies (A. Fritz et al. 2009a, in preparation).

%%%%%%%%%%%%%%%%%%%%% further discussion %%%%%%%%%%%%%%%%%%%%%%%%%%%%%%%

\section{Discussion and Implications}\label{disc}

\subsection{\label{agn}AGNs}

Some previous studies have investigated the possible influence of the
nuclear activity on the evolution of early-type galaxies.
\cite{WULMM04,WUMLM05} find no differences in the evolution of the FP for
BL Lac objects and radio galaxies with a central AGN from normal
early-type galaxies up to $z\sim 0.5$. However, their sample consists
of high-luminosity galaxies, hence it is biased towards high-mass galaxies,
in particular $M>3\times10^{11}M_{\sun}$. 
\cite{vdW05} detected in an FP analysis of field early-type galaxies up
to $z\sim1$ that 4 out of 11 galaxies at $z<0.8$ (36\%) and one out of 16 
for $z>0.9$ show AGN activity. Although their galaxies with AGN are twice
as massive than galaxies without AGN they show similar $M/L$ evolution
and (no blue) rest-frame $(B-I)$ colours. The authors suggest that the
difference between AGN and non-AGN galaxies may become more apparent at
higher redshift $z>0.8$.
A similarly low number of AGN galaxies was detected within 
the sample of \cite{T05}, who identified out of 26 E+S0 X-ray counterparts
nine AGN based on X-ray luminosity (XAGN), 16 LLAGN (see section~\ref{xray}
and Appendix \S~\ref{xrayp} for these group definitions), but only two out
of 13 \oii\ -emitting E+S0 galaxies as secure AGN. The total early-type galaxy
sample of \cite{T05} has a quite low contribution of AGN activity (16\%) and
none of the spectra shows signs of significant SFR, with an
upper limit of $\sim$30 $M_{\sun}$ yr$^{-1}$.

We find three candidates in our field galaxy sample ($\leq$13\%) that
show weak evidence for hosting an AGN (see section~\ref{xray}). Our
observations suggest a very mild evolution in the absolute numbers of AGN
hosts up to $z\approx 0.7$. This is consistent with the results of the
stellar mass function of AEGIS field galaxies \cite{BGN08}, where
the abundance of X-ray selected AGN host galaxies
($L>$10$^{42}$\,ergs\,s$^{-1}$) decreases roughly by a factor of two from
$z\approx 1.2$ to $z\approx 0.5$, but remains relatively flat across the
stellar mass range of $10.35<\log (M/M_{\sun})<11.70$.
However, possible contributions due to LLAGN and/or emission from soft
extended X-ray halos or LMXBs could have influence on the activity of our
individual galaxies as well.
For our field galaxies we measure an average contribution 
of LMXBs to the X-ray emission in the $K$-band of $L_X$(LMXB)$\leq$18\%$\pm$6\%.
The X-ray luminosities in our distant galaxies show a weak excess 
in the soft diffuse X-ray plasma emission, depending on the established 
local $L_X-L_B$ relationship, between $\leq$7\%$\pm$3\% to $\leq$26\%$\pm$15\%
compared to the local reference. The increased luminosities are most likely
originating from a net effect of star formation and/or LLAGN. A more detailed
discussion of these findings and their implications can be found in the
Appendix \S~\ref{xrayp}.

Blue (central) cores in galaxies are not a stringent indicator for a central
AGN power source. For example, within the field early-type galaxy sample of
\cite{T05} only 8\% show signs of central blue rest-frame colours with
$\delta(B-V)\ge0.2$, and only two of these objects (1\%) appear also to host
an AGN based on their high X-ray luminosity ($>$10$^{42}$\,ergs\,s$^{-1}$).
This suggests that the colour information
of the galaxies is not a very powerful technique to distinguish active
from non-active galaxies. As there is no multi-band HST photometry for our
sky fields available, we thus cannot 
investigate the colour gradients in our E+S0 galaxies.

\subsection{\label{quen}Signatures of Star Formation Quenching}

To establish and explore the evolutionary link between E+S0 galaxies at high
redshift and their local descendants, we assume that E+S0 galaxies comprising
passive (red and ``dead'', non-starforming) old stellar populations at
high redshift will only be subject to a pure passive evolution in luminosity
through $z=0$. Our approach is justified by 
the colour evolution of galaxies populating the
red sequence following a pure passive evolution \citep{Bell04,Fab07}. In the
next step, we adopt the luminosity dimming as derived from the evolution of
the FP (see \S~\ref{stpops}) as
$\Delta \ln({M/L_B})=-(1.71\pm0.18)\,z$ and correct for progenitor bias.
This latter effect also compensates for a possible (morphological)
transformation of galaxies which end up as being classified as passive local
E+S0 galaxies. Massive field E+S0 galaxies show a smaller offset from the
$M/L$-relation, whereas less massive ones evolve faster which is supported
by younger stellar populations (see section~\ref{sec:zf}). Independent of
the luminosity of our galaxies, we apply this relationship to all galaxies in
our sample and then investigate a possible dependence of the evolution on
luminosity and the physical mechanisms that drive our measured mass dependent
evolution. The luminosity of the distant E+S0 galaxies is then decreased by a
rate of $\Delta \log L_B=0.74\,z$ to obtain their absolute (rest-frame)
$B$-band magnitude at $z=0$ as
\begin{equation}
M_B(z=0)=M_B(z)+(0.78\times 2.5\times z)=M_B(z)+1.95\,z.
\label{Levp}
\end{equation}
This luminosity is the resulting characteristic brightness at $z=0$ to which the
galaxies would fade, assuming a pure passive evolution. The rate of luminosity
evolution in equation~\ref{Levp} is consistent with the predictions of stellar
population synthesis models that formed the bulk of the stars in E+S0 galaxies
at $z\ga1.5$ \citep{Bell04}.

\begin{figure}
\includegraphics[width=1.0\linewidth]{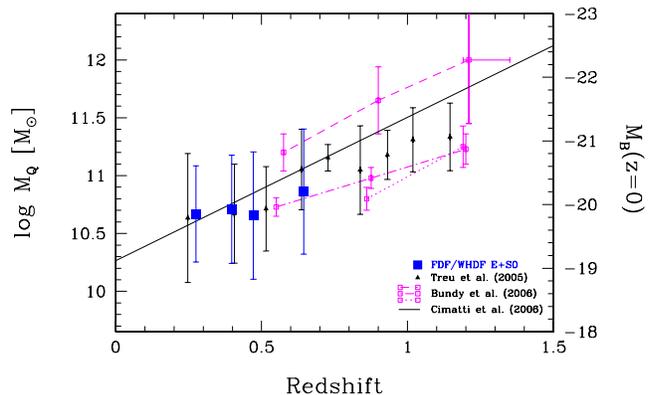}
\caption{\label{fig:mbz}\small{
Evolution of the quenching mass $M_Q$ and rest-frame luminosity (corrected
for passive evolution) for different redshift bins.
Field E+S0 galaxies in the FDF and WHDF are denoted by large squares, 
field E+S0s by Treu et al. (2005) as triangles. 
The evolution of the quenching mass by Bundy et al. (2006) is 
shown for three different sets of field galaxies, for 50\% of
late-type morphologies (dashed), 50\% galaxies with blue
$(U-B)$ colour (dot-dashed), and 50\% with high \oii\ SFR (dotted).
The solid line indicates the passively evolved average LF for E+S0
galaxies by Cimatti et al. (2006) at which about 60\% of
the E+S0 galaxies are already in place. 
There is a weak decrease of $M_Q(z)$ towards lower redshifts, which suggests 
that star formation is efficiently suppressed in more massive galaxies.
}}
\end{figure}

In Figure~\ref{fig:mbz} we compare the evolution 
of the rest-frame luminosity $M_B(z=0)$ (corrected for expected passive
evolution) and stellar mass as a function of redshift for various field
early-type galaxy samples.
Several implications can be drawn from this diagram. 
First, there is a clear trend of $M_B(z=0)$ becoming more luminous with
increasing redshift for the field E+S0 galaxies in the FDF and WHDF
(large squares). 
Our results are in good agreement with the findings for field E+S0 galaxies 
up to $z\sim1$ by \cite{T05}, shown as small triangles together with their
associated errors ($1\sigma$ scatter). These measurements have also been
corrected with equation~\ref{Levp}, but using the evolution as derived 
from their FP.

Moreover, the brightening detected for our galaxies indicates that the most
massive E+S0 galaxies are the first to build up the red sequence whereas less
massive ones join at more recent epochs ($z\la 1$). The observed trend of both
the evolutionary corrected luminosity and total stellar mass is consistent with
that of a characteristic redshift evolution of the so-called ``quenching'' mass
$M_Q\propto(1+z)^{4.5}$ \citep{BEC06}.
The quenching mass limit $M_Q(z)$ is a diagnostic that traces the total
baryonic (stellar) mass scale and a direct by product of the mechanism that
drives downsizing. It is defined as the fraction of star forming
galaxies (usually 1/3 of the total fraction) which contributes to the total
(star forming) mass budget.
Above this mass threshold $M_Q(z)$, star formation in stellar systems is
suppressed and the majority of galaxies have already been
transformed into passive, quiescent galaxies. The
global trend of a downward evolution of the transitional mass with increasing
age of the Universe, where $M_Q(z)$ decreases by a factor of $\sim$5 from
$z\sim 1$, is well reproduced by measurements based on the COMBO-17
and DEEP2 redshift surveys \citep{Bor06,BEC06,HBHE07}. This transitional
(crossover) mass is computed in different redshift bins for 50\% late-type
(spirals and peculiars) morphologies (dashed line in Figure~\ref{fig:mbz}),
50\% galaxies with blue $(U-B)$ colour (dot-dashed line), and 50\% galaxies
with \oii\,-inferred SFRs above 0.2\,$M_{\odot}\,{\rm yr}^{-1}$ (dotted line).
These trends can also be fairly well described by the solid line in
Figure~\ref{fig:mbz}, which indicates the average evolution of the passively
evolved LF for early-type galaxies from the COMBO-17, DEEP2 and SXDS with
redshift \citep{CDR06} at which the bulk ($\approx$60\%) of the E+S0 galaxies
are already in place. While our observed field galaxies are in good agreement
with the model predictions using blue $(U-B)$ color criterion, in particular
for fainter (hence lower mass) individual field galaxies
($M_B(z=0)>-20.5$, $M<10^{11}M_{\sun}$), models with high SFR (as measured from
the \oii\ emission strength) are not consistent with the observations. 
There is a weak increase of the quenching mass $M_Q(z)$ with redshift
which is already apparent at lower redshifts $z<0.7$, which continuous up to
higher redshifts $z\sim1.2$. This trend suggests that star formation is more
effectively suppressed in more luminous (massive) systems.
Strong suppression of star formation could be triggered from AGN feedback
of weak low luminosity (radio-mode) AGN that have red optical colours and are 
situated in very massive dark matter haloes \citep{CSW06,S07}.

Figure~\ref{fig:mbz} suggests further that the evolution is a strong function
of luminosity and mass and the evolution depends on the (evolution corrected)
luminosity and stellar mass range of each specific 
galaxy sample under consideration. Evidence for this trend is also accumulated
by studies of the redshift evolution of the LF, which argue that the 
number density of massive E+S0 galaxies ($M>10^{11}M_{\sun}$) remains constant
out to $z\sim1$ \citep{Bell04,CDR06}. For less massive galaxies 
($M<10^{11}M_{\sun}$), the LF becomes progressively underpopulated with
increasing redshift, which could point to the fact that the progenitors of the
massive local E+S0 galaxies are still experiencing a period of active star 
formation. These galaxies would be too blue to fulfill the requirements in the
LF selection at higher redshift, and could gradually build up the red sequence
of massive E+S0 galaxies by a successive quenching of the star formation at
the less massive end of the LF ($M<10^{11}M_{\sun}$). 
In this picture, a fraction of blue (spiral) galaxies experience a rather slow,
gradual morphological transformation into red, `dead' quiescent E+S0 galaxies
by a progressive quenching of star formation since redshifts $z\la 1.2$.

%%%%%%%%%%%%%%%%%%%%%%%%%%%%%%%%%%%%%%%%%%%%%%%%%%%%%%%%%%%%%%%%%%%%%%%%%%%%%%%%

\section{Conclusions}

Using a combination of high signal-to-noise intermediate-resolution VLT
spectroscopy, deep $UBgRI$ ground--based photometry, and high-resolution
ACS imaging,
we have investigated the properties and evolution of 24 field early-type
(E+S0) galaxies covering redshifts $0.20<z<0.75$ down to a
magnitude limit of $M_B\le-19.30$. The targets were selected 
from the southern FDF and northern WHDF based on 
morphology, luminosity, spectrophotometric type, and
photometric redshift or broad-band colours. Scaling relationships of the
Faber-Jackson relation (FJR), and Kormendy relation as well as the Fundamental
Plane (FP) have been constructed to assess the evolution of the  
stellar populations for the galaxies in our sample. The formation
epoch for the bulk of the stars in these systems was constrained by
comparing the evolution of the $M/L$ ratios to different 
stellar population models. To reveal the origin of the star formation in some
of our galaxies, we have exploited the \textit{Chandra} and
\textit{XMM-Newton} data archives.
The main results of this work can be summarised as follows:
\begin{itemize}
\item The mass-to-light ratio of the field early-type galaxies evolves on
average as $\Delta \ln({M/L_B})=-(1.71\pm0.18)\,z$, which is faster than for
cluster early-type galaxies. This is consistent with field E+S0 galaxies
comprising on average younger stellar populations than those of their cluster
counterparts. 
\item The evolution as derived from the scaling relations of the FJR, the FP
and hence the effective mass-to-light ratio shows a strong dependence on the
mass. Less-massive systems with 
$M<2\times10^{11}M_{\sun}$ 
display a faster $M/L$ evolution of 
$\Delta\nobreak\ln({M/L_B})\nobreak=\nobreak-(1.25\pm0.30)\,z$,
whereas more-massive galaxies with $M\nobreak>\nobreak2\times10^{11}M_{\sun}$ evolve much
slower as $\Delta\nobreak\ln({M/L_B})\nobreak=\nobreak-(0.89\pm0.19)\,z$. This implies that for
the most massive field galaxies ($M\nobreak>\nobreak2\times10^{11}M_{\sun}$)
the majority of their stellar populations have been formed early at
$z_{\rm form}\nobreak=\nobreak3.5\pm1.3$. For less-massive galaxies the evolution in the $M/L$
ratios can be translated into a lower formation redshift of their stars of
$z_{\rm form}=\nobreak1.9\nobreak\pm\nobreak0.5$, which is at variance with the predictions of 
single-burst stellar population models. A fraction of 5\%--10\% in the total
stellar mass of lower-mass galaxies ($M<2\times10^{11}M_{\sun}$) must
have been formed at more recent epochs. At $z=0.4$, an accelerated evolution 
for less-massive galaxies 
($\langle\Delta\nobreak\ln({M/L_B})\rangle\nobreak=\nobreak-0.54\pm0.07$)
than for more-massive ones ($\langle\Delta \ln({M/L_B})\rangle=-0.31\pm0.06$)
is significant on the 2$\sigma$ level.
\item The morphological sub-classes of early-type galaxies, elliptical and
lenticular (S0) galaxies, appear not to represent a homogenous group but
follow different evolutionary tracks. Lenticular galaxies feature on average
a stronger luminosity evolution 
($\langle\Delta \ln({M/L_B})\rangle=-0.24\pm0.10$), 
bluer rest-frame colours and
more diverse stellar populations and hence evolve faster than elliptical
galaxies ($\langle\Delta \ln({M/L_B})\rangle=-0.09\pm0.07$).
However, the differences between elliptical and lenticular galaxies are
only moderate significant and difficult to disentangle from their individual
scatter. Note that for deriving the scatter we took into account random errors
as well as systematic uncertainties (see \S~\ref{eso}). Interestingly, the
scatter appears to be mainly amplified for lenticular galaxies.
\item Evidence for secondary star formation activity in some
of our galaxies is provided by the detection of \oii\ emission or
strong \hd\ Balmer absorption as well as bluer rest-frame $(B-I)$ colour
diagnostics. The \hd\ absorption line strengths in our galaxies implies that
the residual (low level) star formation of the galaxies accounts for
5\% to 10\% in the total stellar mass budget of these systems.
\item We have found three possible candidates that could harbour an central AGN
source. However, the trigger of the star formation in our galaxies is most
likely either stellar origin or the result of AGN feedback from
weak low luminosity (radio-mode) AGN with red optical colours \citep{CSW06,S07}.
\item Over the past $\sim$6~Gyr, there is the trend of a slow
decreasing quenching mass, which characterises a crossover in
stellar mass above star formation in galaxies gets suppressed. This suggest
that our systems experience a gradual suppression in their star formation
processes. Our results favour a scenario where
star formation is not immediately suppressed for less-massive halos and hence
works on longer time scales in lower-mass galaxies.
\end{itemize}

The evolution of the $M/L$ ratio of our field early-type galaxies favours a
downsizing formation scenario, similar to the detected decrease in luminosity
of rapidly star forming field galaxies since $z\sim$1 \citep{CSHC96,GPB96}.
In this `top-down' picture, massive galaxies are 
dominated by quiescent (red, old) stellar populations and
evolve passively on longer timescales.
Less-massive systems, however, exhibit more extended star formation histories
with mass hosting star formation processes decreasing as times evolves.
Both, the mass assembly and star formation
are accelerated in massive stellar systems that are preferably located in high
density environments. In contrast, in less-massive (smaller) systems, these
physical processes work on longer timescales.
Moreover, our results suggest a continuous mass assembly of field early-type
galaxies during the last 5 Gyr. This is supported by recent studies of field
galaxies up to $z\sim1$ \citep{vdW05,T05,BEC06,HBHE07}.

An accelerated evolution in the $M/L$ ratios of the less-massive galaxies with
respect to more-massive ones was also found for \textit{cluster} early-type
galaxies \citep{FZBSD05}.
The lack of a significant age difference between 
field and cluster early-type galaxies as expected from hierarchical formation
models \citep{DKBWSE01,DLKW04}, and the dependence of the evolution on galaxy
mass suggests that environmental effects play a rather minor role in the
formation of early-type galaxies. More important to the evolutionary history of
a stellar system are its internal properties, 
such as chemical composition or element abundance ratios.

In a next step, we will combine our precise FP measurements with an analysis of
the absorption line strengths in our field early-type galaxies. This will
provide strong constraints on the evolution of stellar populations and the
chemical enrichment history of heavy elements in these systems.

%%%%%%%%%%%%%%%%%%%%%%%%%%%%%%%%%%%%%%%%%%%%%%%%%%%%%%%%%%%%%%%%%%%%%%%%%%%%%%%%

\section*{Acknowledgments}

We would like to thank the referee Prof. G.~Gavazzi for insightful comments and
a constructive review which improved the clarity of the manuscript.
We thank the ESO Paranal staff for the professional and efficient assistance
during the spectroscopic observations 
and the FDF Team for the contributions to the first FDF sample analysis,
in particular J.~Fliri and A.~Riffeser (both USM M\"unchen) for
the cosmic ray removal on the HST/ACS images of the FDF.
We are grateful to J.~Heidt (LSW Heidelberg) for taking pre-images of the
WHDF. We also thank Prof. K.~J.~Fricke (G\"ottingen) for encouragement and
Prof. R.~Bender (USM/MPE M\"unchen) for kindly providing his FCQ program.
A.F. would like to acknowledge support from Prof. W.~W. Zeilinger and
the Institut f\"ur Astronomie der Universit\"at Wien where he
was a visiting researcher. 
This work has been supported by the Volkswagen Foundation (I/76\,520).
A.F. acknowledges partial support from grant HST-GO-10826.01 from STScI.
STScI is operated by AURA, Inc., under NASA contract NAS 5-26555.
A.B. acknowledges financial support by the BMBF/DLR (50\,OR\,0301
and 50\,OR\,0404).

%%%%%%%%%%%%%%%%%%%%%%%%%%%%%%%%%%%%%%%%%%%%%%%%%%%%%%%%%%%%%%%%%%%%%%%%%%%%%%%%

%========================= R E F E R E N C E S ===========================

%\bibliography{mnrasmnemonic,literature}

\begin{thebibliography}{}


\bibitem[\protect\citeauthoryear{Appenzeller et al.}{1998}]{App98}
Appenzeller, I., Fricke, K.~J., F\"urtig, W., et al. 1998, 
The Messenger, 94, 1

\bibitem[\protect\citeauthoryear{Balogh et al.}{1997}]{BMYCE97}
Balogh, M.~L., Morris, S.~L., Yee, H. K. C., Carlberg, R. G., \& Ellingson, E. 1997, ApJ, 488, L75

\bibitem[\protect\citeauthoryear{Baugh et al.}{1998}]{BCFL98}
Baugh, C.~M., Cole, S., Frenk, C.~S., \& Lacey, C.~G. 1998, ApJ, 498, 504

\bibitem[\protect\citeauthoryear{Bell \& de Jong}{2001}]{BdJ01}
Bell, E.~F., \& de Jong, R. S., 2001, ApJ, 550, 212 

\bibitem[\protect\citeauthoryear{Bell et al.}{2003}]{BMKW03}
Bell, E., McIntosh, D. H., Katz, N., \& Weinberg, M. D. 2003, ApJS, 149, 289

\bibitem[\protect\citeauthoryear{Bell et al.}{2004}]{Bell04}
Bell, E.~F. et al. 2004, ApJ, 608, 752

\bibitem[\protect\citeauthoryear{Bender}{1990}]{Ben:90}
Bender, R., 1990, A\&A, 229, 441

\bibitem[\protect\citeauthoryear{Bender et al.}{1998}]{BSZ98}
Bender, R., Saglia, R.~P., Ziegler, B., Belloni, P., Bruzual, G., 
Greggio, L., \& Hopp, U. 1998, ApJ, 493, 529

\bibitem[\protect\citeauthoryear{Bender et al.}{2001}]{Ben01}
Bender, R., et al. 2001, in ESO Astrophysics Symp., Deep Fields,
ed. S.~Cristiani, A.~Renzini, \& R.~E.~Williams, Springer, p.~96 

\bibitem[\protect\citeauthoryear{Benson et al.}{2002}]{BEM02}
Benson, A.~J., Ellis, R.~S., \& Menanteau, F. 2002, MNRAS, 336, 564

\bibitem[\protect\citeauthoryear{Bertin \& Arnouts}{1996}]{BA96}
Bertin, E., \& Arnouts, S. 1996,
\newblock A\&AS, 117, 393

\bibitem[\protect\citeauthoryear{Birnboim et al.}{2007}]{Bir07}
Birnboim, Y., Dekel, A., \& Neistein, E. 2007, MNRAS, 380, 339

\bibitem[\protect\citeauthoryear{Borch et al.}{2006}]{Bor06}
Borch, A., et al. 2006, A\&A, 453, 869

\bibitem[\protect\citeauthoryear{Bower et al.}{2006}]{BBM06}
Bower, R.~G., Benson, A.~J., Malbon, R., Helly, J.~C., Frenk, C.~S.,
Baugh, C.~M., Cole, S., Lacey, C.~G. 2006, MNRAS, 370, 645

\bibitem[\protect\citeauthoryear{B\"ohm et al.}{2004}]{BZS04}
B\"ohm, A., Ziegler, B.~L., Saglia, R.~P., et al.~2004, A\&A, 420, 97

\bibitem[\protect\citeauthoryear{B\"ohm \& Ziegler}{2007}]{BZ07}
B\"ohm, A., \& Ziegler, B.~L. 2007, ApJ, 668, 846 

\bibitem[\protect\citeauthoryear{Bruzual \& Charlot}{2003}]{BC03}
Bruzual, G.~A., Charlot, S. 2003, MNRAS, 344, 1000

\bibitem[\protect\citeauthoryear{Bundy et al.}{2006}]{BEC06}
Bundy, K., et al. 2006, ApJ, 651, 120

\bibitem[\protect\citeauthoryear{Bundy et al.}{2008}]{BGN08}
Bundy, K., et al. 2008, ApJ, 681, 931

\bibitem[\protect\citeauthoryear{Chapman et al.}{2004}]{CSBI04}
Chapman, S.~C., Smail, I., Blain, A.~W., \& Ivinson, R.~J. 2004, ApJ, 614, 671

\bibitem[\protect\citeauthoryear{Cimatti et al.}{2006}]{CDR06}
Cimatti, A., Daddi, E., \& Renzini, A. 2006, A\&A, 453, L29

\bibitem[\protect\citeauthoryear{Cole et al.}{2000}]{CLBF00}
Cole, S., Lacey, C.~G., Baugh, C.~M., \& Frenk, C.~S. 2000, MNRAS, 319, 168

\bibitem[\protect\citeauthoryear{Conselice et al.}{2007}]{Con07}
Conselice, C.~J., et al. 2007, MNRAS, 381, 962

\bibitem[\protect\citeauthoryear{Cowie et al.}{1996}]{CSHC96}
Cowie, L.~L., Songaila, A., Hu, E.~M, Cohen, J.~G. 1996, AJ, 112, 839

\bibitem[\protect\citeauthoryear{Croton et al.}{2006}]{CSW06}
Croton, D.~J., et al. 2006, MNRAS, 365, 11

\bibitem[\protect\citeauthoryear{De Lucia et al.}{2004}]{DLKW04}
De Lucia, G., Kauffmann, G., \& White, S.~D.~M., 2004, MNRAS, 349, 1101

\bibitem[\protect\citeauthoryear{De Lucia et al.}{2006}]{DLSWC06}
De Lucia, G., Springel, V., White, S.~D.~M., Croton, D., \&
Kauffmann, G. 2006, MNRAS, 366, 499

\bibitem[\protect\citeauthoryear{Desroches et al.}{2007}]{DQMW07}
Desroches, L.-B., Quataert, E., Ma, C.-P., \& West, A.~A. 2007, MNRAS, 377, 402

\bibitem[\protect\citeauthoryear{Diaferio et al.}{2001}]{DKBWSE01}
Diaferio, A., Kauffmann, G., Balogh, M.~L., 
White, S.~D.~M., Schade, D., \& Ellingson, E. 2001, MNRAS, 323, 999

\bibitem[\protect\citeauthoryear{di Serego Alighieri et al.}{2005}]{SVCL05}
di Serego Alighieri, S., et al. 2005, A\&A, 442, 125

\bibitem[\protect\citeauthoryear{Djorgovski \& Davis}{1987}]{DD87}
Djorgovski, S., \& Davis, M. 1987, ApJ, 313, 59

\bibitem[\protect\citeauthoryear{Dressler}{1980}]{Dre80}
Dressler, A. 1980, ApJ, 236, 351 

\bibitem[\protect\citeauthoryear{Dressler et al.}{1987}]{Dre:87}
Dressler, A., Lynden-Bell, D., Burstein, D., Davies, R.~L., Faber, S.~M.,
Terlevich, R., \& Wegner, G. 1987, ApJ, 313, 42 % (7S), FP

\bibitem[\protect\citeauthoryear{Faber \& Jackson}{1976}]{FJ76}
Faber, S.~M., \& Jackson, R.~E. 1976, ApJ, 204, 668

\bibitem[\protect\citeauthoryear{Faber et al.}{1989}]{FWBDDLT89}
Faber, S.~M., Wegner, G., Burstein, D., Davies, R.~L., Dressler, A.,
Lynden-Bell, D., \& Terlevich, R.~J. 1989, ApJS, 69, 763 

\bibitem[\protect\citeauthoryear{Faber et al.}{2007}]{Fab07}
Faber, S.~M., et al. 2007, ApJ, 665, 265

\bibitem[\protect\citeauthoryear{Fabbiano}{1989}]{Fab89}
Fabbiano, G. 1989, ARA\&A, 27, 87

\bibitem[\protect\citeauthoryear{Ferreras et al.}{2005}]{FLCSM05}
Ferreras, I., et al. 2005, ApJ, 635, 243

\bibitem[\protect\citeauthoryear{Fritz et al.}{2005}]{FZBSD05}
Fritz, A., Ziegler, B.~L., Bower, R.~G., Smail, I., \& Davies, R.~L. 2005, MNRAS, 358, 233

\bibitem[\protect\citeauthoryear{Fritz}{2006}]{F06}
Fritz, A. 2006, PhD thesis, University of G\"ottingen

\bibitem[\protect\citeauthoryear{Fritz et al.}{2006}]{FZBBBSD06}
Fritz, A., Ziegler, B.~L., B\"ohm, A., Balogh, M.~L., Bower, R.~G.,
Smail, I., \& Davies, R.~L. 2006, in {\it The Fabulous Destiny of Galaxies:
Bridging Past and Present}, eds. V. LeBrun, A. Mazure, S. Arnouts \&
D. Burgarella, Paris: Frontier Group, p. 525

\bibitem[\protect\citeauthoryear{Gavazzi et al.}{1996}]{GPB96}
Gavazzi, G., Pierini, D., Boselli, A. 1996, A\&A, 312, 397

\bibitem[\protect\citeauthoryear{Gavazzi et al.}{2002}]{GBSBS02}
Gavazzi, G., Bonfanti, C., Sanvito, G., Boselli, A., Scodeggio, M. 2002, ApJ, 576, 135

\bibitem[\protect\citeauthoryear{Gebhardt et al.}{2003}]{GFKIS03}
Gebhardt, K., Faber, S.~M., Koo, D.~C., et al. 2003, ApJ, 597, 239

\bibitem[\protect\citeauthoryear{Glazebrook et al.}{2004}]{Gla04}
Glazebrook, K., et al. 2004, Nature, 430, 181

\bibitem[\protect\citeauthoryear{Granato et al.}{2004}]{GDZSBD04}
Granato, G.~L., De Zotti, G., Silva, L., Bressan, A.,
\& Danese, L. 2004, ApJ, 600, 580

\bibitem[\protect\citeauthoryear{Heidt et al.}{2003}]{Hei03}
Heidt, J., Appenzeller, I., Gabasch, A., et al. 2003, A\&A, 398, 49

\bibitem[\protect\citeauthoryear{Hopkins et al.}{2007}]{HBHE07}
Hopkins, P.~F., Bundy, K., Hernquist, L., \& Ellis, R.~S. 2007, ApJ, 659, 976

\bibitem[\protect\citeauthoryear{Horne}{1986}]{Horne86}
Horne, K. 1986, PASP, 98, 609

\bibitem[\protect\citeauthoryear{J{\o}rgensen et al.}{1995}]{JFK95b}
J{\o}rgensen, I., Franx, M., \& Kj{\ae}rgaard, P. 1995, MNRAS, 276, 1341

\bibitem[\protect\citeauthoryear{J{\o}rgensen et al.}{1996}]{JFK96}
J{\o}rgensen, I., Franx, M., \& Kj{\ae}rgaard, P. 1996, MNRAS, 280, 167

\bibitem[\protect\citeauthoryear{J{\o}rgensen et al.}{2006}]{JCFBBD06}
J{\o}rgensen, I., Chiboucas, K., Flint, K., Bergmann, M.,
Barr, J., \& Davies, R. 2006, ApJ, 639L, 9

\bibitem[\protect\citeauthoryear{Kauffmann}{1996}]{K96}
Kauffmann, G., 1996, MNRAS, 281, 487

\bibitem[\protect\citeauthoryear{Kauffmann \& Haehnelt}{2000}]{KH00}
Kauffmann, G., \& Haehnelt, M., 2000, MNRAS, 311, 576

\bibitem[\protect\citeauthoryear{Kennicutt}{1992}]{Ken92b}
Kennicutt, R. C., Jr. 1992, ApJ, 388, 310 

\bibitem[\protect\citeauthoryear{Kewley et al.}{2004}]{KGJ04}
Kewley, L. J., Geller, M. J., \& Jansen, R. A. 2004, AJ, 127, 2002

\bibitem[\protect\citeauthoryear{Kim \& Fabbiano}{2003}]{KF03}
Kim, D.-W., \& Fabbiano, G. 2003, ApJ, 586, 826

\bibitem[\protect\citeauthoryear{Kim \& Fabbiano}{2004}]{KF04}
Kim, D.-W., \& Fabbiano, G. 2004, ApJ, 611, 846
		
\bibitem[\protect\citeauthoryear{Kinney et al.}{1996}]{KCBMSS96}
Kinney, A.~L., et al. 1996, ApJ, 467, 38

\bibitem[\protect\citeauthoryear{Kriek et al.}{2006}]{Krie06}
Kriek, M., et al. 2006, ApJ, 649, L71

\bibitem[\protect\citeauthoryear{Le Borgne et al.}{2006}]{GDDSLB06}
Le Borgne, D., et al. 2006, ApJ, 642, 48  

\bibitem[\protect\citeauthoryear{McGrath et al.}{2007}]{MSC07}
McGrath, E.~J., Stockton, A., Canalizo, G. 2007, ApJ, 669, 241

\bibitem[\protect\citeauthoryear{Metcalfe et al.}{2001}]{Met01}
Metcalfe, N., Shanks, T., Campos, A., McCracken, H.~J., \& Fong, R. 2001, 
MNRAS, 323, 779

\bibitem[\protect\citeauthoryear{Monaco et al.}{2006}]{MMBF06}
Monaco, P., Murante, G., Borgani, S., \& Fontanot, F. 2006, ApJ, 652L, 89

\bibitem[\protect\astroncite{M{\"o}ller et~al.}{2001}]{Moe01}
M{\"o}ller, C.~S., Fritze-v.Alvensleben, U., Fricke, K.~J.,
\& Calzetti, D. 2001, Ap\&SS, 276, 799

\bibitem[\protect\citeauthoryear{Noll et al.}{2004}]{NMA04}
Noll, S., et al. 2004, A\&A, 418, 885

\bibitem[\protect\citeauthoryear{O'Sullivan et al.}{2001}]{OSul01}
O'Sullivan, E., Forbes, D. A., Ponman, T. J. 2001, MNRAS, 328, 461

\bibitem[\protect\citeauthoryear{O'Sullivan et al.}{2003}]{OSul03}
O'Sullivan, E., Ponman, T. J., \& Collins, R. S. 2003, MNRAS, 340, 1375

\bibitem[\protect\citeauthoryear{Pahre}{1999}]{P99}
Pahre, M.~A., 1999, ApJS, 124, 127
 
\bibitem[\protect\citeauthoryear{Peng et al.}{2002}]{Pen02}
Peng, C.~Y., Ho, L.~C., Impey, C.~D., \& Rix, H.-W.~2002, AJ, 124, 266

\bibitem[\protect\citeauthoryear{Rusin \& Kochanek}{2005}]{RK05}
Rusin, D. \& Kochanek, C. S. 2005, ApJ, 623, 666

\bibitem[\protect\citeauthoryear{Saglia, Bender \& Dressler}{1993}]{SBD93}
Saglia, R.~P., Bender, R., \& Dressler, A. 1993, A\&A, 279, 75 (SBD93)

\bibitem[\protect\citeauthoryear{Saglia et al.}{1997}]{Sag:97a}
Saglia, R.~P., et al. 1997, ApJS, 109, 79

\bibitem[\protect\citeauthoryear{Salim et al.}{2007}]{S07}	
Salim, S., et al. 2007, ApJS, 173, 267

\bibitem[\protect\citeauthoryear{Salpeter}{1955}]{Sal55}	
Salpeter, E.~E. 1955, ApJ, 121, 161

\bibitem[\protect\citeauthoryear{Sarazin et al.}{2000}]{Sar00}	
Sarazin, C. L., Irwin, J. A., \& Bregman J. N. 2000, ApJ, 544, L101

\bibitem[\protect\citeauthoryear{Sargent et al.}{1977}]{SSBS77}
Sargent, W.~L.~W., Schechter, P.~L., Boksenberg, A., \& Shortridge, K. 1977,
ApJ, 212, 326

\bibitem[\protect\citeauthoryear{S\'{e}rsic}{1968}]{Ser:68}
S\'{e}rsic, J.~L. 1968, Atlas de Galaxias Australes (C\'{o}rdoba, Argentina:
Observatorio Astronomico, Univ. Nac. C\'{o}rdoba)

\bibitem[\protect\citeauthoryear{Simard et al.}{2002}]{SWVSP02}
Simard, L., et al. 2002, ApJS, 142, 1

\bibitem[\protect\citeauthoryear{Sirianni et al.}{2005}]{SJB05}
Sirianni, M., Jee, M.~J., Ben{\'i}tez, N., et al. 2005, PASP, 117, 1049

\bibitem[\protect\citeauthoryear{Somerville \& Primack}{1999}]{SP99}
Somerville, R.~S., \& Primack, J.~R. 1999, MNRAS, 310, 1087

\bibitem[\protect\citeauthoryear{Somerville et al.}{2001}]{SPF01}
Somerville, R.~S., Primack, J.~R., \& Faber, S.~M. 2001, MNRAS, 320, 504

\bibitem[\protect\citeauthoryear{Thomas et al.}{2005}]{TMBO05}
Thomas, D., Maraston, C., Bender, R., de Oliveira, C.~M. 2005, ApJ, 621, 673

\bibitem[\protect\citeauthoryear{Tinsley}{1980}]{Tin80}
Tinsley, B.~M. 1980, Fundamentals of Cosmic Physics, 5, 287

\bibitem[\protect\citeauthoryear{Tonry \& Davis}{1979}]{TD79}
Tonry, J., \& Davis, M. 1979, AJ, 84, 1511	    

\bibitem[\protect\citeauthoryear{Treu et al.}{2001}]{Treu01a}
Treu, T., Stiavelli, M., Bertin, G., Casertano, S., \& M{\o}ller, P. 2001
MNRAS, 326, 237

\bibitem[\protect\citeauthoryear{Treu et al.}{2002}]{TSCMB02}
Treu, T., Stiavelli, M., Casertano, S., M{\o}ller, P., \& Bertin, G. 2002,
ApJ, 564, L13

\bibitem[\protect\citeauthoryear{Treu et al.}{2005}]{T05}
Treu, T., et al. 2005, ApJ, 633, 174

\bibitem[\protect\citeauthoryear{van de Ven et al.}{2003}]{vdV03}
van de Ven, G., van Dokkum, P.~G., Franx, M. 2003, MNRAS, 344, 924

\bibitem[\protect\citeauthoryear{van der Wel et al.}{2005}]{vdW05}
van der Wel, A., Franx, M., van Dokkum, P.~G., Rix, H.-W., Illingworth, G.~D.,
\& Rosati, P. 2005, ApJ, 631, 145

\bibitem[\protect\citeauthoryear{van Dokkum \& Franx}{1996}]{vDF96}
van Dokkum, P.~G., \& Franx, M. 1996, MNRAS, 281, 985

\bibitem[\protect\citeauthoryear{van Dokkum \& Franx}{2001b}]{vD01b}
van Dokkum, P.~G., \& Franx, M. 2001b, ApJ, 553, 90

\bibitem[\protect\citeauthoryear{van Dokkum et al.}{2001a}]{vDFKI01}
van Dokkum, P.~G., Franx, M., Kelson, D.~D., \& Illingworth, G.~D., 2001a,
ApJ, 553, L39

\bibitem[\protect\citeauthoryear{van Dokkum \& Ellis}{2003}]{vDE03}
van Dokkum, P.~G. \& Ellis, R.~S. 2003, ApJ, 592, L53

\bibitem[\protect\citeauthoryear{van Dokkum \& Stanford}{2003}]{vDS03}	
van Dokkum, P.~G., \& Stanford, S.~A. 2003, ApJ, 585, 78

\bibitem[\protect\citeauthoryear{Weisskopf et al.}{2002}]{Weis02}
Weisskopf, M. C., Brinkman, B., Canizares, C., Garmire, G., Murray, S., 
Van Speybroeck, L. P. 2002, PASP, 114, 1

\bibitem[\protect\citeauthoryear{Woo et al.}{2004}]{WULMM04}
Woo, J.~H, Urry, C.~M., Lira, P., van der Marel, R.~P., \& Maza, J.
2004, ApJ, 617, 903

\bibitem[\protect\citeauthoryear{Woo et al.}{2005}]{WUMLM05}
Woo, J.~H, Urry, C.~M., van der Marel, R.~P., Lira, P., \& Maza, J.
2005, ApJ, 631, 762

\bibitem[\protect\citeauthoryear{Worthey \& Ottaviani}{1997}]{WO97}
Worthey, G., \& Ottaviani D.~L. 1997, ApJS, 111, 377

\bibitem[\protect\citeauthoryear{Ziegler et al.}{2002}]{ZBF02} 
Ziegler, B.~L., B\"ohm, A., Fricke, K.~J., et al. 2002, ApJ, 564, L69

\bibitem[\protect\citeauthoryear{Ziegler et al.}{2005}]{ZTBBFM05}
Ziegler, B.~L., Thomas, D., B\"ohm, A., Bender, R.,
Fritz, A., \& Maraston, C. 2005, A\&A, 433, 519

 
\end{thebibliography}

%%%%%%%%%%%%%%%%%%%%%%%%%%%%%%%%%%%%%%%%%%%%%%%%%%%%%%%%%%%%%%%%%%%%%%%%%%%%%%%%

\appendix

\section{Verifications of \boldmath{$\sigma$} Measurements}\label{sigcheck}

\subsection{\label{compfea}Different Absorption Passbands}

\begin{figure} 
\includegraphics[width=1.0\linewidth]{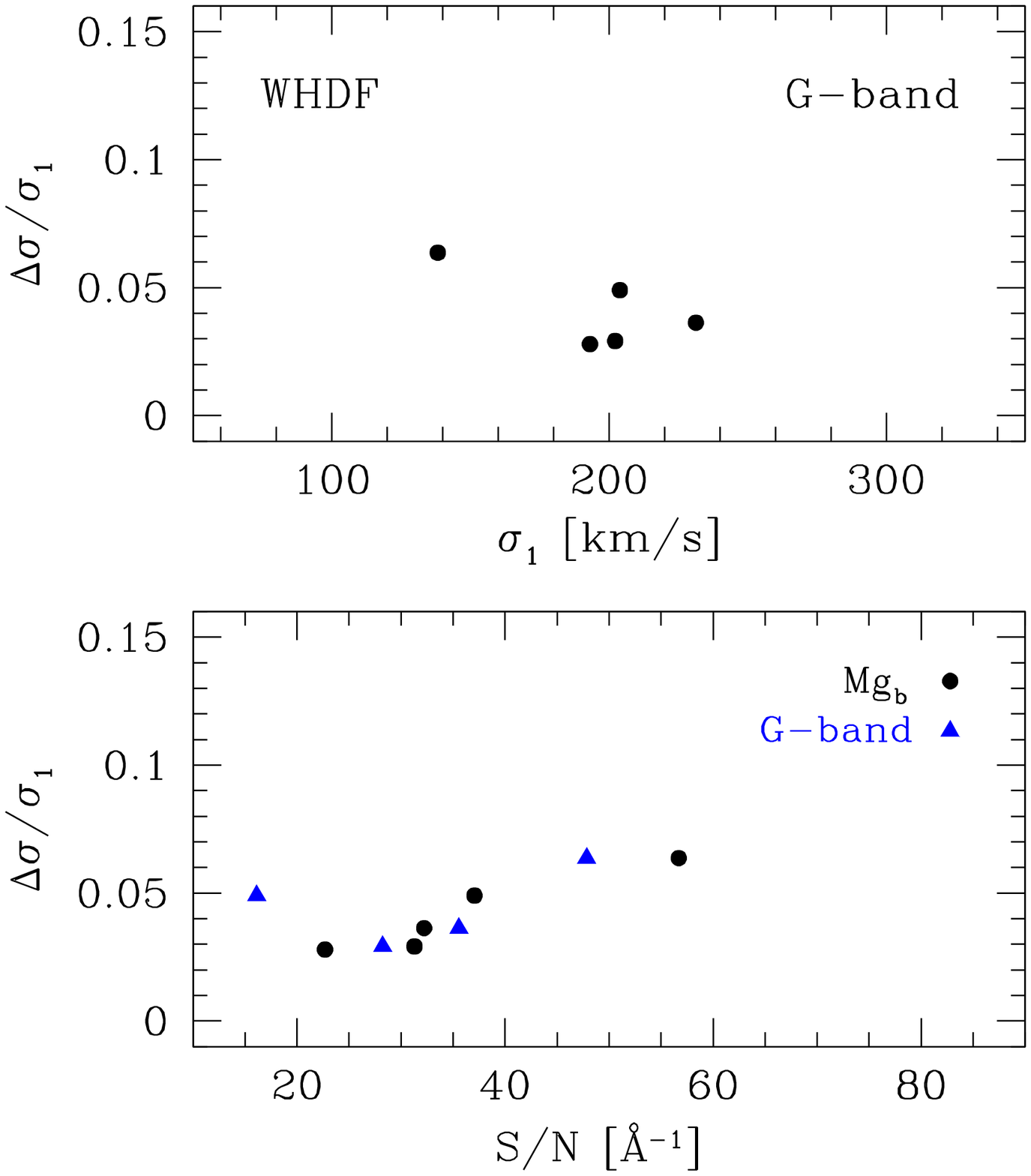}
\caption{\label{cmprdsmgw}\small{Comparison of velocity dispersion measurements
derived from different absorption features. \emph{Upper panel:} relative
uncertainty of velocity dispersions with the G--band as a function of
velocity dispersions $\sigma_{1}$ of early-type field galaxies in the WHDF. 
All galaxies are high quality $\sigma$ measurements. 
\emph{Lower panel:} $\Delta \sigma/\sigma_{1}$ versus the $S/N$ per
\AA. Measurements are splitted according to the $S/N$ of the feature 
(\mgb\ circles, G--band triangles) to visualise possible systematic errors of
low $S/N$ on $\sigma$. For $S/N\ge8$ \AA$^{-1}$, effects of systematic
uncertainties become negligible.}}
\end{figure}

To test the accuracy and the possible internal systematic effects
of velocity dispersion measurements using different absorption features,
galaxies where multiple $\sigma$ determinations could be obtained 
are compared. In case of five field early-type galaxies in the WHDF
$\sigma$ was derived based on the \mgb\ and the G-band feature.
Figure~\ref{cmprdsmgw} shows the formal relative uncertainty 
$\Delta \sigma/\sigma_{1}$ as a function of velocity dispersion (upper panel)
and as a function of the $S/N$ (lower panel) per \AA\ for the early--type
galaxies in the WHDF where both measurements are available. The lower panel is
divided according to the $S/N$ of the feature of interest, $S/N$ values from the
\mgb\ lines are denoted as circles, whereas the $S/N$ based on the G--band is
indicated as triangles. $\Delta \sigma/\sigma_{1}$  was computed as 
the relative difference in velocity dispersion values of the \mgb\
with the G-band line, normalised to the reference velocity dispersions
$\sigma_{1}$. Both features were free from any contamination due to
sky lines and did not suffer from any residual sky background artefacts.
The relative uncertainties cover the range
$0.03\le \Delta \sigma/\sigma_{1}\le 0.06$ with a median of
$\langle\Delta \sigma/\sigma_{1}\rangle=0.05$. The $\sigma$ determinations are
in extremely good agreement and show similar absolute errors between 8 to 18 \kms.
Even for the lowest $S/N$ ratios the formal uncertainty of 5\% is less 
than the typical individual error in the $\sigma$ measurement. 
This limit is also the smallest formal uncertainty that can be reached with
the instrumental resolution and the template stars.
No trend of increasing systematic errors on the $\sigma$ values with low $S/N$
can be found. 
Results from the Monte-Carlo simulations showed that the
effects of systematic errors on $\sigma$ become important at 
$S/N<8$ per \AA\  (cf. \S~\ref{sec:sig}). As this limit is below the
average $S/N$ in the spectra of our low--luminosity galaxies, the formal
uncertainty therefore becomes negligible even for the lowest derived velocity
dispersions in the galaxy samples. A comparison of the absolute velocity
dispersions for the WHDF ellipticals is presented in Figure~\ref{cmpsidou}
in the next section.

\subsection{\label{double}Repeat Observations}

\begin{figure}
\includegraphics[width=1.0\linewidth]{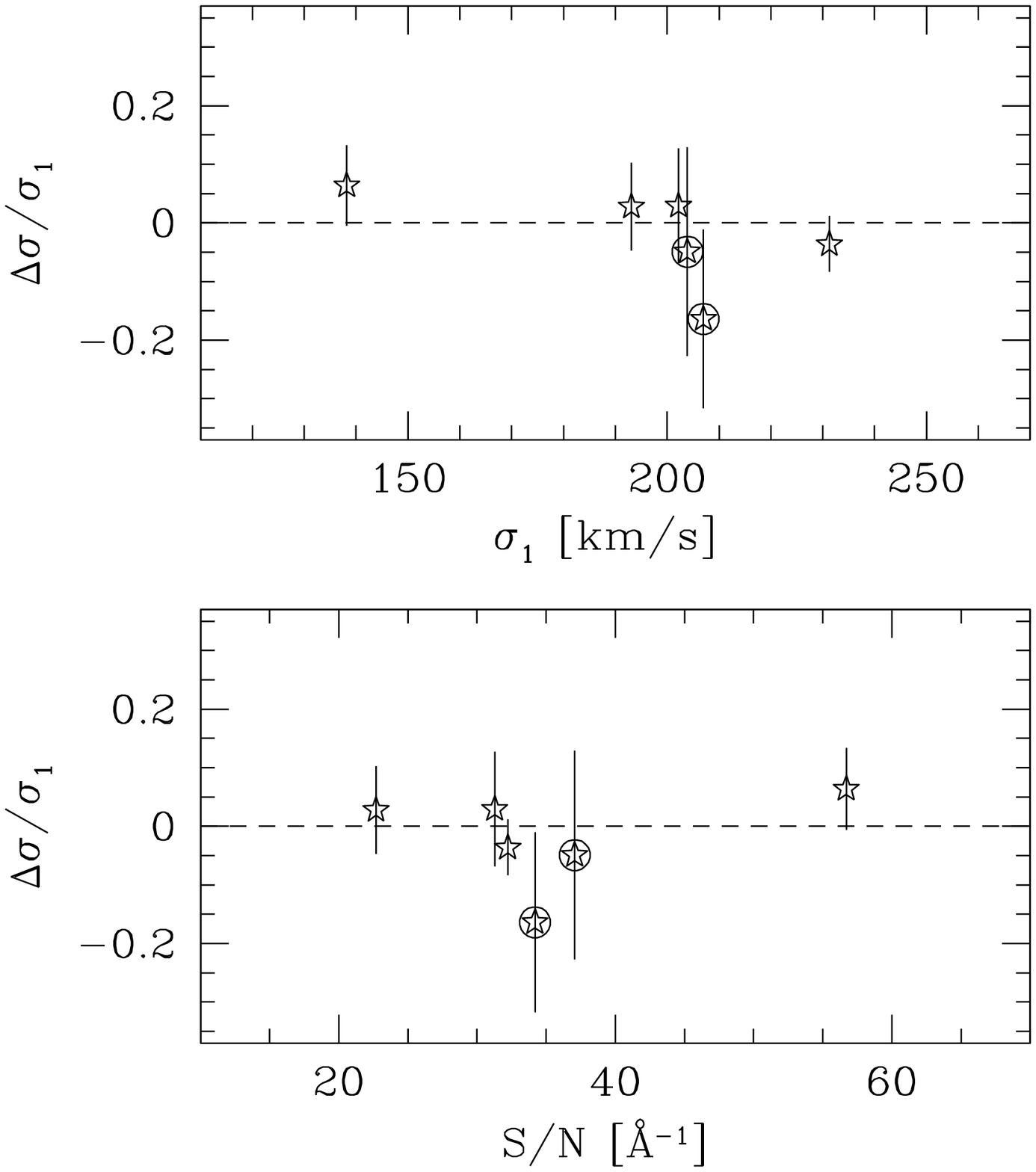}
\caption{\label{cmpsidou}\small{Comparison of velocity dispersion measurements
for FORS2 spectra of field galaxies in the WHDF which have been observed with
two different MOS mask setups and slitlet positions.
One galaxy with four measurements is denoted with two additional large circles.
}}
\end{figure}

For a total of five field galaxies in the WHDF sample which are included in two
or more different MOS masks, it is possible to confirm the internal reliability
of the spectral reduction and analysis. 
A total of nine repeat observations was acquired for the WHDF ellipticals. Two
galaxies were observed thrice (ID\,92 and ID\,111) and ID\,173 four times.
The velocity dispersions have been measured for both
individual setups and the agreement between these two determinations is very
good with a median offset in $\sigma$ at the 4\% level.
Figure~\ref{cmpsidou} compares the velocity dispersion measurements for
the five WHDF field galaxy spectra which were observed with two or up to
four different mask setups.
One galaxy with four repeat observations is denoted with two
additional large circles. In general, velocity dispersions of the secondary
measurements are based on lower $S/N$ spectra and therefore show larger errors.
The final measurement with the smallest error yielded 
$\sigma_{173}=204\pm16$~\kms\  ($S/N=37$), which represents also the
smaller deviation with respect to the reference line.
For the WHDF field ellipticals the relative difference between reference
and secondary measurements is 
$\delta \sigma/\sigma_{1}\nobreak=\nobreak0.041\nobreak\pm\nobreak0.015$
with $23\nobreak\leq S/N\nobreak\leq\nobreak57$.
In order to increase the signal--to--noise in the final spectra, these
repeat observations have been co--added. 
The repeated $\sigma$ measurements in different slitlet configurations
prove that our kinematic measurements are very stable and that
internal effects such as varying spectral wavelength range and slitlet
position on the FORS spectrograph can be neglected.

\section{X-ray Populations}\label{xrayp}

Early-type galaxies in the nearby universe with luminosities typically
$L_X\ga10^{42}$\,ergs\,s$^{-1}$ most likely harbour an AGN source in their
centre \citep{OSul01}. Below this flux limit at $L_X<10^{42}$\,ergs\,s$^{-1}$,
the main contributions to the total X-ray luminosity of a galaxy could be a
variety of sources, such as from low-luminosity AGN (LLAGN), high-mass and
low-mass X-ray (stellar) binaries (LMXBs) and emission fluxes from the diffuse
soft X-ray plasma that surrounds the galaxy ($T_{\rm eff}<1$\,keV).

LMXBs are associated with evolved old stellar populations (A stars) and a
common feature in local bulge systems in the field and globular clusters.
As both E and S0 galaxies comprise homogenous stellar populations, their X-ray
binaries (XRB) populations should also be uniformly distributed
among the galaxy populations. In the past
it was proposed that LMXBs might even dominate the global X-ray emission of
early-type galaxies \citep{Fab89}. However, this issue is hard to quantify as
LMXBs cannot be identified as individual point sources. On the other hand for
fainter X-ray systems observations argue that LMXBs account for a significant
high fraction (from 20\% up to 50\%) of the X-ray emission of the host galaxy
\citep{Sar00,KF03}, which becomes a non-negligible
factor when estimating the dynamical masses of their stellar host systems.

In oder to identify in our sample of galaxies with possible associated
X-ray counterparts (see \S~\ref{xray}) those systems with nuclear activity or
ongoing star formation, an effective way would be to classify our galaxies
schematically based on their measured X-ray luminosity. Using the X-ray flux
measurements for our objects, we can give constraints on the various sources
that contribute to the X-ray emission of our galaxies.
The two main groups of interests are systems with nuclear activity
based on the X-ray luminosity (XAGN) and galaxies with low-luminosity nuclear
activity (LLAGN) and/or ongoing star formation. 

For the old quiescent stellar populations in early-type galaxies, pointlike
X-ray sources are very likely to be entirely resembled by LMXBs. The
contribution due to LMXBs is direct proportional to the stellar mass of a
galaxy. In general, these binary systems ($\sim 30-140$ sources per galaxy)
account for a smaller but non-negligible fraction ($\la$25\%) of the total
galaxy X-ray luminosity \citep{OSul01,KF04}.
Using the relationship between total $B$-band and X-ray 
luminosity by \cite{KF04}
as $L_X$(LMXB)=$(0.90\pm0.50)\times 10^{30}\,L_B/L_{B,\odot}$
(in units of \,ergs\,s$^{-1}$), we measure an average contribution of LMXBs 
of $\leq$17\%$\pm$7\%. The evolution corrected absolute 
$B$-band luminosities of our galaxies fall in the range
$9.95<{\rm log}\,(L_B/L_{B,\odot})<11.30$ with a median of 10.47.
As a comparison, introducing our $K$-band measurements of
$10.68<{\rm log}\,(L_K/L_{K,\odot})<11.79$ (average 11.22) and a
similar relation as above but for the $K$-band luminosity of
$L_X$(LMXB)=$(0.20\pm0.08)\times 10^{30}\,L_K/L_{K,\odot}$ (ergs\,s$^{-1}$), 
we detect an LMXB source contribution of $\leq$18\%$\pm$6\%. 

Local early-type galaxies show a correlation of their soft X-ray
luminosity with their stellar $L_B$ \citep[e.g.,][]{OSul01}. Excluding dwarf
galaxies, brightest cluster/group galaxies and galaxies with AGN hosts,
\cite{OSul01} found for 184 nearby E+S0 galaxies 
log($L_X$)=$(2.17\pm0.11)$log\,$L_B+17.98\pm1.12$. Because of the evolved
stellar populations in the local E+S0 galaxies, any detected excess with
respect to this empirical relation would imply a measure of star formation
processes or LLAGNs in our distant galaxies. Applying the above relationship 
to our sample, we derive an excess in diffuse plasma fraction of 
$\leq$26\%$\pm$15\% compared to the local value. Using instead the relation by
\cite{OSul03} $L_X(0.5-2.0\,{\rm keV})=4\times 10^{41}\,(L_B/10^{11})^{2.7}$,
the fraction of soft X-ray plasma decreases to $\leq$7\%$\pm$3\%. Both
measurements give evidence for a weak excess in the X-ray luminosities in our
distant galaxies ($10^{40}<L_X<10^{42}$) compared to the local reference. 
The increased luminosities are most likely the result of a combination of star
formation and/or LLAGN. However, a possible contribution from LMXBs and soft
diffuse X-ray plasma emission to the X-ray luminosities cannot be ruled out
completely. To clearly distinguish between the various contributions a detailed
analysis of the X-ray spectra is necessary which goes beyond the scope of the
present work.

%%%%%%%%%%%%%%%%%%%%%%%%%%%%%%%%%%%%%%%%%%%%%%%%%%%%%%%%%%%%%%%%%%%%%%%%%%%%%%%%

\normalsize

\bsp
\label{lastpage}

\end{document}